\documentclass[aps,twocolumn,superscriptaddress,groupedaddress,nofootinbib]{emulateapj}  

\usepackage{amssymb,amsmath,amsfonts,amsbsy}
\usepackage{color}
\usepackage{enumerate}
\usepackage{graphicx}
\usepackage{pifont}
\RequirePackage[colorlinks,citecolor=blue,urlcolor=blue]{hyperref}

\usepackage{kotex}

\newcommand{\parallelsum}{\mathbin{\|}}

\begin{document}

\title{Cosmological Parameter Estimation from the Two-Dimensional Genus Topology - Measuring the Expansion History using the Genus Amplitude as a Standard Ruler}

\author{Stephen Appleby}\email{stephen.appleby@apctp.org}
\affiliation{Asia Pacific Center for Theoretical Physics, Pohang, 37673, Korea}
\affiliation{Quantum Universe Center, Korea Institute for Advanced Study, 85 Hoegiro, Dongdaemun-gu, Seoul 02455, Korea}
\author{Changbom Park}
\affiliation{School of Physics, Korea Institute for Advanced Study, 85
Hoegiro, Dongdaemun-gu, Seoul, 02455, Korea}
\author{Sungwook E. Hong (홍성욱)}
\affiliation{Korea Astronomy and Space Science Institute, 776 Daedeokdae-ro, Yuseong-gu, Daejeon 34055, Korea}
\affiliation{Natural Science Research Institute, University of Seoul, 163
Seoulsiripdaero, Dongdaemun-gu, Seoul, 02504, Korea}
\author{Ho Seong Hwang}
\affiliation{Korea Astronomy and Space Science Institute, 776 Daedeokdae-ro, Yuseong-gu, Daejeon 34055, Korea}
\author{Juhan Kim}
\affiliation{Center for Advanced Computation, Korea Institute for Advanced
Study, 85 Hoegiro, Dongdaemun-gu, Seoul, 02455, Korea}
\author{Motonari Tonegawa}
\affiliation{School of Physics, Korea Institute for Advanced Study, 85
Hoegiro, Dongdaemun-gu, Seoul, 02455, Korea}

\begin{abstract}
We measure the genus of the galaxy distribution in two-dimensional slices of the SDSS-III BOSS catalog to constrain the  cosmological parameters governing the expansion history of the Universe. The BOSS catalogs are divided into twelve concentric shells over the redshift range $0.25 < z < 0.6$ and we repeatedly measure the genus from the two-dimensional galaxy density fields, each time varying the cosmological parameters used to infer the distance-redshift relation to the shells. We also indirectly reconstruct the two-dimensional genus amplitude using the three-dimensional genus measured from SDSS Main Galaxy Sample with galaxies at low redshift $z < 0.12$.  We combine the low- and high--redshift measurements, finding the cosmological model which minimizes the redshift evolution of the genus amplitude, using the fact that this quantity should be conserved. Being a distance measure, the test is sensitive to the matter density parameter ($\Omega_{\rm m}$) and equation of state of dark energy ($w_{\rm de}$). We find a constraint of $w_{\rm de} = -1.05^{+0.13}_{-0.12}$, $\Omega_{\rm m} = 0.303 \pm 0.036$ after combining the high- and low--redshift measurements and combining with Planck CMB data. Higher redshift data and combining data sets at low redshift will allow for stronger constraints. 
\end{abstract}

\maketitle

\section{Introduction} 

Topological statistics have a long history of use within cosmology \cite{1990ApJ...352....1G,1991ApJ...378..457P,Mecke:1994ax,Schmalzing:1997aj,Schmalzing:1997uc,Hikage:2006fe,Ducout:2012it,1989ApJ...345..618M,1992ApJ...387....1P,1992ApJ...385...26G}. Theoretical studies of random fields have been undertaken; both Gaussian \citep{1970Ap......6..320D,Adler,Gott:1986uz,Hamilton:1986,Ryden:1988rk,
1987ApJ...319....1G,1987ApJ...321....2W}
  and perturbatively non-Gaussian \citep{Matsubara:1994wn,Matsubara:1994we,Matsubara:1995dv,1988ApJ...328...50M,Matsubara:1995ns,
2000astro.ph..6269M,10.1111/j.1365-2966.2008.12944.x,Pogosyan:2009rg,Gay:2011wz,Codis:2013exa}, and modern techniques are being developed that go beyond the standard Minkowski functional analysis; Minkowski tensors \cite{Beisbart:2001vb,Beisbart:2001gk, Ganesan:2016jdk, Chingangbam:2017sap, Kapahtia:2019ksk, Appleby:2017uvb, Appleby:2018tzk, Kapahtia:2017qrg, K.:2018wpn}, Betti numbers \cite{Park:2013dga,Feldbrugge:2019tal,Pranav:2018lox,Pranav:2018pnu,Pranav:2016gwr,Shivshankar:2015aza,vandeWeygaert:2011ip} and multi-scale analyses of the cosmic web \cite{10.1111/j.1365-2966.2011.18395.x, Codis:2018niz, Kraljic:2019acs}. Previous application of the Minkowski functionals to various modern data sets can be found in \citep{2001ApJ...553...33P,Hikage:2002ki,Hikage:2003fc,Park:2005fk,10.1111/j.1365-2966.2008.14358.x,Gott:2008kk,Choi:2010sx,Zhang:2010tha,Petri:2013ffb,Blake:2013noa,Wiegand:2013xfa,2014ApJ...796...86P,Wang:2015eua,Wiegand:2016ezl,Buchert:2017uup,Sullivan:2017mhr,Hikage_2001,Gott:2006yy}. The advent of cosmological scale, large scale structure data has allowed measurements of the higher point functions induced by gravitational collapse  \cite{Wiegand:2013xfa,Wiegand:2016ezl,Buchert:2017uup,Sullivan:2017mhr} that would be difficult to extract using conventional $N$-point methods.

The genus belongs to the family of Minkowski Functionals. The genus of the matter density field, as traced by galaxies, can be used as a cosmological probe. By measuring the genus curve at different redshifts, one can extract information regarding the parameters governing the expansion history of the Universe. The redshift dependence of the genus amplitude was originally proposed as a standard ruler in \citet{Park:2009ja,doi:10.1111/j.1365-2966.2010.18015.x}. For the $\Lambda$CDM model, the amplitude of the genus curve is related to the slope of the linear matter power spectrum, which does not evolve with redshift. By comparing this quantity at high and low redshift, we should detect no evolution. However, if we select an incorrect cosmological model to infer the distance-redshift relation, then comoving smoothing scales and volumes become systematically incorrect with increasing redshift. This will generate a spurious evolution in the statistic. Hence, by measuring the genus using different cosmological models to infer distance scales, one can find the expansion history that conserves this statistic.

This cosmological test was first proposed in \citet{Park:2009ja}. More recently, the authors have revisited this possibility and applied the method to projected two dimensional galaxy density fields, using all-sky mock galaxy lightcone data \citep{Appleby:2017ahh,Appleby:2018jew}. The analysis presented here provides a conclusion of these works, as we apply the methodology to a combination of low- and high-redshift galaxy catalogs to obtain a constraint on the cosmological parameters $\Omega_{\rm m}$ and dark energy equation of state $w_{\rm de}$. This test was pursued in \cite{Blake:2013noa}, with the first direct application of the method to galaxy data (specifically the WiggleZ survey \cite{Blake:2011wn}). Competitive distance measurements were obtained from three-dimensional Minkowski functional measurements, and issues associated with this measurement (principally sparse sampling) were highlighted. The conclusion of the work was that topology is potentially competitive with Baryon Acoustic Oscillations (BAO) as a standard ruler, however the physics and assumptions that go into the analysis are more involved, as the Minkowski functionals measure the shape of the full extent of the power spectrum in an integrated sense.  

In this work we measure the genus of both the BOSS, LOWZ and CMASS galaxy catalogs \citep{2015ApJS..219...12A} and the SDSS Main Galaxy Sample (SDSS MGS) \citep{2009ApJS..182..543A}. The low redshift SDSS MGS data provides a robust measure of the genus amplitude at low redshift, practically insensitive to the distance-redshift relation. In contrast, the higher redshift BOSS data will be sensitive to our choice of cosmological parameters when inferring distances. If we select an incorrect distance-redshift relation, the genus amplitude extracted from the BOSS data will systematically evolve, relative to the low redshift measurement. The reason for this effect is that an incorrect choice of comoving distance will cause us to select erroneous smoothing scales and effective areas, meaning that we will be measuring the slope of the matter power spectrum at different scales as a function of redshift. As the matter power spectrum is not scale invariant, this will manifest as an evolving genus amplitude.

The principal challenge when using the genus as a standard ruler is that we must compare high redshift measurements to low redshift counterparts. However, the low redshift Universe is restricted in volume and the statistical uncertainty provides the dominant limitation on parameter constraints. To mitigate this problem, we measure the genus of the full three-dimensional field at low redshift. We then convert the three-dimensional measurement into a constraint on the theoretical expectation of the two-dimensional genus amplitude.

The paper will proceed as follows. In Section \ref{sec:theory} we discuss some of the issues associated with using the genus amplitude as a standard ruler, and our method of extracting this quantity from galaxy data. We briefly review the extraction of the genus from two-dimensional shells of BOSS data in Section \ref{sec:obs}. In Section \ref{sec:3D} we detail the data, mask, mock catalogs and systematics associated with SDSS MGS measurement of the three-dimensional genus. The conversion from three dimensional measured genus to the theoretical expectation value of the two-dimensional genus amplitude is explained in Section \ref{sec:conv}. Finally in Section \ref{sec:constraints} we place constraints on cosmological parameters, then close with a discussion in Section \ref{sec:discuss}. 

This work is a companion to \cite{Appleby:2020pem}, which uses the absolute value of the genus amplitude (rather than its evolution with redshift, as in this work) to place constraints on the shape of the matter power spectrum. We discuss the relation between the two approaches in Section \ref{sec:discuss}.

\section{Genus amplitude as a standard ruler}
\label{sec:theory}
  
The two-dimensional genus of a perturbatively non-Gaussian field without boundary is given by the so-called Edgeworth expansion \cite{Matsubara:1994we,2000astro.ph..6269M,Pogosyan:2009rg,Gay:2011wz,Codis:2013exa}

\begin{eqnarray} \nonumber & &   g_{\rm 2D}(\nu_{\rm A}) = A_{\rm G}^{(\rm 2D)} e^{-\nu_{\rm A}^{2}/2} \left[ H_{1}(\nu_{\rm A})+ \left[ {2 \over 3} \left( S^{(1)} - S^{(0)}\right) \times \right. \right. \\
\label{eq:mat1} & & \quad \left. \left. H_{2}(\nu_{\rm A}) +  
  {1 \over 3} \left(S^{(2)} - S^{(0)}\right)H_{0}(\nu_{\rm A}) \right] \sigma_{0} + {\cal O}(\sigma_{0}^{2}) \right] , \end{eqnarray} 

\noindent where $A_{\rm G}^{(\rm 2D)}$ is the amplitude

\begin{eqnarray} 
\label{eq:ag} & & A_{\rm G}^{(\rm 2D)} \equiv  {1 \over 2(2\pi)^{3/2}} {\sigma_{1}^{2} \over \sigma_{0}^{2}} , \end{eqnarray}

\noindent and the skewness parameters $S^{(0)}, S^{(1)}, S^{(2)}$ are related to the three point cumulants and will not be used here. $\sigma_{1}$ and $\sigma_{0}$ are defined as integrals over the power spectrum, in this work smoothed with Gaussian kernels of comoving scale $R_{\rm G}$

\begin{eqnarray}
\label{eq:s02} & & \sigma_{0}^{2} = {1 \over (2\pi)^{2}}\int d^{2} k_{\perp} e^{-k_{\perp}^{2}R_{\rm G}^{2}} P_{\rm 2D}(k_{\perp},z)   , \\
\label{eq:s12} & & \sigma_{1}^{2} = {1 \over (2\pi)^{2}}\int d^{2} k_{\perp} k_{\perp}^{2} e^{-k_{\perp}^{2}R_{\rm G}^{2}} P_{\rm 2D}(k_{\perp},z), 
  \end{eqnarray}

\noindent and the projected two-dimensional power spectrum $P_{\rm 2D}(k_{\perp},z)$ is related to its full three-dimensional counterpart according to

\begin{equation}\label{eq:p2d} P_{\rm 2D}(k_{\perp},z) = {2 \over \pi} \int dk_{\parallelsum} P_{\rm 3D}\left(k,z\right) {\sin^{2} [k_{\parallelsum} \Delta] \over k_{\parallelsum}^{2} \Delta^{2}}  , \end{equation} 

\noindent  where $\Delta$ is the comoving thickness of the two-dimensional slices of the field. ${\vec k}_{\perp}$ and $k_{\parallelsum}$ are the wave numbers perpendicular and parallel to the line of sight respectively. The three-dimensional power spectrum of the density field that is traced by galaxies is the sum of the redshift-space distorted matter field and a shot noise contribution

\begin{equation}\label{eq:p3df} P_{\rm 3D} (k,k_{\parallelsum},z) = b^{2} \left( 1 + \beta {k_{\parallelsum}^{2} \over k^{2}}\right)^{2} P_{\rm m}(z, k) + P_{\rm SN} ,  \end{equation}

\noindent where $P_{\rm m}(z,k)$ is the matter power spectrum at redshift $z$, $P_{\rm SN}$ is the shot noise power spectrum $P_{\rm SN} =1/\bar{n}$, where $\bar{n}$ is the number density of galaxies. We introduce $\beta = f/b$, $b$ is the linear galaxy bias and $f$ is the growth factor. The quantity $\nu_{A}$ is the density threshold such that the excursion set has the same area fraction as a corresponding Gaussian field - 

\begin{equation}\label{eq:afrac} f_{A} = {1 \over \sqrt{2\pi}} \int^{\infty}_{\nu_{A}} e^{-t^{2}/2} dt , \end{equation}

\noindent where $f_{A}$ is the fractional area of the field above $\nu_{A}$. This choice of $\nu_{\rm A}$ parameterization eliminates the non-Gaussianity in the one-point function \citep{1987ApJ...319....1G,1987ApJ...321....2W,1988ApJ...328...50M}.

For the case of a Gaussian field, the genus amplitude is a measure of the shape of the linear matter power spectrum $P_{\rm m}(z,k)$, which is a conserved quantity for the $\Lambda$CDM model and certain generalisations (such as $w$CDM, assuming dark energy perturbations are negligible). If we use an incorrect cosmological model to infer the distance-redshift relation, then we get the smoothing scale $R_{\rm G}$ and volume occupied by galaxy data systematically wrong at different redshifts. Hence we will measure the shape of the power spectrum at different scales when using an incorrect expansion history. As a result, the genus amplitude that we extract from the data will spuriously evolve with redshift if we get the expansion history wrong. A low redshift measurement will represent the `true' genus amplitude having little dependence on the cosmology adopted, against which high redshift measurements can be compared. This effect was predicted in \citep{Park:2009ja} and explicitly measured using mock galaxies in \cite{Appleby:2018jew}.

In reality a number of small systematic effects are present in real galaxy data that generate redshift evolution of this statistic. The primary sources of contamination are as follows, listed in order of severity 

\begin{enumerate}
\item{We bin galaxies into redshift shells and apply a mass cut to fix the number density of tracers at each redshift to be constant, thus fixing a constant shot noise power spectrum $P_{\rm SN}$ in each shell. In contrast, the amplitude of the matter power spectrum $P_{\rm m}(z,k)$ decreases with redshift. It follows that the relative importance of the shot noise contribution in ($\ref{eq:p3df}$) will increase with redshift, which will manifest as an increasing genus amplitude at higher $z$. This effect depends on $R_{\rm G}$ relative to the mean galaxy separation $\bar{r}$, and is negligible for $R_{\rm G} \gg \bar{r}$.}
\item{Linear redshift space distortion decreases the amplitude of the two-dimensional genus by around $\sim 9\%$, roughly constant over the redshift range $0 < z < 0.7$. However, it also introduces a mild $\sim 1\%$ redshift dependent evolution, decreasing the amplitude with increasing redshift. This is due to the redshift dependence of $\beta(z)$ in equation ($\ref{eq:p3df}$).}
\item{Non-linear gravitational evolution will typically act to decrease the genus amplitude with decreasing redshift, which is an ${\cal O}(\sigma_{0}^{2})$ effect (so-called gravitational smoothing \cite{1989ApJ...345..618M,1991ApJ...378..457P,2005ApJ...633....1P}). }
\end{enumerate} 

The magnitude of each of these effects depends on the number density of galaxies, the smoothing scales perpendicular and parallel to the line of sight ($R_{\rm G}$ and $\Delta$) and the area of the data. In Appendix A we use mock galaxy lightcone data to examine these effects in isolation, and argue that for the data and smoothing scales used in this work, no significant redshift evolution of the genus amplitude will be induced.

To briefly summarise the results in Appendix A : We take constant comoving scale $R_{\rm G} = 20 {\rm Mpc}$ to Gaussian smooth the data perpendicular to the line of sight, and comoving slice thickness $\Delta = 80 \, {\rm Mpc}$ along the line of sight. At these scales, the redshift space distortion and shot noise effects both introduce an evolution of the genus amplitude of order $\sim 1\%$ over the redshift range $0 < z < 0.7$. Shot noise/redshift space distortion causes the genus amplitude to increase/decrease with increasing $z$. The two competing effects effectively cancel for the particular galaxy sample considered in this work. Furthermore, the mean galaxy separation of the two-dimensional projected fields is approximately  $\bar{r} \simeq 15 \, {\rm Mpc}$, smaller than $R_{\rm G} = 20 \, {\rm Mpc}$. This makes the non-Gaussianity of the shot noise contribution small. 

The non-Gaussian gravitational corrections to the amplitude are small. We quantify this statement by measuring the next-to-leading-order correction term $a_{3}H_{3}(\nu_{\rm A})$, finding it to be $\sim {\cal O}(1\%)$ at the scales probed. Non-Gaussian corrections are suppressed when the area fraction threshold $\nu_{\rm A}$ is used to define the excursion set as opposed to the standard threshold $\nu$. We find no evidence of evolution of $a_{3}$ over the range $0.25 < z < 0.6$ relevant to the BOSS data. 

Numerical systematic effects also exist. The area of our data slices decreases at low redshifts for fixed solid angle, and the excursion set regions at high $|\nu|$ are more difficult to be sampled in a smaller area. Whenever the excursion set is poorly sampled, the genus amplitude will generically be biased high. To eliminate this bias, we must only measure the genus curve over a range of threshold values $-\nu_{0} < \nu < \nu_{0}$ for which the excursion set is well sampled at all redshifts. We vary the threshold limit $\nu_{0}$ to check that the data provides an unbiased measurement of the genus curve. The range $|\nu_{A}| < 2.5$ is well represented within our shells, so we measure the genus curve over this range.

\section{Observational Data $0.25 < \lowercase{z} < 0.6$} 
\label{sec:obs}

Our treatment of the high redshift data -- SDSS-III Baryon Oscillation Spectroscopic Survey (BOSS) \citep{2000AJ....120.1579Y} -- has been described in detail in \cite{Appleby:2020pem}. To briefly review, we bin the galaxies into $N_{\rm z} = 12$ shells of comoving thickness $\Delta = 80 \, {\rm Mpc}$, 6/6 from the LOWZ and CMASS data, over the range $0.25 < z < 0.6$. We apply a mass cut to fix the number density as $\bar{n} = 6.25 \times 10^{-5} \, ({\rm Mpc})^{-3}$ within each shell. With this choice, the shot noise contribution to the field is large but the clustering signal is dominant for the smoothing scales adopted in this work. The galaxies are weighted to account for observational systematics. Specifically, the following weight was applied to each galaxy in the LOWZ and CMASS sample 
\begin{equation} w_{\rm tot} = w_{\rm systot}\left( w_{\rm cp} + w_{\rm noz} -1 \right)  \end{equation} 
\noindent where $w_{\rm cp}$ is the correction factor to account for the subsample of galaxies that are not assigned a spectroscopic fibre, $w_{\rm noz}$ is for the failure in the pipeline to assign redshifts due to certain galaxies, and $w_{\rm systot}$ represents non-cosmological fluctuations in the CMASS target density due to stellar density and seeing.

The redshift bin limits are presented in Table \ref{tab:1}; these were derived using the Planck cosmological parameters $w_{\rm de} = -1$, $\Omega_{\rm m} = 0.307$ to define slices of constant comoving thickness $\Delta=80 \, {\rm Mpc} $. We should vary these limits and re-bin the galaxies each time we vary the cosmology in the distance redshift relation. However, because the genus amplitude is insensitive to $\Delta$ for thick slices, we can fix these limits throughout without biasing our results. We provide evidence to support this statement in Appendix B.

\begin{table}
\begin{center}
 \begin{tabular}{|| c | c  ||}
 \hline
 LOWZ & CMASS  \\
 \hline
 \, $0.250 <  z \leq 0.271$ \, & \, $0.453 < z \leq 0.476$ \, \\
 \, $0.271 < z \leq 0.292$ \, & \, $0.476 < z \leq 0.500$ \, \\
 \, $0.292 < z \leq 0.313$ \, & \, $0.500 < z \leq 0.524$ \, \\
 \, $0.313 < z \leq 0.334$ \, & \, $0.524 < z \leq 0.548$ \, \\
 \, $0.334 < z \leq 0.356$ \, & \, $0.548 < z \leq 0.573$ \, \\
 \, $0.356 < z \leq 0.378$ \,  & \, $0.573 < z \leq 0.598$  \, \\ 
 \hline
\end{tabular}
\caption{\label{tab:1} The redshift limits of the LOWZ and CMASS shells used in this work.}
\end{center} 
\end{table}

 HEALPix\footnote{http://healpix.sourceforge.net} \citep{Gorski:2004by} is used to bin the galaxies into pixels on the unit sphere. A galaxy number density field $\delta_{i,j} \equiv (n_{i,j}-\bar{n}_{j})/\bar{n}_{j}$ is defined, where $1 \leq j \leq  N_{\rm z}$ denotes the redshift bin (of which there are $N_{\rm z}=12$ in total) and $1 \leq i \leq N_{\rm pix}$ is the pixel identifier on the unit sphere. $\bar{n}_{j}$ is the mean number of galaxies contained within a pixel at each redshift shell, and $n_{i,j}$ is the number of galaxies contained within pixel $i$ in redshift slice $j$. We use $N_{\rm pix} = 12 \times 512^{2}$ pixels. The survey geometry and veto masks \cite{Reid:2015gra} were then used to generate a binary healpix map : $\Theta_{i} =1$ if the survey  angular selection function in the $i^{\rm th}$ pixel is larger than some cutoff $\Theta_{\rm cut} = 0.8$ and $\Theta_{i} = 0$ otherwise, where $i$ runs over $N_{\rm pix}$ pixels. The $\Theta_{i}$ mask was applied to the galaxy field $\delta_{i, j}$. 
 
 We smooth the two-dimensional density fields, and the $\Theta_{i}$ mask, in each shell using angular scale $\theta_{\rm G} = R_{\rm G}/d_{\rm cm}(z_{j},\Omega_{\rm m}, w_{\rm de})$, where $R_{\rm G} = 20 {\rm Mpc}$ is the comoving smoothing scale and $d_{\rm cm}(z_{j},\Omega_{\rm m}, w_{\rm de})$ is the comoving distance to the center of the $j^{\rm th}$ redshift shell.  Defining $\tilde{\Theta}_{i,j}$ and $\tilde{\delta}_{i,j}$ as the smoothed mask and density fields, we re-define $\tilde{\delta}_{i,j} = 0$ if $\tilde{\Theta}_{i,j} < \Theta_{\rm cut}$ and $\tilde{\delta}_{i,j} \to \tilde{\delta}_{i,j}/\tilde{\Theta}_{i,j}$ otherwise. Finally, we re-apply the original unsmoothed $\Theta_{i}$ mask. This procedure eliminates regions close to the boundary, where the field may not be well reconstructed. In Appendix C of \cite{Appleby:2020pem} we explicitly show that our masking procedure, and method of genus extraction, provides an unbiased estimate of the genus, and we direct the reader to this paper for further details. The important underlying point is that we are extracting the genus per unit area, which is a local quantity and hence can be estimated in an unbiased manner from a cut-sky galaxy sample.
 
 Finally we divide the genus by the total area of the data $A_{j} = 4\pi f_{\rm sky} d_{\rm cm}^{2}(z_{j},\Omega_{\rm m}, w_{\rm de})$, where $f_{\rm sky}$ is the fractional area of the data on the sky. The genus is reconstructed using the method described in \cite{Schmalzing:1997uc,Appleby:2018jew}, which provides an unbiased estimate of the full sky genus from an observed patch.

We measure the genus for $200$ values of the threshold $\nu_{\rm A}$, equi-spaced over the range $-2.5 < \nu_{\rm A} < 2.5$, then take the average over every four values to obtain $N_{\nu_{\rm A}}=50$ measurements. We label the measured values $g_{j}^{n}$, where $j$ runs over the redshift shells and $1 \leq n \leq N_{\nu_{\rm A}}$ over the $N_{\nu_{\rm A}}=50$ thresholds. We then extract the genus amplitudes $A^{(\rm 2D)}_{j}$ by minimizing the following $\chi^{2}$ functions at each redshift -- 

\begin{equation}\label{eq:ch2d} \chi^{2}_{j} =  \sum_{n=1}^{N_{\nu_{\rm A}}}\sum_{m=1}^{N_{\nu_{\rm A}}} \Delta g^{n}_{j} \Sigma_{n,m}^{-1}(z_{j}) \Delta g^{m}_{j} , \end{equation} 

\noindent with respect to the parameters $A_{j}^{\rm (2D)}, a_{0, j}, a_{2, j}, a_{3, j}$, where 

\begin{eqnarray} \nonumber & &  \Delta  g^{n}_{j} = g_{j}^{n} - A_{j}^{\rm (2D)}  e^{-\nu_{{\rm A},n}^{2}/2} \left[ a_{0, j} H_{0}(\nu_{{\rm A},n}) + \right. 
\\ \label{eq:herm2d} & & \qquad \left.  H_{1}(\nu_{{\rm A},n}) + a_{2, j} H_{2}(\nu_{{\rm A},n}) + a_{3, j} H_{3}(\nu_{{\rm A},n}) \right] , \end{eqnarray}

\noindent and $\Sigma_{n,m}(z_{j})$ are the covariance matrices associated with $\Delta g^{n}_{j}$. $\Sigma_{n,m}(z_{j})$ are obtained using the patchy mock galaxy catalogs \citep{2016MNRAS.456.4156K,2016MNRAS.460.1173R,2014MNRAS.439L..21K,10.1093/mnras/stv645} -- further information on the covariance matrices used in our analysis can be found in \cite{Appleby:2020pem}. 

The measured genus values $g_{j}^{n}$ are functions of the distance-redshift relation, and hence the cosmological parameters $(\Omega_{\rm m}, w_{\rm de})$. This parameter sensitivity enters in the definition of the angular smoothing scale $\theta_{\rm G} = R_{\rm G}/d_{\rm cm}(z_{j},\Omega_{\rm m}, w_{\rm de})$ and the area occupied by the data $A_{j} = 4\pi f_{\rm sky} d_{\rm cm}^{2}(z_{j},\Omega_{\rm m}, w_{\rm de})$. We repeat our measurement of $g_{j}^{n}$ and minimization of ($\ref{eq:herm2d}$) for each cosmological parameter set. We fix $h=0.677$ to its Planck value throughout, where $H_{0} = 100 h \, {\rm km}\, {\rm s^{-1}} \, {\rm Mpc}^{-1}$.

\begin{table}
\begin{center}
 \begin{tabular}{||c  c ||}
 \hline
 Parameter \, & Fiducial Value \\ [0.5ex] 
 \hline\hline
 $\Omega_{\rm m}$ & $0.307$   \\ 
 $h$ & $0.677$   \\
 $w_{\rm de}$ & $-1$ \\
 $\Delta$ & $80 {\rm Mpc}$   \\
 $R_{\rm G}$ & $20 {\rm Mpc}$ \\ 
  \hline 
\end{tabular}\label{tab:ini}
\caption{\label{tab:ii}Fiducial parameters used to fix the slice thickness, and the fiducial parameters used to calculate the genus in this work. $\Delta$ is the thickness of the two dimensional slices of the density field, and $R_{\rm G}$ is the Gaussian smoothing scale used in the two-dimensional planes perpendicular to the line of sight.  }
\end{center} 
\end{table}

\begin{figure}[b!]
  \includegraphics[width=0.5\textwidth]{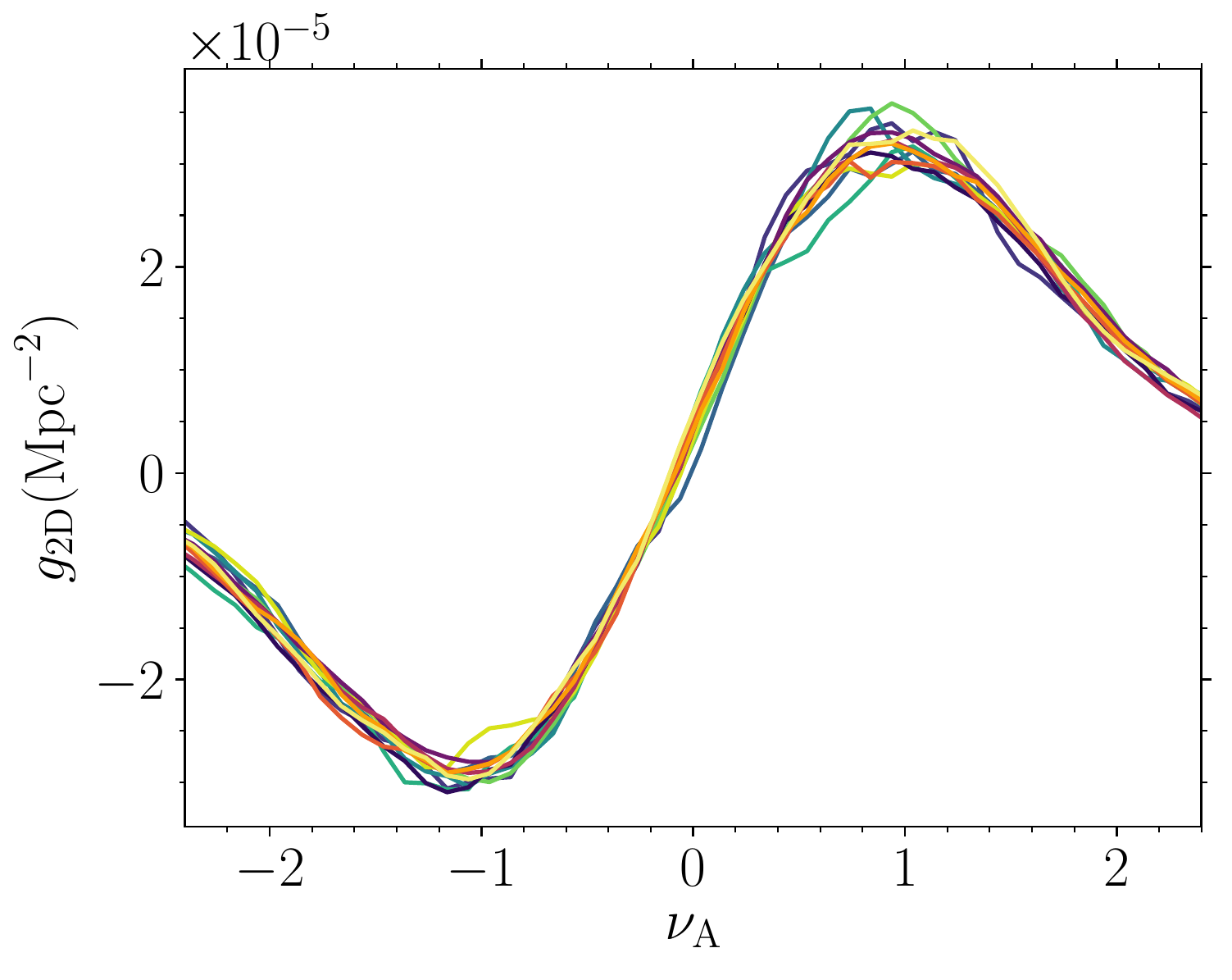}
  \includegraphics[width=0.5\textwidth]{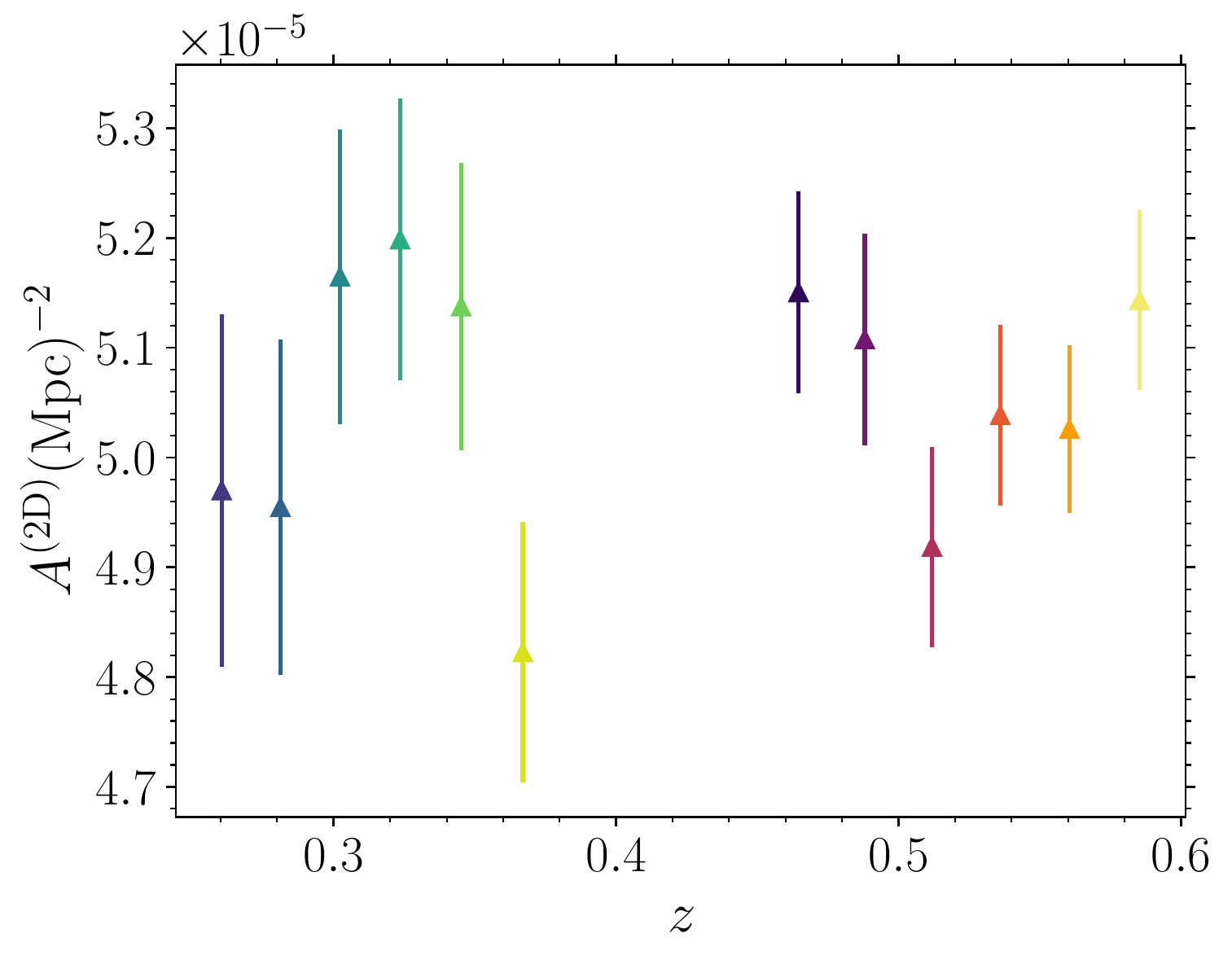}
  \caption{[Top panel] Twelve, two-dimensional genus curves obtained from the BOSS LOWZ and CMASS data, as a function of $\nu_{\rm A}$.  [Bottom panel]  Two-dimensional genus amplitude measurements derived from the $N_{\rm z} = 12$ genus curves presented in the top panel. The same color scheme is applied in both panels.  }
  \label{fig:4}
\end{figure}

Figure \ref{fig:4} exhibits the two-dimensional genus curves [top panel] and the corresponding amplitudes $A^{(\rm 2D)}_{j}$ [bottom panel] extracted from the $N_{\rm z} = 12$ LOWZ and CMASS data shells \citep{Appleby:2020pem}. The genus curves and amplitudes are functions of the assumed cosmological model, and in this figure we have taken a $\Lambda$CDM model with parameters given in Table \ref{tab:ii}. The genus amplitude is reconstructed to accuracy $\sim 3\%$ and $\sim 1.5\%$ in the LOWZ/CMASS data respectively, and we present the best fit amplitudes, $1 \, \sigma$ error bars and reduced $\chi^{2}$ values in Table \ref{tab:amps}. Three of the redshift bins present relatively poor fits with a $\chi^{2}$ per degree of freedom $> 1.5$ : two in the LOWZ data and one CMASS slice. This could indicate that the mocks are under-predicting the true statistical uncertainty, possibly lacking cosmic variance. A theoretical understanding of the statistical uncertainty of the Minkowski functionals is currently lacking, as no prediction for their covariance is available. 

However, the amplitudes extracted from the slices are all consistent. We test this by performing a simple linear regression to the best fit $A^{(2D)}$ data points, assuming the data are uncorrelated. We find a p-value of $p=0.83$ for the null hypothesis that the slope of the linear fit is consistent with zero, indicating no statistically significant redshift evolution of the genus amplitude over the redshift range $0.25 < z < 0.6$. This is expected from theoretical arguments, but provides an important consistency check on our analysis.

\begin{table}
\begin{center}
 \begin{tabular}{|| c  c  c ||}
 \hline
 Redshift & $A_{\rm 2D} \times 10^{5} {\rm Mpc}^{-2}$ & $\chi^{2}/{\rm DoF}$ \\
 \hline
 0.26 & $4.97 \pm 0.16$ &  1.49  \\
 0.28 & $4.95 \pm 0.15$ &  1.12 \\
 0.30 & $5.16 \pm 0.13$ &  1.53 \\
 0.32 & $5.19 \pm 0.13$ &  1.58  \\
 0.35 & $5.14 \pm 0.13$ &  1.02   \\
 0.37 & $4.82 \pm 0.12$ &  1.20  \\ 
 \hline 
 0.46 & $5.15 \pm 0.09$ & 1.34 \\ 
 0.49 & $5.11 \pm 0.10$ & 1.08 \\
 0.51 & $4.92 \pm 0.09$ & 1.63 \\ 
 0.54 & $5.04 \pm 0.08$ & 0.96 \\
 0.56 & $5.03 \pm 0.08$ & 1.15 \\ 
 0.59 & $5.14 \pm 0.08$ & 1.41 \\
 \hline
\end{tabular}
\caption{\label{tab:amps}The mean and $1 \, \sigma$ uncertainty of the genus amplitudes extracted from the six LOWZ and CMASS shells.  The third column is the reduced $\chi^{2}$ value of the fit (46 degrees of freedom).  }
\end{center} 
\end{table}

\section{Low Redshift Data $0 < \lowercase{z} < 0.12$} 
\label{sec:3D}

To test the expansion history, we also require an accurate measurement of the genus at low redshift, which should be practically insensitive to the distance-redshift relation. This would provide an anchor, a measurement of the shape of the linear matter power spectrum against which high redshift genus curves can be compared. 

However, two-dimensional slices at low redshifts have very small areas and suffer from curvature effects. To overcome this limitation we use the three-dimensional local galaxy distribution in the SDSS MGS and apply a Gaussian smoothing over a smaller scale. The measured three-dimensional genus will be used to estimate the two-dimensional genus amplitude. In the following sections, we describe in detail our method -- the theory underlying the three-dimensional genus, the galaxy data used, the mask, how we remove systematics from the genus amplitude using mock galaxy catalogs and how we infer the two-dimensional genus amplitude from the three-dimensional data.

\subsection{Theory -- Expectation value of three-dimensional genus}

The genus per unit volume of a three dimensional, Gaussian random field as a function of threshold $\nu$ is given by \citep{10.1143/PTP.76.952, Adler, Gott:1986uz, Hamilton:1986}

\begin{eqnarray} \label{eq:gg} & &  g_{\rm 3D}(\nu) = {1 \over 4\pi^{2}} \left({\Sigma_{1}^{2} \over 3\Sigma_{0}^{2}}\right)^{3/2} \left(1 - \nu^{2} \right) e^{-\nu^{2}/2} , \\
\nonumber & & \Sigma_{0}^{2} = \langle \delta^{2}_{\rm 3D} \rangle  ,   \qquad  \Sigma_{1}^{2} = \langle |\nabla \delta_{\rm 3D} |^{2} \rangle  , \end{eqnarray} 

\noindent where $\Sigma_{0,1}$ are the two-point cumulants of the three-dimensional field, related to the power spectrum as 

\begin{eqnarray}
\label{eq:s03} & & \Sigma_{0}^{2} = \int d^{3} k e^{-k^{2}\Lambda_{\rm G}^{2}} P_{\rm 3D}(k)   , \\
\label{eq:s13} & & \Sigma_{1}^{2} = \int d^{3} k e^{-k^{2}\Lambda_{\rm G}^{2}} k^{2} P_{\rm 3D}(k)   , \\
  \end{eqnarray} 

\noindent where we have smoothed with a Gaussian kernel of width $\Lambda_{\rm G}$. The genus amplitude is given by  

\begin{equation}\label{eq:g1}  A_{\rm G}^{(\rm 3D)} = {1 \over 4\pi^{2}} \left({\Sigma_{1}^{2} \over 3\Sigma_{0}^{2}}\right)^{3/2} . \end{equation} 

\noindent The leading order non-Gaussian expansion of the genus, in terms of the $\nu_{\rm A}$ threshold convention, is given by \citep{Matsubara:1994we, 2000astro.ph..6269M,Pogosyan:2009rg, Gay:2011wz,Codis:2013exa}

\begin{eqnarray} \nonumber & &  g_{\rm 3D}(\nu_{\rm A}) = A_{\rm G}^{(\rm 3D)} e^{-\nu_{\rm A}^{2}/2} \left[H_{2}(\nu_{\rm A}) + \left[ \left(S^{(1)} - S^{(0)}\right) \times \right. \right. \\
\label{eq:mats3d} & & \quad \left. \left. H_{3}(\nu_{\rm A}) + \left(S^{(2)} - S^{(0)}\right) H_{1}(\nu_{\rm A})\right] \Sigma_{0} + {\cal O}(\Sigma_{0}^{2}) \right] . \end{eqnarray}

\noindent As for the two-dimensional genus, the amplitude (coefficient of $H_{2}$ Hermite polynomial) is not modified by the non-Gaussian effect of gravitational collapse to linear order in the  $\Sigma_{0}$ expansion ($\ref{eq:mats3d}$).

\subsection{Data}
\label{sec:3DSDSS}

To extract the genus of the low redshift matter density, we use the seventh data release of the main galaxy catalog of the SDSS DR7 \citep{2009ApJS..182..543A}. Specifically, we adopt the Korea Institute for Advanced Study Value Added Galaxy Catalog (KIAS VAGC) \citep{articleyyc,2005AJ....129.2562B,2008ApJ...674.1217P}. The KIAS catalog supplements redshifts from other existing galaxy redshift catalogs -- the updated Zwicky catalog \citep{1999PASP..111..438F}, the IRAS Point Source Catalog Redshift Survey \cite{Saunders:2000af}, the Third Reference Catalogue of Bright Galaxies \citep{1991rc3..book.....D}, and the Two Degree Field Galaxy Redshift Survey \citep{2001MNRAS.328.1039C}. 

The KIAS VAGC contains $593,514$ redshifts of SDSS main galaxies in the $r$-band Petrosian magnitude range $10 < r_{\rm p} < 17.6$. Details of the selection criteria, classification schemes and angular selection functions can be found in \citep{articleyyc}. To maximize the area to boundary ratio of the data, we remove the three southern stripes and Hubble deep field region. 

\begin{figure}[b!]
  \includegraphics[width=0.5\textwidth]{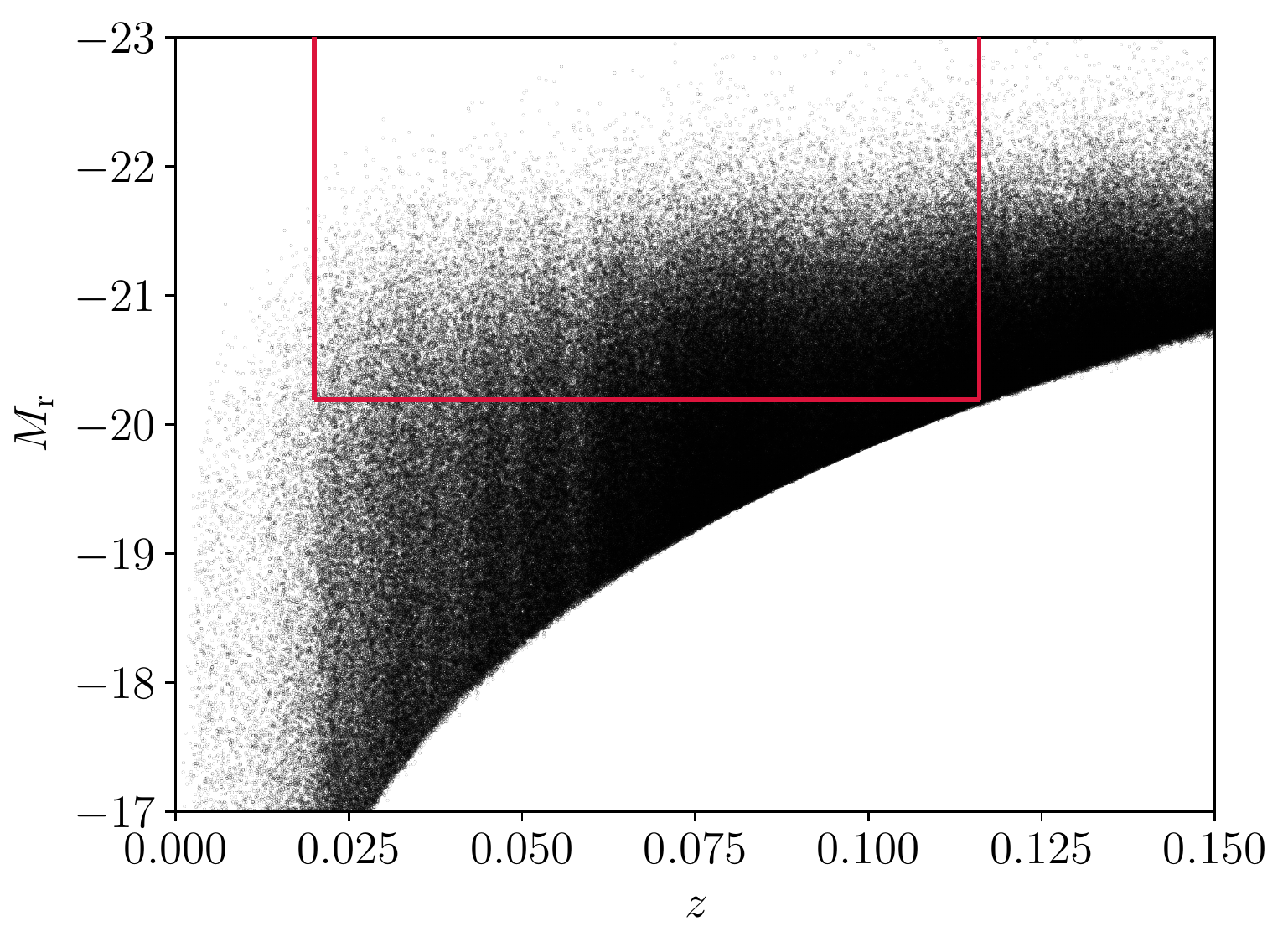}
  \includegraphics[width=0.5\textwidth]{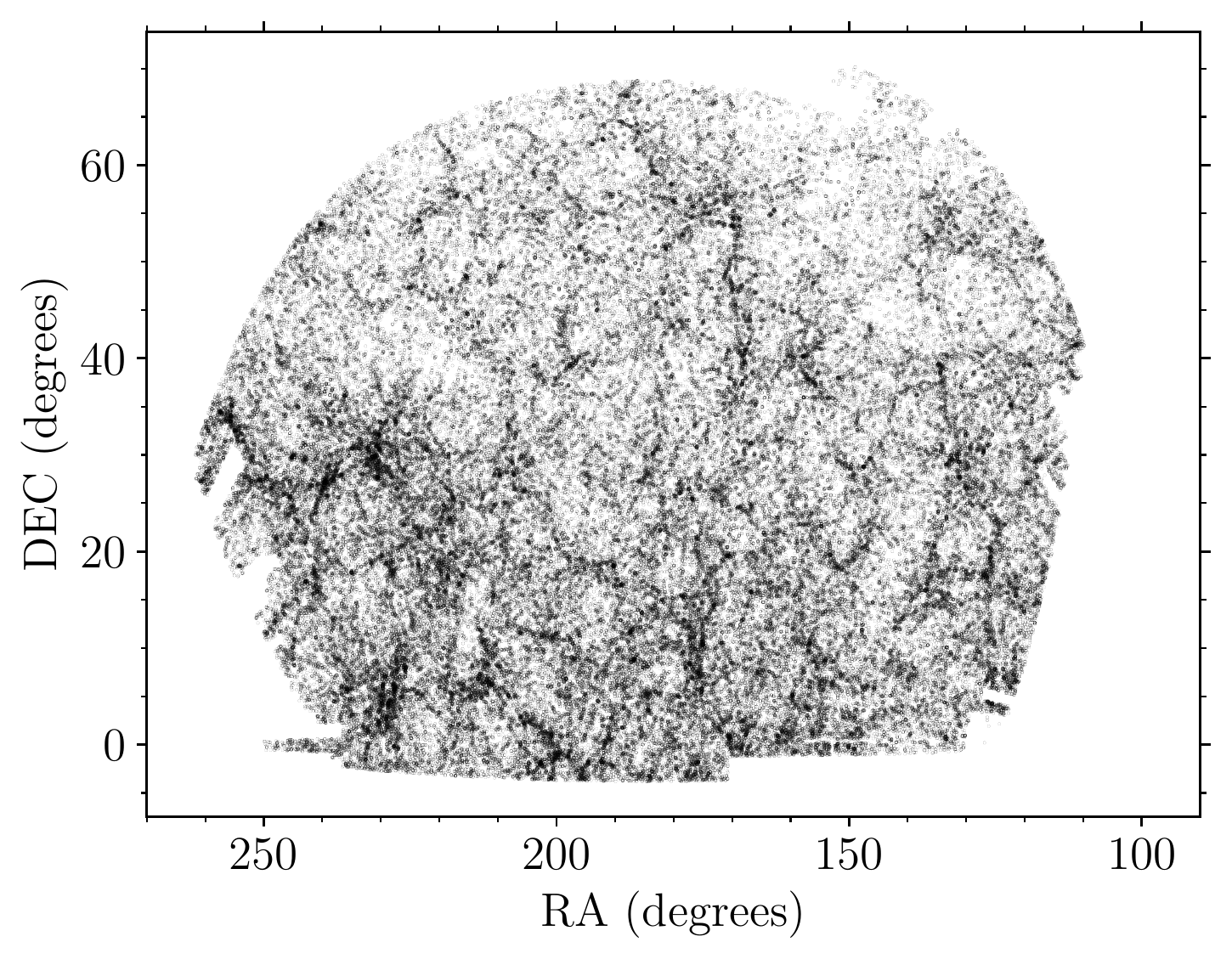}
  \caption{[Top panel] The absolute, $r$-band magnitude of the SDSS MGS galaxies as a function of redshift. The solid red lines indicate the boundaries of our volume limited sample with $0.02 < z < 0.116$ and $M_{\rm r} < -20.19$. [Bottom panel] The angular distribution of the volume limited sample of galaxies on the sky; declination vs right ascension (in degrees).  }
  \label{fig:sdss}
\end{figure}

The catalog provides angular positions, redshifts and absolute, $r$-band magnitudes normalised to the $z=0.1$ epoch, calculated from extinction corrected AB fluxes and an evolution correction $E(z) = 1.6(z-0.1)$ \citep{Tegmark:2003uf}. All magnitudes and colors are corrected to the redshift $z=0.1$ epoch. Following \cite{Choi:2010sx}, we apply a magnitude cut $M_{\rm r} < -20.19+ 5 \log h$ to generate a volume limited sample over the redshift range $0.02 < z < 0.116$, with a mean galaxy separation of $r_{\rm gal} = \bar{n}_{\rm gal}^{-1/3} = 8.3 {\rm Mpc}$, where $\bar{n}_{\rm gal}$ is the mean galaxy number density within the volume. The redshift range was selected to ensure a maximal number of galaxies are used in the analysis. The galaxies are presented as a function of redshift and absolute magnitude in Figure \ref{fig:sdss} (top panel), and the angular distribution of all galaxies used in this work are presented in the bottom panel. Note that in this Figure and in what follows the factor $5 \log h$ will be dropped in the expression of $M_{\rm r}$.

To convert the galaxy catalog into a three-dimensional density field, we construct a regular three-dimensional $N_{\rm pix}^{3} = 512^3$ pixel lattice in a cube of side length $L_{\rm box} = 750 {\rm Mpc}$ and use cosmological parameters given in Table \ref{tab:ii} to infer the distance-redshift relation. We bin the galaxies into pixels using the Cloud-in-Cell scheme, generating a three-dimensional number density field $\delta_{ijk} = (n_{ijk}-\bar{n})/\bar{n}$, where $\bar{n}$ is the average number of galaxies within the unmasked pixels and $1 \leq i,j,k \leq N_{\rm pix}$ subscripts are pixel labels. The galaxies are weighted via the angular selection function during this binning procedure. 

The angular selection function constitutes a set of weights as a function of angular position on the sky - $w(\theta,\phi)$. It is defined in this work as $w = 0$ when outside the survey geometry or inside a bright star mask and $0 < w \leq 1$ when inside the survey geometry. This function represents the survey completeness as a function of position on the sky. Because we weight the galaxies according to the angular selection function, we convert $w$ into a binary field with $w = 1$ if $w > w_{\rm cut}$ and $w=0$ otherwise, where $w_{\rm cut}$ was selected as $w_{\rm cut} = 0.8$. Using the fiducial distance-redshift relation, we define $w_{ijk}$ as the projection of the angular selection function into a $512^{3}$, three-dimensional pixel cube of the same dimensions as $\delta_{ijk}$. 

We smooth the three-dimensional density field $\delta_{ijk}$ with a Gaussian kernel of width $\Lambda_{\rm G} = 8.86 {\rm Mpc}$ (this value is $R_{G} = 6 {\rm Mpc}/h$, following \cite{Choi:2010sx}), and also smooth the projected selection function $w_{ijk}$ with the same kernel, defining the smoothed counterparts as $\tilde{\delta}_{ijk}$ and $\tilde{w}_{ijk}$. We then redefine $\tilde{\delta}_{ijk} = 0$ for all pixels in which $\tilde{w}_{ijk} < 0.9$ and $\tilde{\delta}_{ijk} = \tilde{\delta}_{ijk}/\tilde{w}_{ijk}$ if $\tilde{w}_{ijk} \ge  0.9$. We then re-apply the original mask and set $\tilde{\delta}_{ijk} = 0$ if $w_{ijk} = 0$. This eliminates all data in the vicinity of the survey boundary. 

From the masked field we reconstruct the three dimensional genus, by generating iso-field triangulated meshes and calculating the Gaussian curvature at the triangle vertices. Details of the method can be found in \citep{Appleby:2018tzk}.  We calculate the genus as a function of $\nu_{\rm A}$, where $\nu_{\rm A}$ is the threshold chosen to match the volume fraction of a Gaussian random field. We select $200$, $\nu_{\rm A}$ threshold values over the range $-2.5 < \nu_{\rm A} < 2.5$, then take the average of every four values to obtain the genus at $N_{\nu_{\rm A}} = 50$, $\nu_{\rm A}$ values equi-spaced over this range. The resulting measured genus values are presented in Figure \ref{fig:sdss_g3d} (top panel, red points). 

To extract the genus amplitude from the measurements, we fit a Hermite polynomial expansion to the data points by minimizing the following $\chi^{2}$ function 

\begin{equation}\label{eq:chb} \chi^{2} =  \Delta g^{\rm T} \Gamma^{-1} \Delta g , \end{equation}

\noindent where 

\begin{eqnarray} \nonumber  & & \Delta g_{i}  = g_{i} -  A^{(\rm 3D)} e^{-\nu_{{\rm A},i}^{2}/2} \left[ a_{0} H_{0}(\nu_{{\rm A},i}) +a_{1} H_{1}(\nu_{{\rm A},i}) +  \right. \\
\label{eq:herma} & & \qquad \qquad  \left. H_{2}(\nu_{{\rm A},i}) +a_{3} H_{3}(\nu_{{\rm A},i}) +a_{4} H_{4}(\nu_{{\rm A},i})  \right] . \end{eqnarray} 

\noindent $A^{(\rm 3D)}, a_{0}, a_{1}, a_{3}, a_{4}$ are free parameters to be constrained via the minimization of ($\ref{eq:chb}$), the $i$ subscript denotes the $i^{\rm th}$, $\nu_{\rm A}$ threshold bin and $g_{i}$ are the measured genus values. In the fitting procedure we include the leading order $a_{1},a_{3}$ Hermite polynomial coefficients and the next-to-leading order even Hermite polynomial contributions $a_{0}$, $a_{4}$. Introducing additional Hermite polynomials does not significantly modify the fit. 

The covariance matrix $\Gamma_{ij}$ is obtained from mock galaxy catalogs, as described in the following section.

\subsection{Mock Galaxy Catalogs}

Mock galaxy catalogs are generated using Horizon Run 4 (HR4) \cite{Kim:2015yma}. Horizon Run 4 is a cosmological scale dark matter simulation in which $N = 6300^{3}$ particles in a volume $V = (3150 {\rm Mpc}/h)^{3}$ are evolved using a modified GOTPM scheme\footnote{For a description of the original GOTPM code, please see \cite{Dubinski:2003fq}. A description of the modifications introduced in the Horizon Run project can be found at https://astro.kias.re.kr/~kjhan/GOTPM/index.html.}. The initial conditions are obtained using second order Lagrangian perturbation theory \cite{L'Huillier:2014dpa}, and the cosmological parameters used are $h=0.72$, $n_{\rm s} = 0.96$, $\Omega_{\rm m} = 0.26$, $\Omega_{\rm b} = 0.048$. We use the $z=0$ snapshot box to create mock galaxy catalogs, using the HR4 cosmological parameters to infer distances. Details of the numerical implementation, and the method by which mock galaxies are constructed can be found in \cite{Hong:2016hsd}. The mock galaxies are defined using the most bound halo particle galaxy correspondence scheme, and the survival time of satellite galaxies post merger is estimated via the merger timescale model described in \cite{Jiang:2007xd}. The snapshot box is decomposed into $N_{\rm r}=360$ non-overlapping volumes, and mock galaxy catalogs are constructed from each region, with the same number density, redshift range and survey geometry as the data.

From each mock sample we repeat our analysis ; bin the galaxies into a regular cubic pixel lattice, smooth the resulting number density field with a Gaussian of scale $\Lambda_{\rm G} = 8.86 {\rm Mpc}$, apply the smoothed mask $\tilde{w}_{ijk}$ then unsmoothed binary mask $w_{ijk}$, then extract the genus from $\tilde{\delta}_{ijk}$.

The result is a set of genus measurements $g^{(\rm 3D)}_{i, m}$, where $1 \leq i \leq N_{\nu_{\rm A}}$ runs over $N_{\nu_{\rm A}} = 50$, $\nu_{\rm A}$ bins uniformly sampled in the range $-2.5 < \nu_{\rm A} < 2.5$ and $1 \leq m \leq N_{\rm r}$ runs over the randomly sampled realisations. We measured the genus for $200$ values and averaged every fourth point to arrive at the $N_{\nu_{\rm A}} = 50$ values in each mock sample. The covariance matrix is constructed as 

\begin{equation} \label{eq:cov3d} \Gamma_{ij} = {1 \over N_{\rm r} - 1 }\sum_{m=1}^{N_{\rm r}} \left(g^{(\rm 3D)}_{i,m} - \langle g^{(\rm 3D)}_{i} \rangle  \right) \left(g^{(\rm 3D)}_{j,m} - \langle g^{(\rm 3D)}_{j} \rangle  \right) , \end{equation}

\noindent where $\langle g_{i}^{(\rm 3D)} \rangle$ is the average value of the genus in the $i^{\rm th}$ threshold bin. In Figure \ref{fig:sdss_g3d} (bottom panel) we exhibit the covariance matrix $\Gamma_{ij}$ extracted from the mock realisations. We note the strong correlation between genus values measured at different thresholds. A similar covariance matrix was numerically extracted from mock data in \cite{Blake:2013noa}.

\subsection{Results -- Three-Dimensional Genus of SDSS MGS}
\label{sec:res3d}

\begin{figure}
  \includegraphics[width=0.48\textwidth]{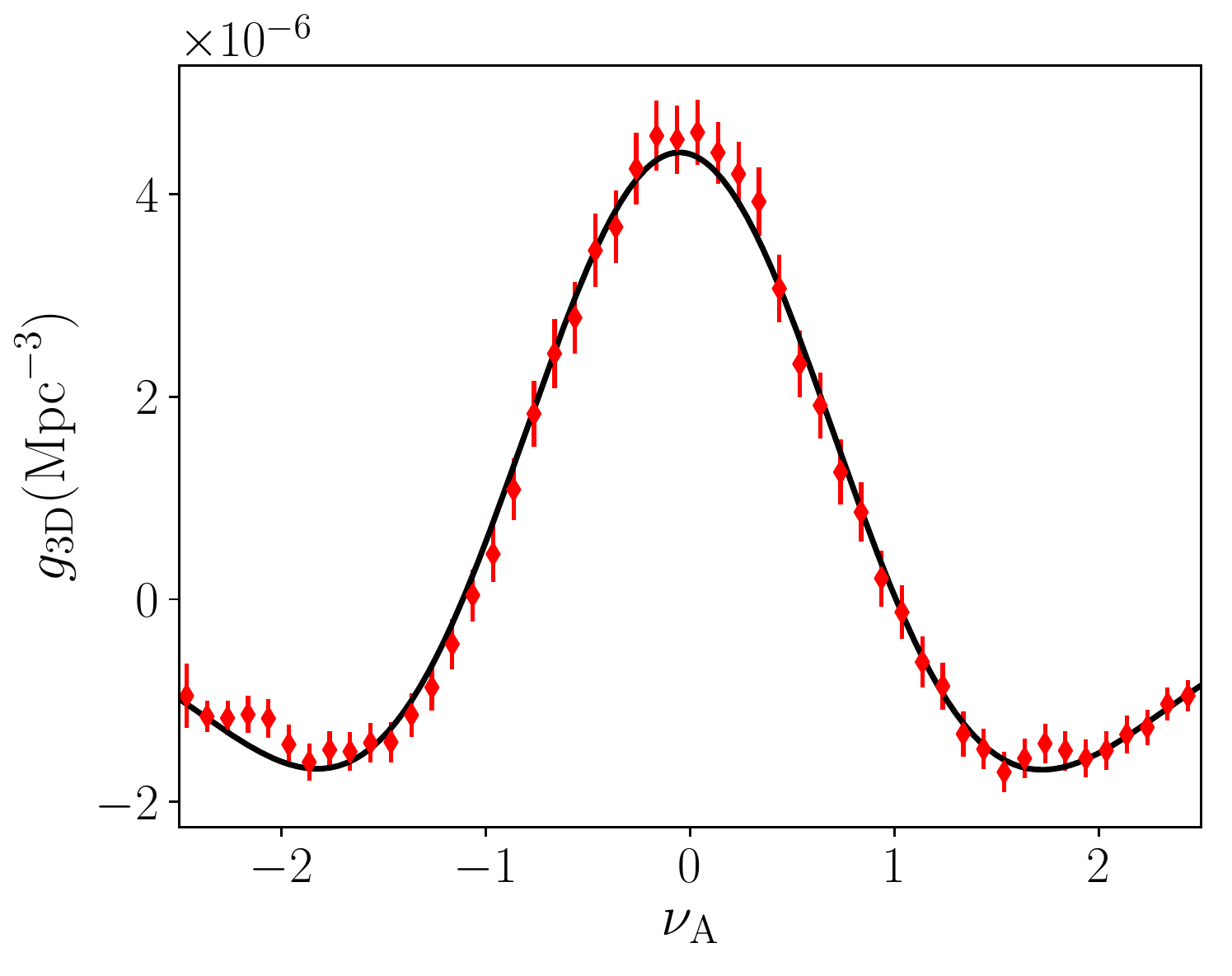}
    \includegraphics[width=0.48\textwidth]{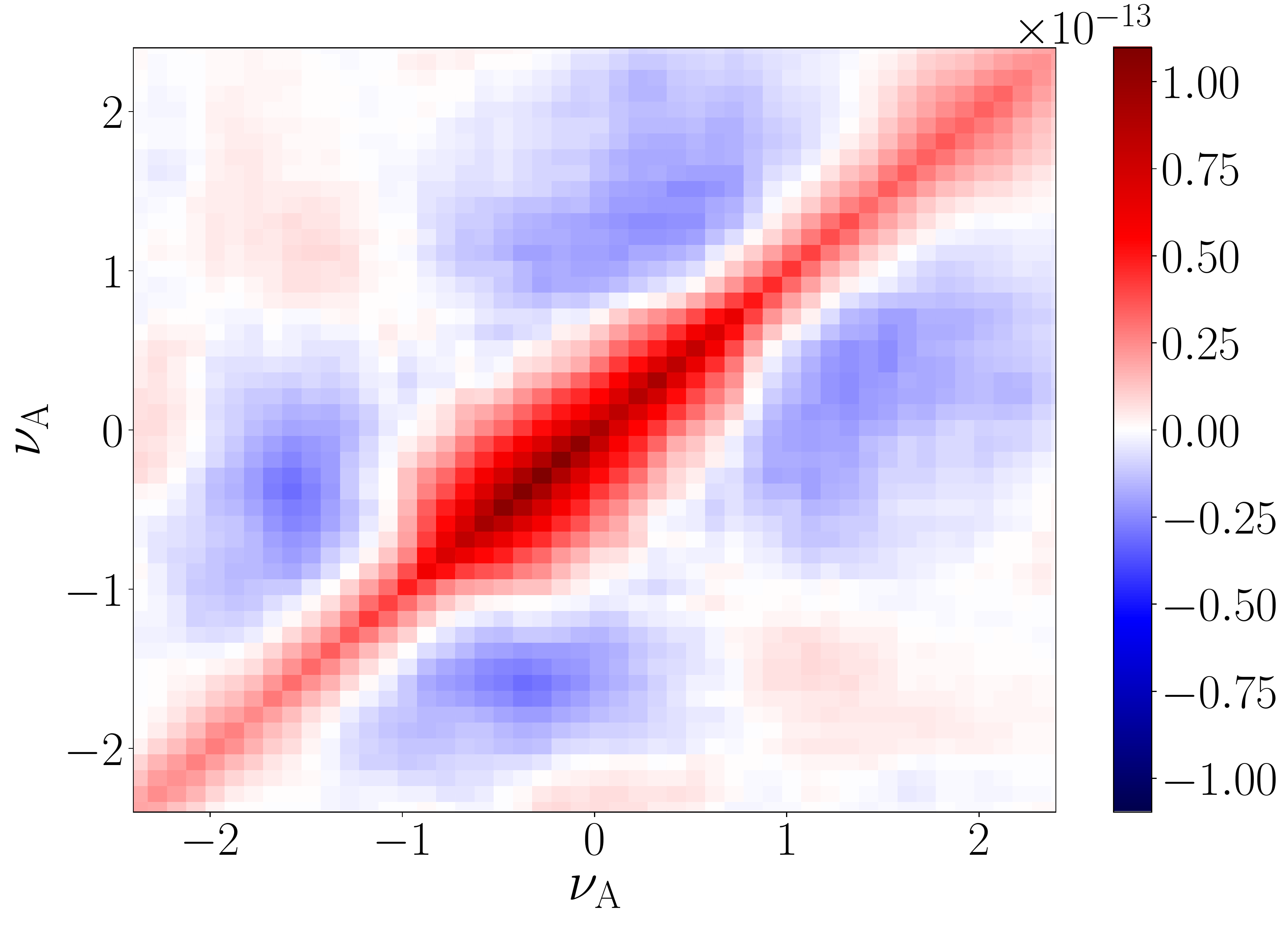}
  \caption{[Top panel] The genus curve measured from the SDSS MGS using a Gaussian smoothing length of $\Lambda_{\rm G} = 8.86 {\rm Mpc}$. The red points correspond to measured values, and the error bars are from the diagonal components of the covariance matrix ($\ref{eq:cov3d}$). The black solid line is the best fit curve reconstruction ($\ref{eq:herma}$) with parameters given in the first row of table \ref{tab:parm_herm}. [Bottom panel] The covariance matrix $\Gamma_{ij}$. Bins separated by $\Delta \nu_{\rm A} < 0.25$ are strongly correlated (red), and bins at larger separations present anti-correlation (blue). }
  \label{fig:sdss_g3d}
\end{figure}

In Figure \ref{fig:sdss_g3d} (top panel) we exhibit the genus measured from the SDSS MGS (red points), and the best-fit curve reconstruction $g^{(\rm th)}(\nu_{\rm A})$ (black solid line). The error bars are the square root of the diagonal elements of $\Gamma_{ij}$. After minimizing the $\chi^{2}$ function ($\ref{eq:chb}$), in Table \ref{tab:parm_herm} (first row)  we present the best fit and uncertainty on the $(A^{(\rm 3D)}, a_{0}, a_{1}, a_{3}, a_{4})$ parameters. We also present a fit including just $(A^{(\rm 3D)}, a_{1}, a_{3})$, and $(A^{(\rm 3D)})$ only for comparison. If we regard equation ($\ref{eq:herma}$) as an expansion in $\sigma_{0}$, then $a_{1},a_{3}$ should be of order $a_{1,3} \sim {\cal O} (\sigma_{0})$ and $a_{0}, a_{4} \sim {\cal O}(\sigma_{0}^{2})$. At the scales adopted in this work, the higher order terms are large, which indicates that the field is non-linear. In spite of this, all three amplitude measurements are consistent. However, the second and third rows yield a significantly worse $\chi^{2}$.

The genus amplitude presented in the first row of Table \ref{tab:parm_herm}; $A^{(\rm 3D)} = 4.040 \times 10^{-6} {\rm Mpc}^{-3}$, and uncertainty $\Delta A^{(\rm 3D)} = 0.197 \times 10^{-6} {\rm Mpc}^{-3}$, will be used as the low redshift genus amplitude measurement from the SDSS MGS. This low redshift data point will be used to complement the higher redshift, two-dimensional BOSS measurements.

\begin{table*}
\begin{center}
 \begin{tabular}{|| c  c  c  c  c  c ||}
 \hline
 $A^{(\rm 3D)} \times 10^{6} (\rm Mpc^{-3})$ & $a_{0}$ & $a_{1}$ & $a_{3}$ & $a_{4}$ & $\chi^{2}$ \\ [0.5ex] 
 \hline\hline
 $4.040 \pm 0.197$ & $0.095 \pm 0.016$ & $-0.009 \pm 0.025$ & $0.042 \pm 0.019$ & $-0.006 \pm 0.014$ &  $63.9$ \\ 
 $4.167 \pm 0.135 $ & - & $-0.015 \pm 0.026$ & $0.026 \pm 0.018$ & - &  $117.6$  \\ 
 $4.084 \pm 0.130 $ & - & - & - & - & $124.0$  \\ 
 \hline
\end{tabular}
\caption{\label{tab:parm_herm}Best fit Hermite polynomial coefficients for the three-dimensional genus curve extracted from the SDSS MGS. The top row is the full fitting function used in this work. In the second row we set $a_{0} = a_{4} =  0$ and in the third row we fix $a_{0} = a_{1} = a_{3} = a_{4} = 0$, and fit a Gaussian curve to the points. }
\end{center} 
\end{table*}

\begin{figure}
  \includegraphics[width=0.48\textwidth]{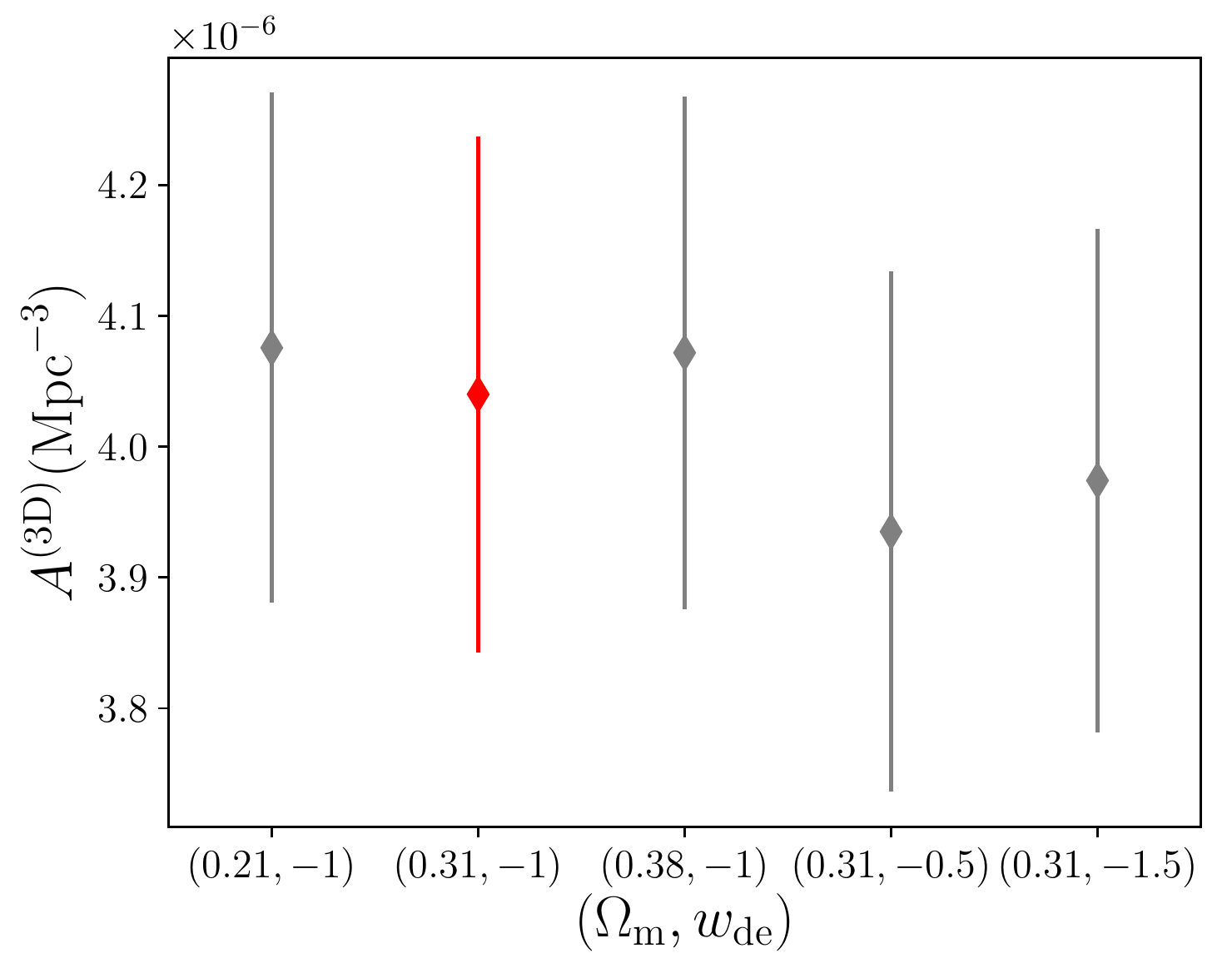}
  \caption{The genus amplitude as measured from the SDSS MGS, assuming five different cosmological models to infer the distance redshift relation. The measured amplitude of the low redshift sample is effectively insensitive to our choice. We select a fiducial cosmology $(\Omega_{\rm m}, w_{\rm de}) = (0.307, -1)$ to infer $A^{(\rm 3D)}$ (red diamond). }
  \label{fig:ps1}
\end{figure}

The measured amplitude $A^{(\rm 3D)}$ of the SDSS MGS is effectively insensitive to cosmological parameters. We confirm that reasonable variation of cosmological parameters does not affect the measured value of $A^{(\rm 3D)}$ in Figure \ref{fig:ps1}. We select five parameter sets $(\Omega_{\rm m},w_{\rm de})= (0.21,-1), (0.31,-1), (0.38,-1), (0.31,-0.5), (0.31,-1.5)$ to infer the distance redshift relation, construct the density field from the galaxy positions and measure the genus amplitude $A^{(\rm 3D)}$ by minimizing the $\chi^{2}$ function ($\ref{eq:chb}$). The resulting amplitudes and uncertainties are presented in Figure \ref{fig:ps1}. We find no significant change in the measured genus amplitude if we use different cosmological parameters to infer the distance redshift relation, as expected at low redshift $z < 0.12$. For this reason, we fix $A^{(\rm 3D)} = 4.040 \pm 0.197  \times 10^{-6} {\rm Mpc}^{-3}$, corresponding to the red data point in Figure \ref{fig:ps1}.

\section{Three- to Two-dimensional Genus Amplitude}
\label{sec:conv}

Our intention is to combine the SDSS MGS and BOSS genus measurements, and find the cosmology that minimizes the evolution of the two-dimensional genus amplitude. However, to directly compare these results, we must convert the three dimensional genus amplitude measurement from the SDSS MGS to a corresponding effective two-dimensional amplitude. To do so, we perform the following steps -- 

\begin{enumerate} 
\item{Correct the measured three-dimensional genus amplitude for gravitational smoothing and non-linear redshift space distortion with a correction factor obtained from simulations.}
\item{Using the now corrected, real space amplitude, perform a cosmological parameter search by comparing this value to its Gaussian expectation value. The result is a set of parameter constraints on  $(\Omega_{\rm c}h^{2}$, $n_{\rm s})$ which determine the shape of the linear power spectrum.}
\item{Use the best fit cosmological parameters $(\Omega_{\rm c}h^{2}$, $n_{\rm s})$ to infer the two-dimensional theoretical expectation of the genus amplitude 

\begin{equation}\label{eq:agauss} A^{(\rm 2D)}_{{\rm G}} = {1 \over 2(2\pi)^{3/2}} { \int k_{\perp}^{3} e^{-k_{\perp}^{2}R_{\rm G}^{2}} P_{\rm 2D}(k_{\perp}) d k_{\perp} \over \int  k_{\perp} e^{-k_{\perp}^{2}R_{\rm G}^{2}} P_{\rm 2D}(k_{\perp}) d k_{\perp} } . \end{equation}

Where the two-dimensional power spectrum $P_{\rm 2D}$ is related to the three dimensional matter power spectrum according to equation ($\ref{eq:p2d}$), and we use the three-dimensional power spectrum ($\ref{eq:p3df}$) with $\bar{n}=6.25 \times 10^{-5} \, {\rm Mpc}^{-3}$, $b=2$, $R_{\rm G} = 20 {\rm Mpc}$, $\Delta = 80 {\rm Mpc}$; the values relevant to the BOSS data. The end result is an inferred two-dimensional genus amplitude, based on the SDSS MGS data.  }
\end{enumerate} 

\noindent In the following subsections we discuss each point in turn.

\subsection{Systematics removal} 
\label{sec:3Dsys}

Before comparing $A^{(\rm 3D)}$ to its Gaussian expectation value, we must account for non-linear effects. The most significant systematics that must be corrected are non-linear gravitational evolution and redshift space distortion. At the smoothing scale $\Lambda_{\rm G} = 8.86 {\rm Mpc}$, the effect of redshift space distortion on the genus amplitude will not be well approximated by the linear Kaiser approximation \cite{Choi:2013eej}. Therefore, to eliminate its effect we directly compare the three-dimensional genus measured from the simulations in real and redshift space, and use the difference between these measurements as a correction factor to be applied to the SDSS genus amplitude measurement.

To model these systematic corrections, we use the KIAS Multiverse simulations; a set of five cosmological scale, dark matter only simulations. Each is generated from a different cosmological model in which $\Omega_{\rm m}$ and $w_{\rm de}$ are varied \citep{2017ApJ...843...73S,Park:2019mvn,10.1093/mnras/staa566,article_moto}. Since our low redshift genus measurement will be practically insensitive to the value of $w_{\rm de}$, we use three of the simulations with cosmological parameters $(\Omega_{\rm m},w_{\rm de}) = (0.21,-1)$, $(0.26, -1)$ and ($0.31, -1)$ with all other cosmological parameters fixed as $\Omega_{\rm b} = 0.044$, $h = 0.72$, $n_{\rm s} = 0.96$. Each simulation comprises $N_{p} = 2048^3$ dark matter particles in a $1024^{3} h^{-3} {\rm Mpc}^{3} = 1422^{3} {\rm Mpc}^{3}$ box, gravitationally evolved using a modified GOTPM code which uses the Poisson equation 

\begin{equation} \nabla^{2} \Psi = 4 \pi G a^{2} \bar{\rho}_{\rm m}\delta_{\rm m} \left( 1 + {D_{\rm de} \over D_{\rm m}}{\Omega_{\rm de}(a) \over \Omega_{\rm m}(a)}\right) . \end{equation}

\noindent The same random number sequence was used to generate the initial condition for each simulation at $z=99$, to eliminate cosmic variance when comparing different models. The power spectrum was normalised such that the rms of the matter fluctuation, smoothed with top hat $8 h^{-1} {\rm Mpc}$ and linearly evolved to $z=0$, is  $\sigma_{8} = 0.794$.

\begin{figure}
  \includegraphics[width=0.48\textwidth]{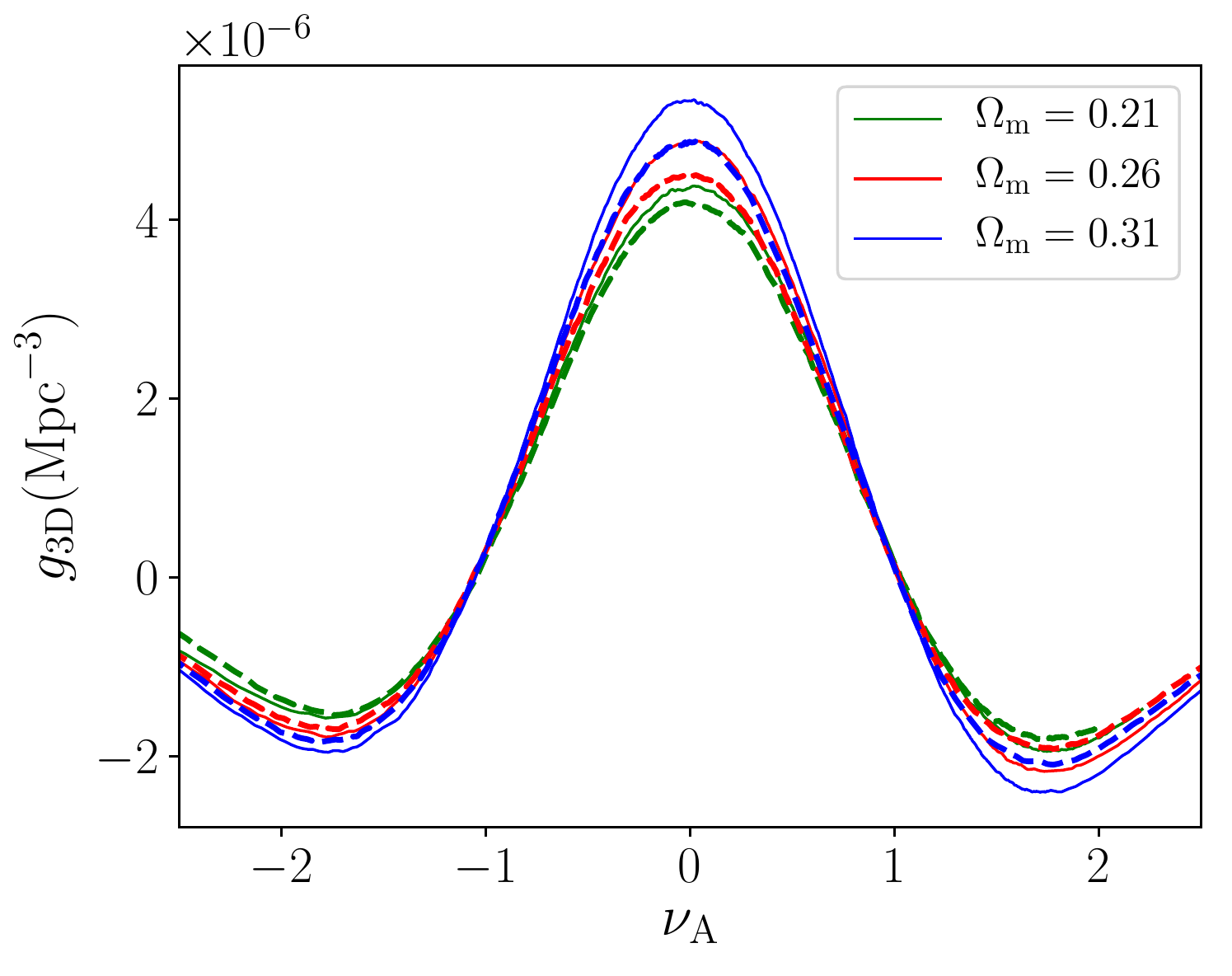}
  \includegraphics[width=0.48\textwidth]{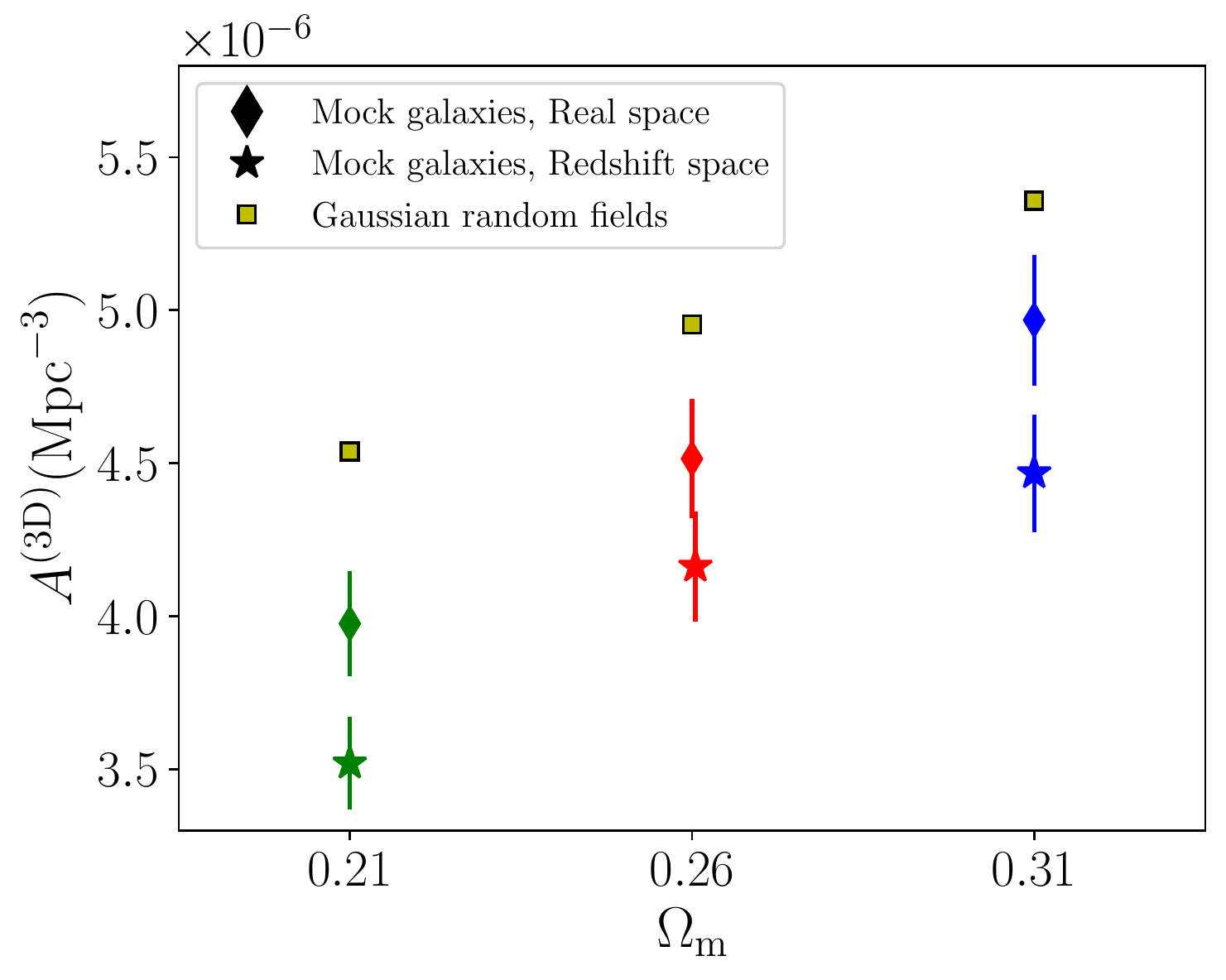}
  \caption{[Top panel] Measured genus curves as a function of $\nu_{\rm A}$ for three multiverse simulations with $\Omega_{\rm m} = 0.21, 0.26, 0.31$ (green, red, blue lines). The solid lines are real space measurements, dashed are redshift space. [Bottom panel] The genus amplitudes extracted from the top panel. The diamonds/stars represent the real/redshift space measurements respectively. The yellow squares represent the prediction for a Gaussian field for the given $\Omega_{\rm m}$. The real space, mock galaxy amplitudes are lower than the Gaussian prediction due to gravitational smoothing, and the redshift space values are still lower due to the effect of redshift space distortion.}
  \label{fig:3DRSD}
\end{figure}

The genus curves of the Multiverse $z=0$ snapshot boxes in real and redshift space are presented in the top panel of Figure \ref{fig:3DRSD}, for three different cosmological models $\Omega_{\rm m} = 0.21, 0.26, 0.31$. We use the entire box with periodic boundary conditions to make these measurements -- that is we apply no mask in this subsection.  For each simulation, we fix the number density of the mock galaxies such that the mean separation is $\bar{r} = \bar{n}^{-1/3}_{\rm gal} = 8.33 \, {\rm Mpc}$, by applying a mass cut. The solid/dashed lines correspond to real/redshift space mock galaxy catalogs, and green/red/blue corresponds to $\Omega_{\rm m} = 0.21, 0.26, 0.31$ respectively. In all cases one can observe an amplitude drop due to the effect of redshift space distortion. 

In the bottom panel, we exhibit amplitude measurements extracted from the genus curves in the top panel. The green/red/blue color scheme is the same as for the top panel, and diamonds/stars correspond to real/redshift space measurements of the genus amplitude. We denote the real/redshift space genus amplitudes as $A^{(\rm 3D)}_{\rm real}$ and $A^{(\rm 3D)}_{\rm rsd}$ respectively. The fractional difference between the redshift and real space amplitude measurements -- $a_{\rm rsd}^{(\rm 3D)} \equiv A^{(\rm 3D)}_{\rm rsd}/A^{(\rm 3D)}_{\rm real}$ -- is $a_{\rm rsd}^{(\rm 3D)} = 0.89, 0.92, 0.91$ for $\Omega_{\rm m} = 0.21, 0.26, 0.31$ respectively. The effect of redshift space distortion is a $\sim 10\%$ effect on the genus amplitude at these scales, and is only weakly dependent on cosmological parameters. Specifically, $a_{\rm rsd}^{(\rm 3D)}$ exhibits no significant, systematic dependence on $\Omega_{\rm m}$.  We use the $\Omega_{\rm m} = 0.26$ simulation and take $a_{\rm rsd}^{(\rm 3D)} = 0.92$ in what follows, correcting the measured $A^{(\rm 3D)}$ amplitude by a factor of $(1 - \Delta_{\rm rsd})^{-1}$ with $\Delta_{\rm rsd} = 0.08$. This factor converts $A^{(\rm 3D)}$ to real space.

To account for non-linear gravitational evolution, we compare the measurement of the three-dimensional genus amplitude of the multiverse simulations in real space to the Gaussian expectation value ($\ref{eq:g1}$), where we use the linear matter power spectrum plus shot noise 

\begin{equation} P_{\rm 3D}(k) = b_{\rm sdss}^{2} P_{\rm m}(k) + P_{\rm SN, sdss} , \end{equation} 

\noindent to generate the cumulants $\Sigma_{0,1}$. We use the SDSS MGS number density $\bar{n} = 1.7 \times 10^{-3} {\rm Mpc}^{-3}$ for the shot noise power spectrum $P_{\rm SN, sdss } = 1/\bar{n}$ and galaxy bias $b_{\rm sdss} = 1.5$ \citep{Howlett:2014opa,Ross:2014qpa}. We have already converted $A^{(\rm 3D)}$ to real space using the correction factor $\Delta_{\rm rsd}$. 

In Figure \ref{fig:3DRSD} (bottom panel) we exhibit the genus amplitude of the $z=0$, real space Multiverse simulation snapshot boxes (green, red and blue diamonds), and the corresponding Gaussian expectation value ($\ref{eq:g1}$) with the same cosmological parameters (yellow squares, labeled `GRF').

The Gaussian expectation values are systematically higher than the genus measured from each simulation box -- this highlights the `gravitational smoothing' effect of non-linear gravitational collapse. The effect is $a_{\rm gr} \equiv A^{(\rm 3D)}_{\rm real}/ A^{(\rm 3D)}_{\rm G} =  0.88,0.90, 0.92$ for the $\Omega_{\rm m} = 0.21, 0.26, 0.31$ simulations respectively. To directly compare the measured genus amplitude from the SDSS MGS to the corresponding Gaussian expectation value, we correct $A_{\rm G}^{(\rm 3D)}$ by a factor of $(1 - \Delta_{\rm gr})^{-1}$ with $\Delta_{\rm gr} = 0.10$.

After correcting the measured genus amplitude $A^{(\rm 3D)}$ to account for non-linear redshift space distortion and gravitational smoothing, the next step is to compare $A^{(\rm 3D)}$ to the expectation value ($\ref{eq:g1}$) to obtain a set of parameter constraints. We minimize the simple $\chi^{2}$ function 

\begin{equation}\label{eq:chi3D} \chi^{2} = {[(1-\Delta_{\rm rsd} - \Delta_{\rm gr})^{-1} A^{(\rm 3D)} - A^{(\rm 3D)}_{\rm G}(\Omega_{\rm c}h^{2},n_{\rm s})]^{2} \over \sigma_{\rm 3D}^{2} } , \end{equation}

\noindent where $A^{(\rm 3D)}_{\rm G}$ is the Gaussian expectation value of the three-dimensional genus curve ($\ref{eq:g1}$), and is sensitive to $\Omega_{\rm c}h^{2}$, $n_{\rm s}$ and weakly to $\Omega_{\rm b}h^{2}$. As the dependence on $\Omega_{\rm b}h^{2}$ is very weak, we fix this parameter to its Planck best fit value $\Omega_{\rm b}h^{2} = 0.0222$ \cite{Aghanim:2018eyx}. $\sigma_{\rm 3D} =0.197 \times 10^{-6} {\rm Mpc}^{-3}$ is the statistical uncertainty on $A^{(\rm 3D)}$.

\begin{figure}
  \includegraphics[width=0.48\textwidth]{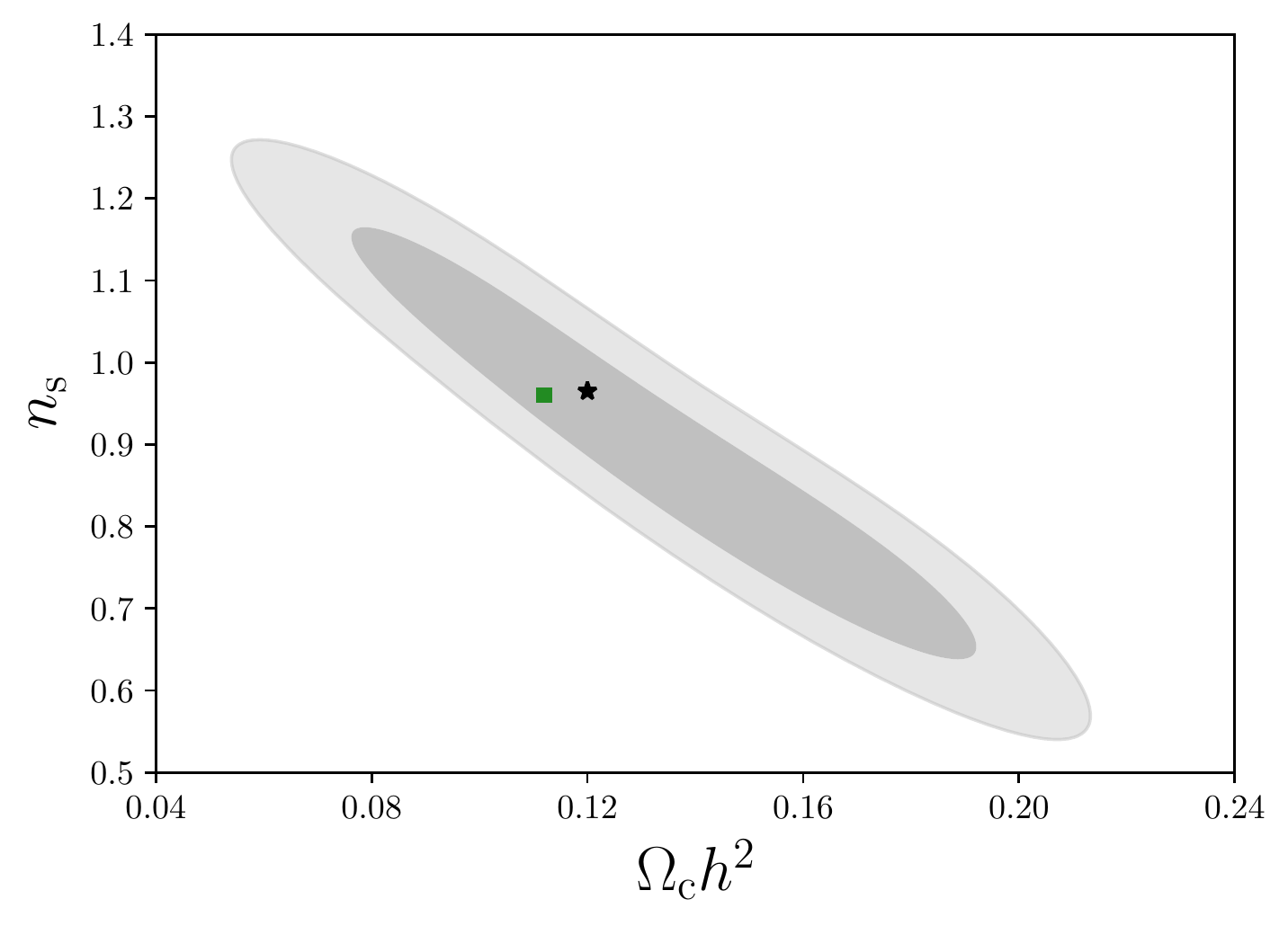} 
  \caption{The $1,2 -\sigma$ contours in the $n_{\rm s}$-$\Omega_{\rm c}h^{2}$ plane obtained by minimizing the chi square function ($\ref{eq:chi3D}$). The black star is the Planck best fit value of these parameters and green square the cosmological model used to infer the non-linear corrections to the measured genus curve.  }
  \label{fig:3Dcon}
\end{figure}

In Figure \ref{fig:3Dcon} we present the two-dimensional $1,2-\sigma$ contours in the $n_{\rm s}$, $\Omega_{\rm c}h^{2}$ plane, obtained by performing an MCMC parameter search, minimizing ($\ref{eq:chi3D}$). We observe a strong degeneracy between $n_{\rm s}$ and $\Omega_{\rm c}h^{2}$, as both can vary the degree of small scale power and hence increase/decrease the genus amplitude. The Planck best fit is shown as a black star, and the cosmological model of the Multiverse simulation used to make the non-linear redshift space distortion and gravitational smoothing corrections $\Delta_{\rm rsd}$, $\Delta_{\rm gr}$ is presented as a green square. Both are within the $1-\sigma$ contour.

In the next step of our analysis, we convert these constraints to a measure of the two-dimensional genus amplitude $A^{(\rm 2D)}_{\rm G}$.

\subsection{Conversion from cosmological parameters to $A^{(\rm 2D)}_{\rm G}$}

\begin{figure}
  \includegraphics[width=0.48\textwidth]{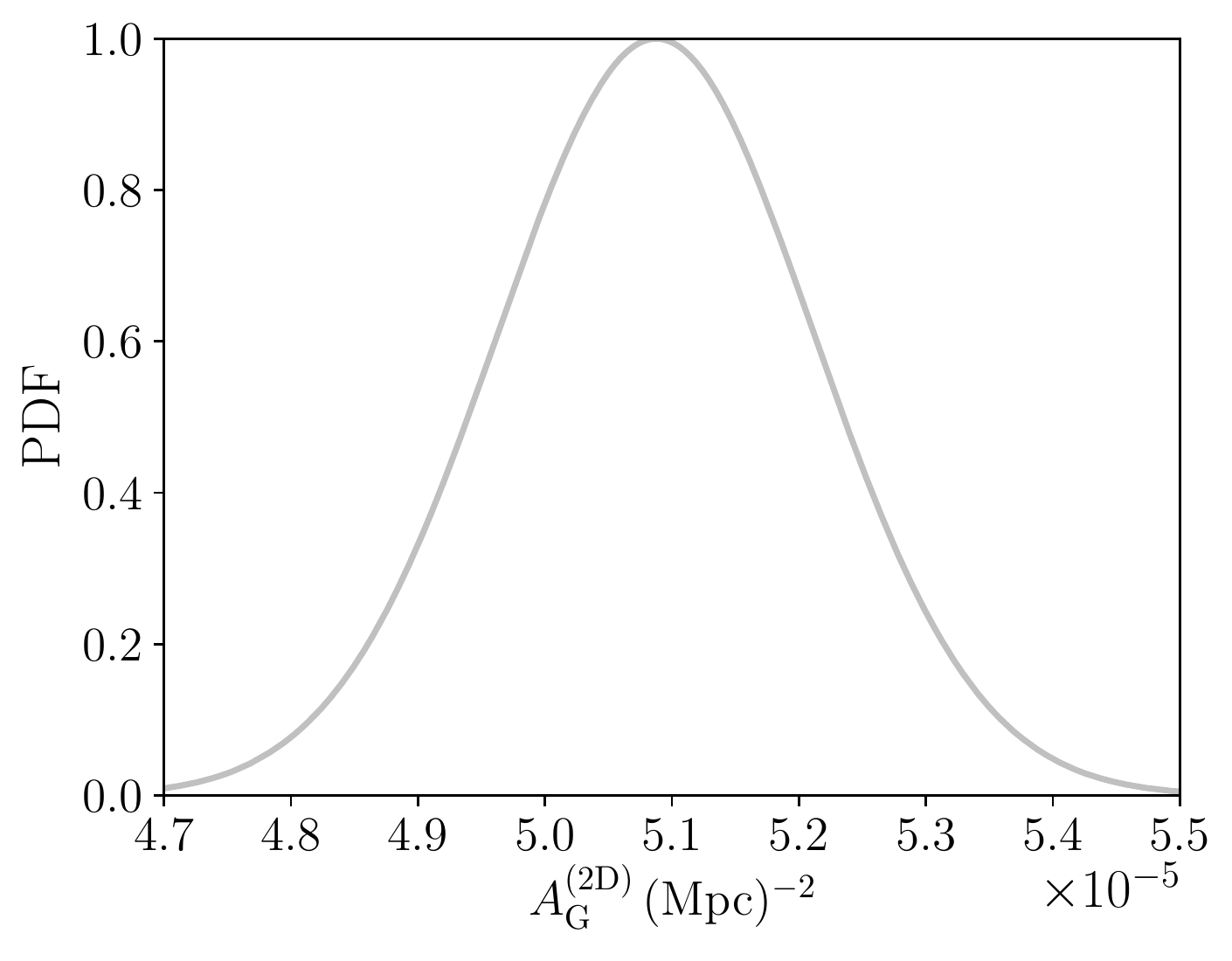}
  \includegraphics[width=0.48\textwidth]{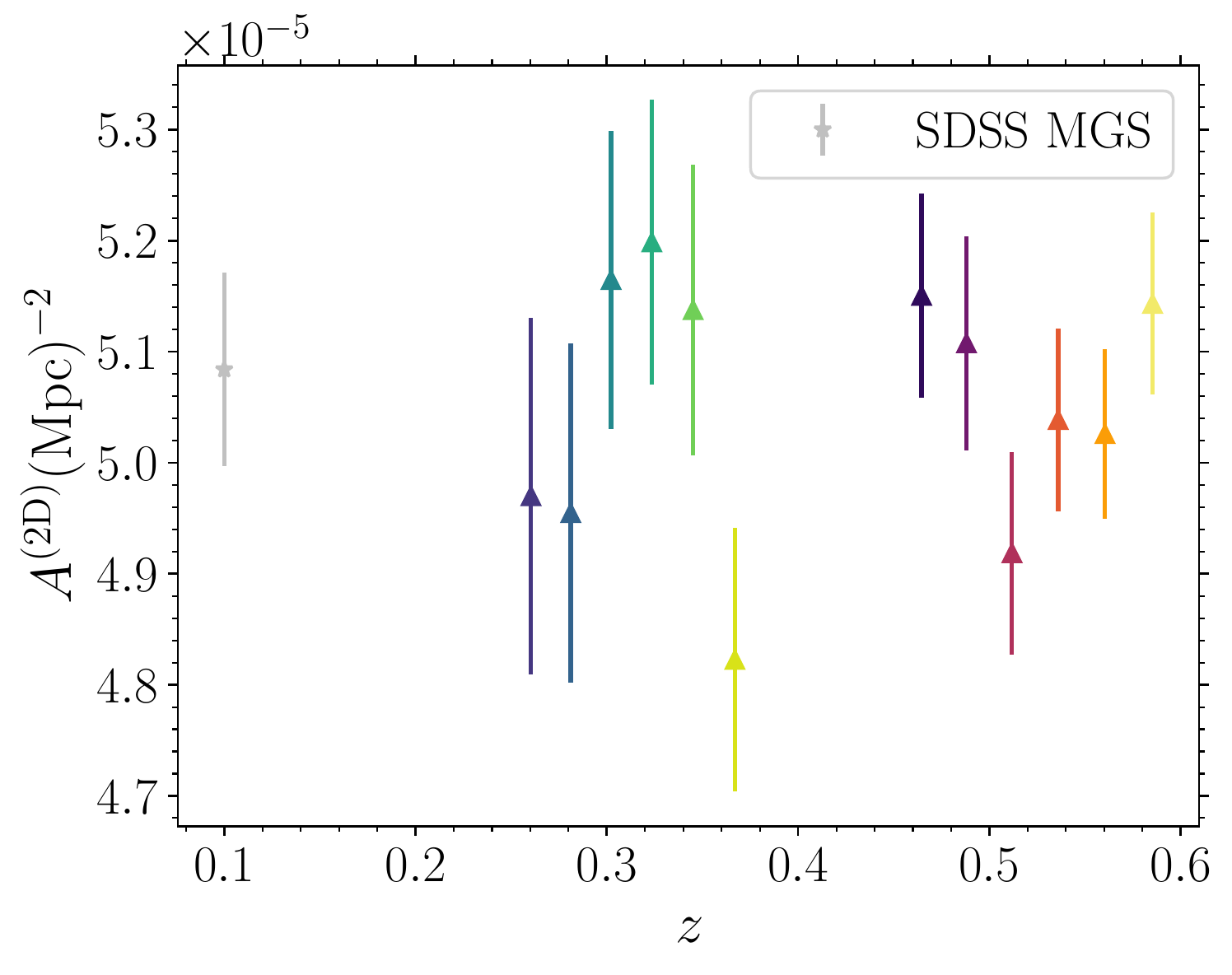}
  \caption{[Top panel] The probability distribution of the amplitude of the two-dimensional genus $A_{\rm G}^{(\rm 2D)}$ obtained by minimizing the $\chi^{2}$ function ($\ref{eq:chi3D}$). [Bottom panel] The two-dimensional genus measurement $A_{\rm G}^{(\rm 2D)}$ inferred from the low redshift SDSS MGS is presented as a silver star, along with the BOSS data points over the redshift range $0.25 < z <0.6$. The BOSS points are derived assuming a Planck cosmology. }
  \label{fig:3D2D}
\end{figure}

Finally, we transform from the cosmological parameters $\Omega_{\rm c}h^{2}$, $n_{\rm s}$ to a prediction for the two dimensional genus amplitude. To do so, we transform each parameter set and corresponding $\chi^{2}$ value $(\Omega_{\rm c}h^{2}, n_{\rm s},\chi^{2})$ from the previous section to $(A^{(\rm 2D)}_{\rm G}, \chi^{2})$ by inserting $\Omega_{\rm c}h^{2}$, $n_{\rm s}$ into the definition of the theoretical expectation of the two-dimensional genus amplitude ($\ref{eq:ag}$) using ($\ref{eq:s02}-\ref{eq:p3df}$), taking $b=2$ and $\bar{n} = 6.25\times 10^{-5} \, {\rm Mpc}^{-3}$ suitable for the BOSS galaxy sample, and smoothing scales $\Delta = 80 \, {\rm Mpc}$, $R_{\rm G} = 20 \, {\rm Mpc}$. The result is a one-dimensional probability distribution function for $A^{(\rm 2D)}_{\rm G}$, as inferred from the three-dimensional measurement. We present the resulting probability distribution in Figure \ref{fig:3D2D} (top panel). From this we infer the best fit and $1\sigma$ uncertainty on the two-dimensional genus amplitude as $A^{(\rm 2D)}_{\rm G} = 5.084 \pm 0.087 \times 10^{-5} \, ({\rm Mpc})^{-2}$.

To review, the $A^{(\rm 3D)}$ measurement provides a constraint of the shape of the linear matter power spectrum. We have used the best fit values and uncertainties on the parameters $\Omega_{\rm c}h^{2}$ and $n_{\rm s}$ to infer the best fit and uncertainty on the theoretical expectation value of $A^{(\rm 2D)}_{\rm G}$ at low redshift. In the bottom panel of Figure \ref{fig:3D2D} we present the two-dimensional genus measurement inferred from the SDSS MGS (silver star) and those directly measured from the BOSS data (multi-coloured data points). We take the SDSS MGS measurement to lie at $z=0.1$. We assume that the low redshift measurement of $A^{(\rm 2D)}_{\rm G}$ is insensitive to variations of the expansion history in what follows, and treat it as a constant $A^{(\rm 2D)}_{\rm G} = 5.084 \pm 0.087 \times 10^{-5} \, ({\rm Mpc})^{-2}$.

In this section, we have pursued a rather complex path to extracting $A^{(\rm 2D)}_{\rm G}$ from the low-redshift data. However, the reasoning behind our method lies in maximizing the constraining power of the data. For our method to provide a reasonable constraint, we must minimize the statistical uncertainty of the low-redshift measurement as far as possible, and this required us to select smaller smoothing scales than the high redshift data. The smallest smoothing scales that we can adopt at low and high-redshift are fixed by the mean galaxy separation of the SDSS MGS and BOSS catalogs; $\bar{r}_{\rm sdss} \sim 8.3 {\rm Mpc}$ and $\bar{r}_{\rm BOSS} \sim 25 {\rm Mpc}$ respectively. We cannot smooth below these scales without introducing unknown non-Gaussian systematics due to shot noise. This philosophy motivated our choice of $\Lambda_{\rm G} = 8.86 {\rm Mpc}$ ($= 6 \, {\rm Mpc}/h$) and $\Delta = 80 {\rm Mpc}$, $R_{\rm G} = 20 {\rm Mpc}$.

Given the different number densities, bias factors and smoothing scales used in the analysis of the SDSS MGS and BOSS data, the only logical approach to relate the two is to infer the theoretical expectation $A^{(\rm 2D)}_{\rm G}$ using one of the data sets, and proceed to compare this value to the second. To apply this method, one must carefully correct for any non-linear effects such as gravitational collapse using simulations.

\section{Parameter Constraints} 
\label{sec:constraints}

We are now able to combine the low- and high-redshift measurements to constrain the distance-redshift relation. To do so, we minimize the following $\chi^{2}$ function 

\begin{equation}\label{eq:chirz} \chi^{2} = \sum_{k=1}^{N_{\rm z}}\sum_{j=1}^{N_{\rm z}} {\bf p}_{j}{\rm cov^{-1}}_{jk}{\bf p}_{k} , \end{equation}

\noindent where ${\bf p}_{j} = A^{(\rm 2D)}_{j}/A^{(\rm 2D)}_{\rm G} - 1$ and ${\rm cov}_{jk} = (\sigma_{{\rm BOSS},j}^{2} + \sigma_{\rm SDSS}^{2})\delta_{jk}$; the covariance matrix is the sum of the statistical uncertainties on the BOSS and SDSS measurements. We have used a diagonal covariance matrix for our analysis, assuming that $A^{\rm (2D)}_{\rm G}$ represents the unbiased, central theoretical expectation value of the genus amplitude to which we compare our measured genus values to. A direct comparison between measured values of any statistic at high and low redshift would introduce correlation between ${\bf p}_{j}$ components. However, our analysis has used the measured low redshift data to infer the theoretical expectation value of the genus amplitude. A direct comparison of $A^{(\rm 2D)}_{\rm G}$ posterior probability distributions inferred from the SDSS data and twelve BOSS shells, for each parameter set used to infer the distance redshift relation, would provide a more rigorous statistical comparison. However, such a procedure is computationally intractable so we make the simplifying assumption that the high redshift shells are drawn from a PDF with central value given by the SDSS LRG value of $A^{\rm (2D)}_{\rm G}$.  

The second implicit assumption  with our choice of diagonal covariance matrix is that we have neglected large-wavelength correlations between the SDSS LRG and BOSS galaxy samples. When generating the covariance matrices for the two-dimensional genus measurements of the BOSS data, we found no statistically significant correlation between neighbouring shells. This indicates that the cross correlation of the genus measurements is negligible.

\begin{figure}
  \includegraphics[width=0.45\textwidth]{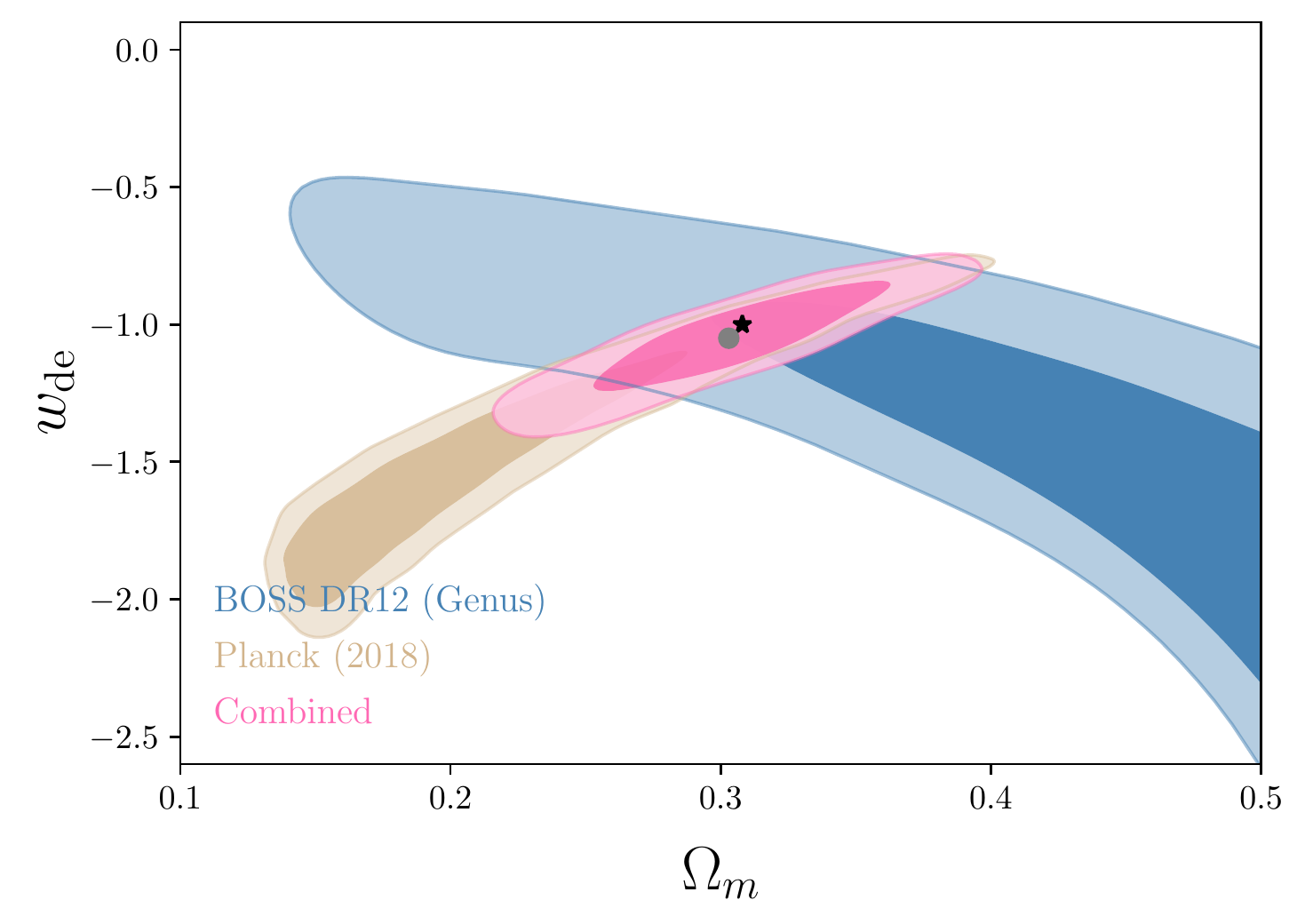}
  \includegraphics[width=0.45\textwidth]{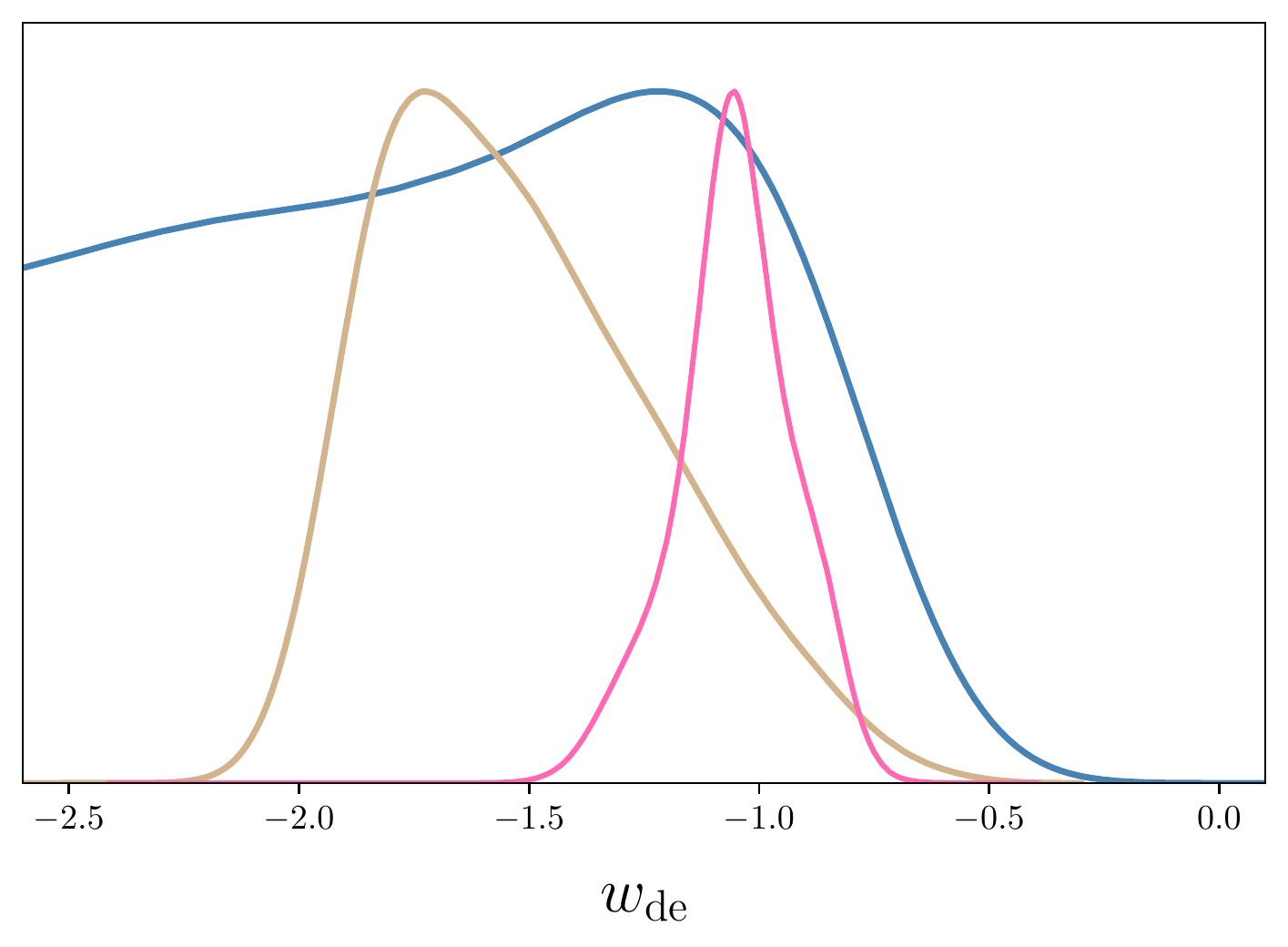}
  \caption{[Top panel] Two-dimensional $68,95\%$ contours in the $(\Omega_{\rm m},w_{\rm de})$ plane obtained by minimizing the $\chi^{2}$ function ($\ref{eq:chirz}$) and using the genus amplitude as a standard ruler (blue contours). The tan contours are the marginalised constraints in the $w_{\rm de}$-$\Omega_{\rm m}$ plane obtained from Planck temperature data \citep{Aghanim:2018eyx}, and the pink contours are the result of combining the genus and Planck $\chi^{2}$ functions in quadrature. The black star is the $\Lambda$CDM Planck best fit, and grey circle the best fit of the combined genus + Planck  (pink) contour.  [Bottom panel] The marginalised one-dimensional probability distribution functions of $w_{\rm de}$. The colour scheme is the same as in the top panel. }
  \label{fig:dz}
\end{figure}

We fix $h=0.677$ and vary $\Omega_{\rm m}$, $w_{\rm de}$. For each parameter set  $\Omega_{\rm m}$, $w_{\rm de}$, we estimate the distance to the centers of the $j$ redshift shells using $d_{\rm cm}(z_{j},\Omega_{\rm m}, w_{\rm de})$ and reconstruct the genus curves using angular smoothing scales $\theta_{{\rm G}, j} = R_{\rm G}/d_{\rm cm}(z_{j},\Omega_{\rm m}, w_{\rm de})$ and effective area of the data $A_{j} = 4\pi f_{\rm sky} d_{\rm cm}^{2}(z_{ j},\Omega_{\rm m}, w_{\rm de})$. After measuring the genus curves for the given expansion history, we calculate the $\chi^{2}$ function ($\ref{eq:chirz}$). The low redshift measurement $A_{\rm G}^{(\rm 2D)}$ is assumed to be independent of input cosmological model, as elucidated in section \ref{sec:res3d}. Performing a MCMC exploration of the two-dimensional parameter space, the resulting $1,2-\sigma$ contours (blue) are presented in Figure \ref{fig:dz}. The tan contour is the $w$CDM parameter constraint obtained from the Planck 2018 temperature data. If we combine the two data sets, we obtain a combined constraint on $\Omega_{\rm m}$ and $w_{\rm de}$ (pink contours)\footnote{Specifically, we used publicly available $w$CDM,  \href{https://wiki.cosmos.esa.int/planck-legacy-archive/index.php/Cosmological_Parameters}{MCMC chains} from the Planck collaboration \citep{Aghanim:2018eyx}, combining ($\ref{eq:chirz}$) and the Planck MCMC likelihoods in quadrature.}.  The marginalised parameter constraints for $\Omega_{\rm m},w_{\rm de}$ are presented in Table \ref{tab:dz}. 

The degeneracy between $\Omega_{\rm m}$ and $w_{\rm de}$, exhibited in the blue contour in Figure \ref{fig:dz}, has been found previously \citep{Park:2009ja,Appleby:2018jew}. The Planck temperature data presents an almost orthogonal contour in the $w_{\rm de}$-$\Omega_{\rm m}$ plane, so by combining these two data sets we can obtain a $\sim 15\%$ constraint on the equation of state of dark energy, $w_{\rm de} = -1.05^{+0.13}_{-0.12}$. 

The sensitivity of our test to $\Omega_{\rm m}$ and $w_{\rm de}$ is relatively weak as we are restricted to redshifts $z < 0.6$. A higher redshift measurement will improve the constraints considerably. Although the constraint is modest, the $\Lambda$CDM expansion history is consistent with the data over the redshift range considered. The constraint from the genus arises almost entirely from the combination of SDSS MGS low-redshift and CMASS data points; the LOWZ data have error bars that are too large and lie at a redshift that is too low to make a strong contribution. 

The results are not sensitive to the absolute value of the genus amplitude -- we are extracting information from the difference between different redshift bins. The absolute value also contains information, related to the shape of the matter power spectrum, as discussed further in a companion paper \cite{Appleby:2020pem}. 

The derived parameter constraints have been obtained under the assumption that the genus amplitude is a conserved quantity. For non-standard gravity or dark matter models, the matter power spectrum can possess redshift and scale dependent corrections. Similarly, we have assumed that dark energy perturbations do not significantly affect the shape of the matter power spectrum at low redshift.

\begin{center}
\begin{table}
 \begin{tabular}{||c  c  c ||}
 \hline
  Data \, & \, $\Omega_{\rm m}$ \, & \, $w_{\rm de}$  \\ [0.5ex] 
 \hline\hline
 Genus (BOSS+MGS) \, & \, $0.507^{+0.104}_{-0.126}$ \,  &  \, $-2.24^{+1.07}_{-1.14}$    \\ 
  \, & \, & \\
 \begin{tabular}{@{}c@{}}Genus (BOSS+MGS)  \\ + Planck (2018) \end{tabular}   \, & \, $0.303\pm 0.036$ \, & \,  $-1.05^{+0.13}_{-0.12}$  \\
 \hline
\end{tabular}
\caption{\label{tab:dz}Parameter best fit and $1-\sigma$ uncertainties, obtained by minimizing the $\chi^{2}$ function ($\ref{eq:chirz}$). After combining our genus likelihood with Planck 2018 temperature data, we obtain the second row.}
\end{table}
\end{center}

\section{Discussion}
\label{sec:discuss} 

In this work, we have obtained constraints on $w_{\rm de}$ and $\Omega_{\rm m}$ from the tomographic analysis of the two-dimensional slices of observed large-scale galaxy distribution. The amplitudes of the two-dimensional genus curves are measured in a series of concentric slices of density fields derived from the SDSS BOSS data. The amplitude at low-redshift is derived from the three-dimensional genus of the SDSS MGS data, and combined to find the cosmological parameters minimizing the redshift evolution of the genus.  In doing so, we arrive at a constraint of $w_{\rm de}=-2.24^{+1.07}_{-1.14}$, or $w_{\rm de} = -1.05^{+0.13}_{-0.12}$, $\Omega_{\rm m} = 0.303 \pm 0.036$ if we combine our analysis with Planck temperature data \citep{Aghanim:2018eyx}. The parameter constraints arising solely from the genus statistic are particularly weak; this is due to the strong degeneracy between parameters and also the limited statistical power that we are able to employ. The presence of shot noise fundamentally restricts our ability to reconstruct the density field from the galaxy point distribution, as we must smooth on scales of at least the mean galaxy separation \cite{Kim:2014axe,Blake:2013noa,Appleby:2017ahh}. In contrast, methods such as the Alcock-Paczynski (AP) test \cite{Li:2016wbl} (see also \cite{Li:2017nzs,Zhang:2018jfu,Li:2018nlh,Park:2019mvn,Zhang:2019jsu}) employ information from very small scales, eliminating non-perturbative, non-linear systematics using simulations. In addition, the AP test does not require the application of mass cuts to generate uniform data samples with redshift, as we are forced to. As a result, \cite{Li:2016wbl} were able to obtain tight parameter constraints on $w_{\rm de}$ and $\Omega_{\rm m}$ using the same BOSS data. For the genus to be competitive with other statistics, we must first learn how to model and remove observational systematics.

Beyond sampling noise, another dominant limitation of the method is in the comparison of high- and low-redshift measurements, as the low-redshift data are subject to large statistical uncertainty. This is the dominant contribution to the parameter uncertainties. The only way to evade this issue is to smooth the data on smaller scales, but in doing so we are increasingly exposed to non-linear physics. In this work we corrected the low-redshift, three-dimensional genus amplitude by factors of $\Delta_{\rm rsd} = 0.08$ and $\Delta_{\rm gr} = 0.10$ to account for redshift space distortion and gravitational collapse. These values were inferred from simulations. Better theoretical understanding of the non-linear regime and its impact on the genus curve will be necessary in the future to improve our analysis. Similarly, a better understanding of the effect of shot noise will allow us to probe smaller scales; in the current work we regard the mean galaxy separation $\bar{r}$ of a catalog to be a hard limit below which we are subjected to unknown non-Gaussian corrections. As the low redshift SDSS MGS is more dense than the BOSS catalog, we were able to smooth the former on smaller scales and thus extract more information. 

In a companion paper \cite{Appleby:2020pem}, we measured the genus curves of two-dimensional shells of the BOSS data and directly compared their amplitudes to the Gaussian expectation value. As we smooth the BOSS data with large scales $\Delta = 80 \, {\rm Mpc}$, $R_{\rm G} = 20 \, {\rm Mpc}$, we did not apply any non-linear correction factors to our measurements, and were able to use the Kaiser formula to estimate the effect of redshift space distortion. In \cite{Appleby:2020pem} we placed constraints on cosmological parameters that determine the shape of the linear matter power spectrum; $\Omega_{\rm c}h^{2}$ and $n_{\rm s}$. The information extracted in that work came from the absolute value of the genus amplitude. In the present analysis, we measure the redshift evolution of the genus amplitude, irrespective of its absolute value. One can interpret the two approaches as a measure of the initial condition/transfer function of the dark matter perturbations and a test of the expansion history respectively. In \cite{Appleby:2020pem}, we fixed the distance-redshift relation using the Planck 2018 best fit cosmology \cite{Aghanim:2018eyx} and measured the genus curves of the BOSS data a single time. We were able to do this as we restricted our analysis to the $\Lambda$CDM model, and the constraints obtained in this work are considerably weaker than those obtained in \cite{Appleby:2020pem}. The redshift evolution test considered here is {a measure of distance, and hence is principally sensitive to $\Omega_{\rm m}$ and the equation of state of dark energy}. 

To improve the parameter constraints, a number of avenues remain open. We can combine different low-redshift data sets, increasing the effective volume and reducing the statistical uncertainty. We can calculate analytically the non-linear corrections due to gravitational smoothing and redshift space distortion, which will provide a better understanding of the non-linear effects that we must account for on small scales. In addition, we can apply our method to high-redshift data, such as Lyman break galaxies. We expect that a high redshift data point will provide a significantly improved constraint on the expansion history. As the distance between observer and data increases, the effect of choosing an incorrect cosmology becomes more pronounced. 

Finally, the three-dimensional Minkowski Functionals contain more information than their two-dimensional counterparts, and a complete analysis of the three-dimensional field will be forthcoming. In this work, and throughout a series of papers \citep{Appleby:2017ahh,Appleby:2018jew,Appleby:2020pem}, we have focused on the two-dimensional genus, extracted from shells of the three-dimensional galaxy distribution. The reasoning behind this choice is two-fold. First, the BOSS galaxy catalog is relatively sparse, and we mitigate this issue by taking thick slices along the line of sight. Binning galaxies in this way is a smoothing choice, so we can interpret our approach as anisotropic smoothing perpendicular and parallel to the line of sight. Smoothing on larger scales parallel to the line of sight allows us to use linear redshift space distortion physics, which is important as non-linear redshift space distortion effects on topological statistics are not yet well understood. Second, in future work we intend to compare our results with higher redshift photometric redshift catalogs, which will require galaxies to be binned into thick shells. An understanding of how photometric redshift uncertainty modifies our analysis must be further explored before this comparison can be made.

\section*{Acknowledgement}

SAA is supported by an appointment to the JRG Program at the APCTP through the Science and Technology Promotion Fund and Lottery Fund of the Korean Government, the Korean Local Governments in Gyeongsangbuk-do Province and Pohang City and by a KIAS Individual Grant QP055701 via the Quantum Universe Center at Korea Institute for Advanced Study. SEH was supported by Basic Science Research Program through the National Research Foundation of Korea funded by the Ministry of Education (2018\-R1\-A6\-A1\-A06\-024\-977). We thank the Korea Institute for Advanced Study for
providing computing resources (KIAS Center for Advanced
Computation Linux Cluster System).

Funding for SDSS-III has been provided by the Alfred
P. Sloan Foundation, the Participating Institutions,
the National Science Foundation, and the U.S. Department
of Energy Office of Science. The SDSS-III web
site is http://www.sdss3.org/. SDSS-III is managed by
the Astrophysical Research Consortium for the Participating
Institutions of the SDSS-III Collaboration including
the University of Arizona, the Brazilian Participation
Group, Brookhaven National Laboratory, Carnegie Mellon
University, University of Florida, the French Participation
Group, the German Participation Group, Harvard
University, the Instituto de Astrofisica de Canarias, the
Michigan State/Notre Dame/JINA Participation Group,
Johns Hopkins University, Lawrence Berkeley National
Laboratory, Max Planck Institute for Astrophysics, Max
Planck Institute for Extraterrestrial Physics, New Mexico
State University, New York University, Ohio State
University, Pennsylvania State University, University of
Portsmouth, Princeton University, the Spanish Participation
Group, University of Tokyo, University of Utah,
Vanderbilt University, University of Virginia, University
of Washington, and Yale University.

The massive production of all MultiDark-Patchy mocks for the BOSS Final Data Release has been performed at the BSC Marenostrum supercomputer, the Hydra cluster at the Instituto de Fısica Teorica UAM/CSIC, and NERSC at the Lawrence Berkeley National Laboratory. We acknowledge support from the Spanish MICINNs Consolider-Ingenio 2010 Programme under grant MultiDark CSD2009-00064, MINECO Centro de Excelencia Severo Ochoa Programme under grant SEV- 2012-0249, and grant AYA2014-60641-C2-1-P. The MultiDark-Patchy mocks was an effort led from the IFT UAM-CSIC by F. Prada’s group (C.-H. Chuang, S. Rodriguez-Torres and C. Scoccola) in collaboration with C. Zhao (Tsinghua U.), F.-S. Kitaura (AIP), A. Klypin (NMSU), G. Yepes (UAM), and the BOSS galaxy clustering working group.

Some of the results in this paper have been derived using the healpy and HEALPix package

\section*{Appendix A -- Systematic Effects}

In Section \ref{sec:theory} we listed three effects that can introduce a small evolution in the genus amplitude. In the Appendix we consider each point in turn and in isolation, to confirm that we have all known systematics under control. 

We use all sky, mock galaxy lightcone data from the Horizon Run 4 dark matter simulation project to perform these tests. We direct the reader to \cite{Kim:2015yma,Hong:2016hsd} for information on the simulation and mock galaxy catalogs. We use the lightcone data over the range $0.15 < z < 0.7$, creating $N=20$ shells and applying mass cuts to generate constant number density samples, exactly as we did for the BOSS data. We bin the galaxies into shells of thickness $\Delta = 80 {\rm Mpc}$ and smooth perpendicular to the plane with comoving scale $R_{\rm G} \, {\rm Mpc}$. The simulation was performed using a flat $\Lambda$CDM cosmology with parameters $h=0.72$, $\Omega_{\rm m} = 0.26$, $\sigma_{8} = 0.794$, $n_{\rm s} = 0.96$.

\subsection*{1 - S\lowercase{hot Noise}}

Shot noise is the single largest systematic associated with information extraction using the genus curve \cite{Kim:2014axe}. There are two issues associated with this phenomenon -- it is non-Gaussian and it can potentially introduce a redshift evolution of the genus amplitude. 

First regarding the non-Gaussianity. As a simple approximation, we have corrected for shot noise by adding a constant white noise contribution to the total power spectrum; $P_{\rm SN} = 1/\bar{n}$. In reality, the noise is a Poisson process (roughly speaking), but when writing the genus in terms of a Hermite polynomial expansion as in ($\ref{eq:mat1}$) we have implicitly assumed that the field is drawn from a perturbatively Gaussian distribution. As a Poisson distribution possesses a different moment generating function compared to a Gaussian, we can expect that shot noise will introduce modifications to the shape of the genus curve. This was observed in both \cite{Kim:2014axe} and \cite{Appleby:2018jew}. If shot noise becomes significant, then the shape of the genus will not be well represented by the first few Hermite polynomials and we lose the interpretation of the genus amplitude as the ratio of second order cumulants of the perturbatively Gaussian field that we are trying to measure. In short, the field that we measure is non-Gaussian due to both gravitational collapse and the nature in which it is sampled. However, only the gravitational non-Gaussianity is treated in the expansion ($\ref{eq:mat1}$). This issue is suppressed if we smooth on scales larger than the mean galaxy separation, in which case the shot noise effect can be approximately represented by the white noise term $P_{\rm SN}$. This remains an imperfect approximation except in the limit $R_{\rm G} \gg \bar{r}$.

\begin{figure}
  \includegraphics[width=0.45\textwidth]{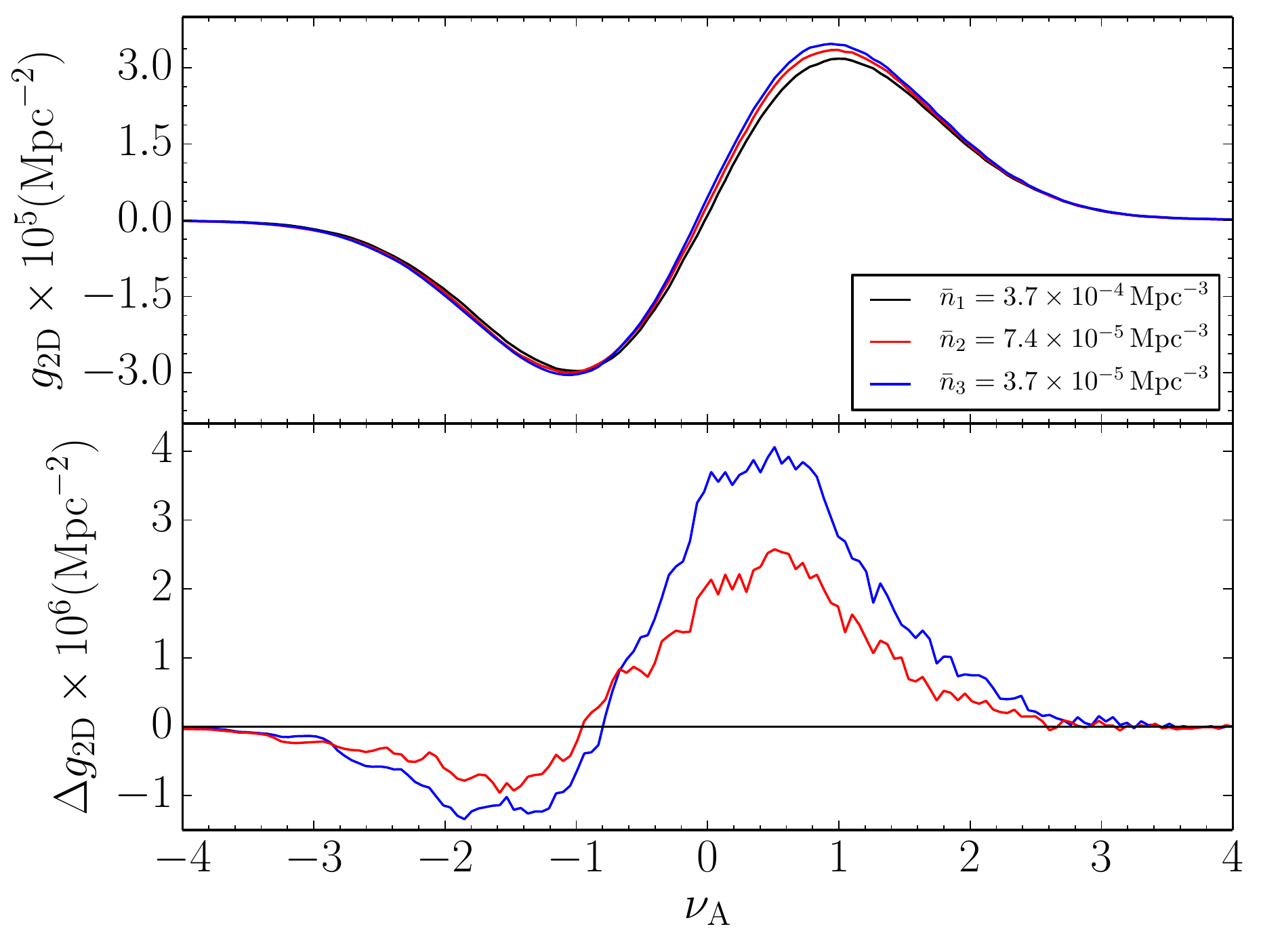}
  \includegraphics[width=0.45\textwidth]{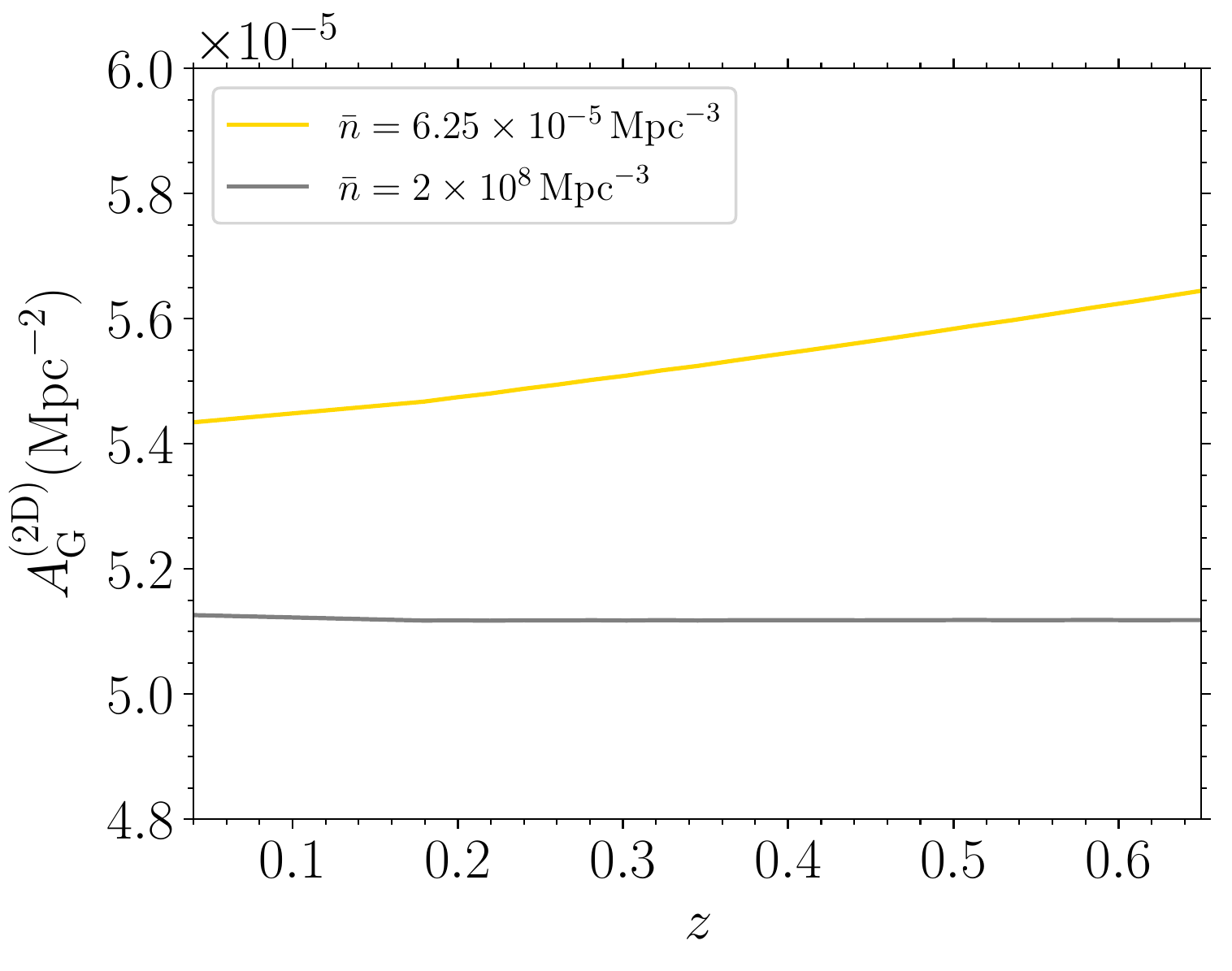}
    \includegraphics[width=0.45\textwidth]{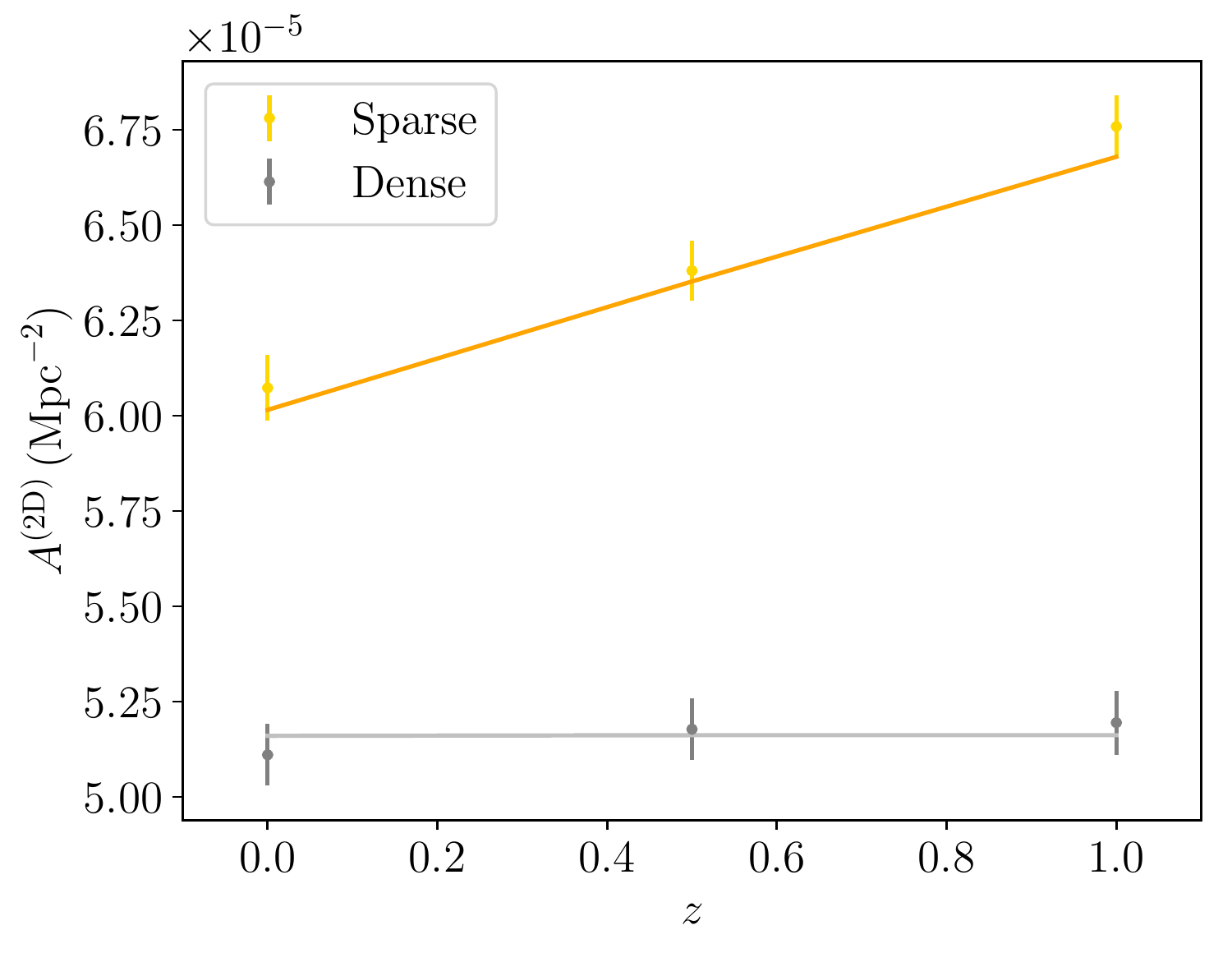}
  \caption{[Top panel] The average genus curves extracted from all-sky mock galaxy lightcone data, taking three mass cuts to fix a constant number density in each shell $\bar{n}_{1} = 3.7 \times 10^{-4} {\rm Mpc}^{-3}$ (black), $\bar{n}_{2} = 7.4 \times 10^{-5} {\rm Mpc}^{-3}$ (red), $\bar{n}_{3} = 3.7 \times 10^{-5} {\rm Mpc}^{-3}$ (blue). We also exhibit the difference $\Delta g_{\rm 2D}$ between the sparse samples and the dense catalog.   [Middle panel]  The Gaussian prediction for the genus amplitude in real space, for a model with no shot noise (grey) and with the fiducial number density $\bar{n} = 6.25 \times 10^{-5} {\rm Mpc}^{-3}$ (yellow). Shot noise introduces significant evolution in the genus amplitude. [Bottom panel] Genus amplitudes extracted from two-dimensional slices of dark matter particle snapshot boxes for a sparse $N=64^3$ (gold points) and dense $N=512^3$ (grey points) sample. The solid lines are the Gaussian prediction. }
  \label{fig:app1}
\end{figure}

To present the non-Gaussianity induced by the shot noise sampling, we measure the genus of two-dimensional shells of the Horizon Run 4 mock galaxy lightcone in real space, fixing the smoothing scales $\Delta = 80 {\rm Mpc}$, $R_{\rm G} = 20 {\rm Mpc}$ and applying three mass cuts to the data to fix the number density of our galaxy sample as $\bar{n}_{1} = 3.7 \times 10^{-4} {\rm Mpc}^{-3}$, $\bar{n}_{2} = 7.4 \times 10^{-5} {\rm Mpc}^{-3}$ and $\bar{n}_{3} = 3.7 \times 10^{-5} {\rm Mpc}^{-3}$. We assume that the most dense sample has a shot noise contribution that is suppressed, as the mean galaxy separation is much lower than the smoothing scale $R_{\rm G} = 20 {\rm Mpc}$. We label the genus curves $g^{(1)}_{\rm 2D}(\nu_{\rm A})$, $g^{(2)}_{\rm 2D}(\nu_{\rm A})$ and $g^{(3)}_{\rm 2D}(\nu_{\rm A})$ respectively, where $1,2,3$ superscripts denote the number density cuts $\bar{n}_{1,2,3}$. We repeat our measurement for twenty concentric non-overlapping shells and take the average genus curve to show the effect of shot noise. In Figure \ref{fig:app1} (top panel) we exhibit the average genus curves for $\bar{n}_{1}$ (black), $\bar{n}_{2}$ (red) and $\bar{n}_{3}$ (blue). Below we also present the difference between the genus curves $\Delta g_{\rm 2D}^{(2)} = g^{(2)}_{\rm 2D} - g^{(1)}_{\rm 2D}$ (blue) and $\Delta g_{\rm 2D}^{(3)} = g^{(3)}_{\rm 2D} - g^{(1)}_{\rm 2D}$ (red) respectively. Clearly the difference between the genus curves is not simply an amplitude shift -- the shape of the residual curve is both shifted and distorted. This is due to the non-Gaussian nature of the sampling, and is most significant in the most sparse sample $\bar{n}_{3}$. For the case $\bar{n}_{2}$, these effects are less pronounced but still present. This indicates that our treatment of shot noise using the white noise contribution $P_{\rm SN}$ is imperfect. This effect will be further studied by the authors in the future. 

However, even if the effect of shot noise can be represented by a white noise term $P_{\rm SN} = 1/\bar{n}$, it can still generate a redshift evolution in the genus amplitude. This is because we are fixing $P_{\rm SN}$ to be constant at each redshift, but the matter power spectrum has a decreasing amplitude with increasing redshift. Hence the shot noise term increases in significance to the past, and will manifest as an increasing genus amplitude with increasing $z$.

To show the hypothetical redshift evolution of the genus amplitude, in Figure \ref{fig:app1} (middle panel) we present the theoretical expectation $A_{\rm G}^{(\rm 2D)}$ as a function of redshift, using ($\ref{eq:ag}$) with power spectrum ($\ref{eq:p3df}$) and taking the real space power spectrum (that is, setting $\beta=0$ in equation ($\ref{eq:p3df}$)). We plot the amplitude assuming negligible shot noise, setting an arbitrarily high hypothetical number density $\bar{n} = 2 \times 10^{8} {\rm Mpc}^{-3}$ (grey line) and the fiducial number density of the BOSS catalog used in this work $\bar{n} = 6.25 \times 10^{-5} {\rm Mpc}^{-3}$ (yellow line), both with constant linear galaxy bias $b=2$. The grey line represents the idealised case and as expected is constant; in this instance the genus is a measure of the shape of the linear matter power spectrum. The yellow curve represents a hypothetical sparse galaxy catalog with constant large scale galaxy bias -- the shot noise contribution causes the genus amplitude to evolve with redshift.

To confirm this behaviour, we take simulated dark matter particle snapshot boxes of volume $V = (1024 {\rm Mpc}/h)^{3}$ at $z=0, 0.5, 1$ and sampled $512^3$ (dense) and $64^3$ (sparse) particles randomly. We then construct flat, two-dimensional slices of thickness $\Delta = 80 \, {\rm Mpc}$ and smoothed in the plane with $R_{\rm G} = 20 \, {\rm Mpc}$ Gaussian kernel. The we extract the genus from these fields and then the genus amplitudes. The results are presented in the bottom panel of Figure \ref{fig:app1}. The points/error bars are the mean and standard deviation of $15$ slices of the snapshot boxes, with the yellow/grey points corresponding to the sparse and dense samples respectively. The solid yellow/grey lines are the Gaussian expectation value for the given number density (and bias factor $b=1$, as we are using dark matter particles). The behaviour of the middle panel is reproduced, the sparse sample exhibits a systematic evolution in redshift, but the dense sample is conserved.  

In this work we have fixed the galaxy bias of the BOSS galaxies to be constant, $b=2$. If the bias evolves with redshift, then this must also be taken into account when assessing the effect of shot noise. The net effect depends on the relative amplitude of the galaxy power spectrum -- hence $b^{2}(z) D^{2}(z) A_{\rm s}$ -- and the shot noise term $P_{\rm SN}$, where $D^{2}(z)$ is the growth rate and $A_{\rm s}$ is the primordial amplitude.

\subsection*{2 - R\lowercase{edshift Space Distortion}}

\begin{figure}
  \includegraphics[width=0.45\textwidth]{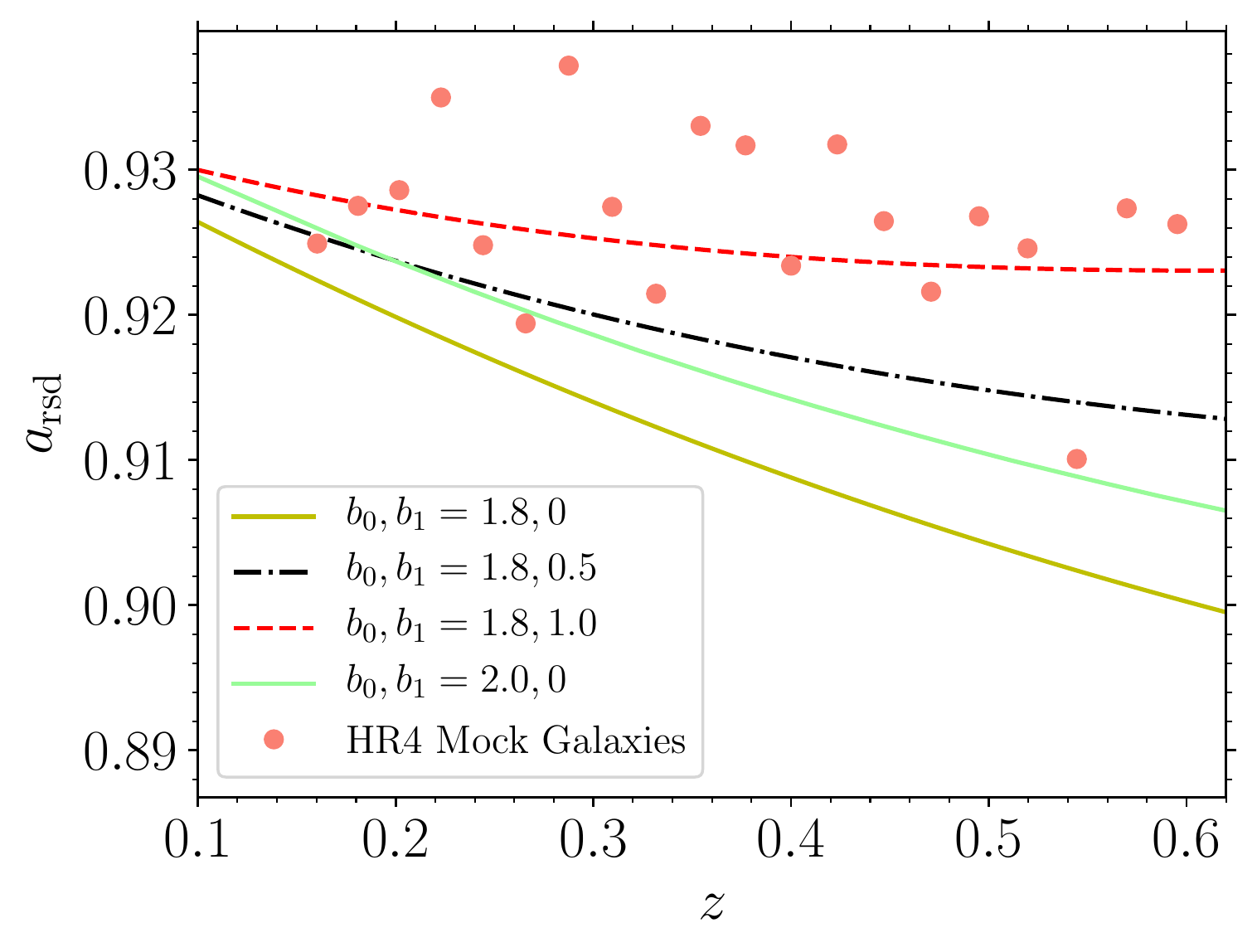}
  \includegraphics[width=0.45\textwidth]{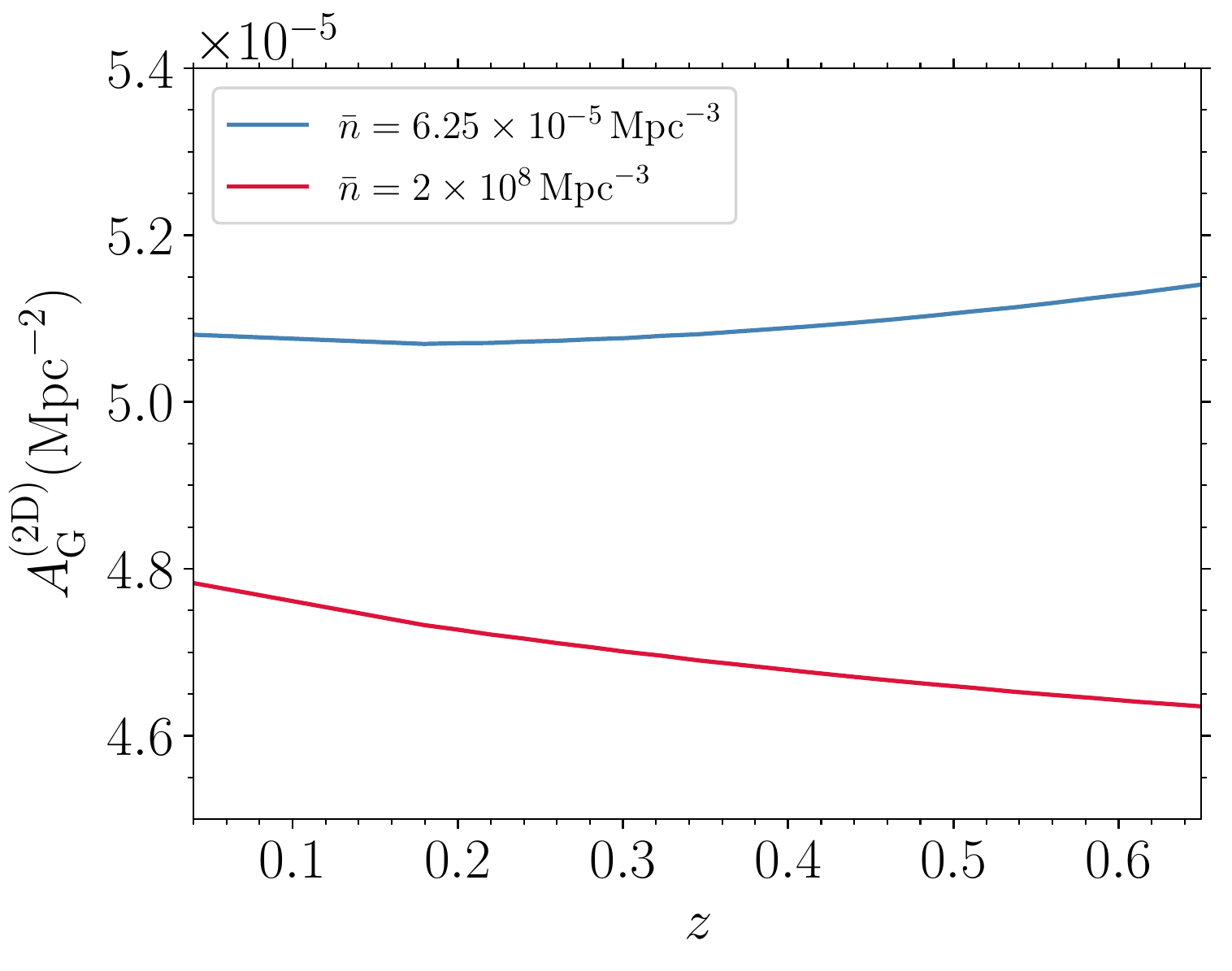}
  \includegraphics[width=0.45\textwidth]{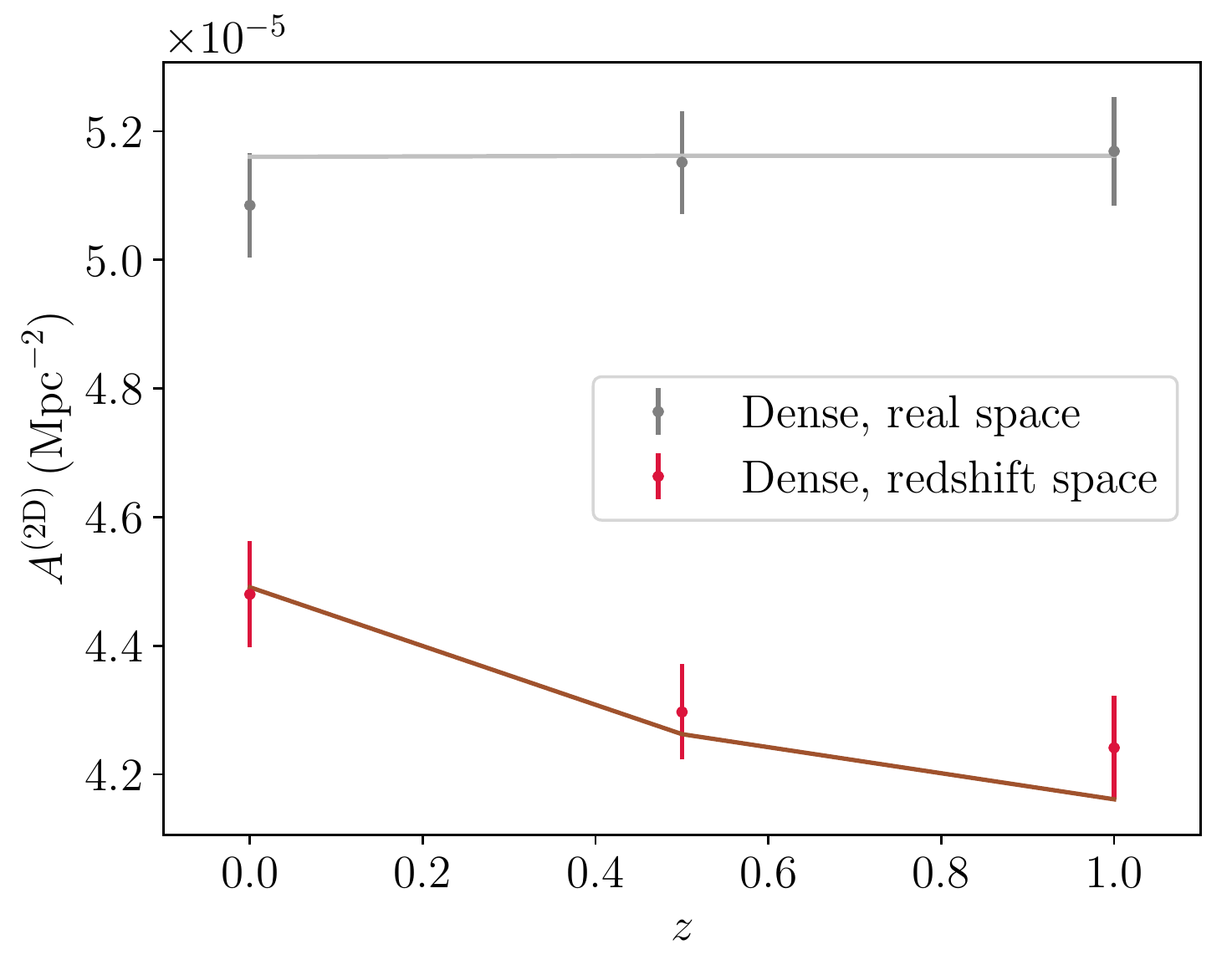}
  \caption{[Top panel] The ratio of genus amplitudes $A_{\rm G}^{(\rm 2D)}$ as measured in redshift and real space, for different bias models $b(z) = b_{0} + b_{1}z$. The green line corresponds to the fiducial values used in this work $b_{0} = 2$, $b_{1}=0$ in the galaxy bias model $b(z) = b_{0} + b_{1}z$. The red points are the values of $a_{\rm rsd}$ inferred from the mock galaxies from the Horizon Run 4 lightcone in real and redshift space. Generally, the effect of linear redshift space distortion is to decrease the genus amplitude by $\sim 8\%$ and introduce a weak redshift dependence.  [Middle panel] The expectation value of the genus amplitude in redshift space, taking $b=2$, $\bar{n} = 2 \times 10^{8} {\rm Mpc}^{-3}$ (red) and $\bar{n} = 6.25 \times 10^{-5} {\rm Mpc}^{-3}$ (blue).  The measured genus amplitude from the BOSS data should trace the blue curve. [Bottom panel] The genus amplitude extracted from dark matter snapshot boxes for a dense sample $512^3$ in real (grey) and redshift (red) space.  }
  \label{fig:app2}
\end{figure}

The effect of redshift space distortion is to decrease the genus amplitude by $\sim 8\%$, and introduce a mild redshift dependence. To show this effect, in Figure \ref{fig:app2} (top panel) we present the ratio of $A_{\rm G}^{(\rm 2D)}$ redshift and real ($\beta=0$) space, obtained from the theoretical expectation ($\ref{eq:ag}$) assuming negligible shot noise (that is, fixing $\bar{n} = 2 \times 10^{8} {\rm Mpc}^{-3}$) and different galaxy bias values. The green solid line is the fiducial, constant galaxy bias used in this work $b=b_{0}$. We also exhibit $a_{\rm rsd}$ for different linear galaxy bias models $b(z) = b_{0} + b_{1}z$; $(b_{0},b_{1}) = (1.8,0), (1.8,0.5), (1.8,1)$ (yellow solid, black dash-dot and red dashed lines respectively). We also present the values of $a_{\rm rsd}$ inferred from the Horizon Run 4 all-sky mock galaxy shells as pale red points. For our fiducial choice $b=2$, the genus amplitude decreases by $\sim {\cal O}(8\%)$ and decreases with increasing redshift when measured in redshift space (green line). For different bias factors, the redshift dependence of $a_{\rm rsd}$ can change significantly (cf yellow, black, red lines).  

In the middle panel of the figure, we plot $A_{\rm G}^{(\rm 2D)}$ in redshift space for two different number densities -- $\bar{n} = 2 \times 10^{8} {\rm Mpc}^{-3}$ (red) and fiducial number density used in this work $\bar{n} = 6.25 \times 10^{-5} {\rm Mpc}^{-3}$ (blue), fixing $b=2$. In the absence of shot noise, the genus amplitude decreases with redshift (red line), however as shown in the previous section shot noise acts to increase the genus amplitude with $z$. The net effect is that redshift space distortion decreases and shot noise increases $A_{\rm G}^{(\rm 2D)}$, with the result that the measured genus amplitude of the galaxy catalogs should remain approximately constant over the redshift range considered in this work $0.1 < z < 0.7$. Specifically, the measured genus amplitude should follow the blue curve in the bottom panel. We stress however, that this argument is sensitive to the galaxy sample. Different bias factors, number densities and redshift ranges will not necessarily yield a genus amplitude that is conserved with redshift. 

We again confirm our hypothesis that redshift space distortion introduces a mild dependence of the genus amplitude on redshift by extracting $A^{(\rm 2D)}$ from slices of dark matter particle data from our simulation. We take $z=0,0.5,1$ snapshot boxes of volume $V = (1024 {\rm Mpc}/h)^{3}$, sub-sample $512^3$ particles from the full data then perturb the particles along the $x_{3}$ direction according to their velocities to create plane, parallel redshift space distorted slices. We then extract the two-dimensional genus amplitude from slices using $\Delta = 80 {\rm Mpc}$, $R_{\rm G} = 20 {\rm Mpc}$ as before. The results are exhibited in the bottom panel of Figure \ref{fig:app2}. The grey/red points and error bars are the mean and standard deviation of $15$ slices of the snapshot boxes in real/redshift space respectively. We observe no evolution in real space, but a systematic decrease in the genus amplitude in redshift space.

Future dense and high redshift galaxy catalogs will not suffer from the many of the issues discussed in this work. For these data, the shot noise contribution will be significantly reduced. In this case, the correct course of action would be to correct the measured genus amplitudes by a multiplicative factor to convert them to real space \cite{1996ApJ...457...13M}, after which they should be conserved with redshift. This procedure was undertaken for mock galaxies in \cite{Appleby:2018tzk}.

\subsection*{3 - G\lowercase{ravitational Smoothing}}

It is well known that higher order corrections in the non-Gaussian expansion of the genus curve $\sim {\cal O}(\sigma_{0}^{2})$ will modify the genus amplitude ; empirically it has been observed that the genus amplitude decreases on small scales compared to the Gaussian expectation value when measured from galaxy catalogs \cite{1989ApJ...345..618M,1991ApJ...378..457P,2005ApJ...633....1P}. To test the magnitude of this effect, we measure the coefficient of the $a_{3}$ Hermite polynomial expansion of the genus curve -- 

\begin{equation} g_{\rm 2D} \simeq a_{1} e^{-\nu_{\rm A}^{2}/2} \left[ a_{0}H_{0} + H_{1} + a_{2} H_{2} + a_{3}H_{3} \right] . \end{equation} 

\noindent According to the non-Gaussian perturbative expansion of the genus \cite{Matsubara:1994we,2000astro.ph..6269M,Pogosyan:2009rg,Gay:2011wz,Codis:2013exa}, $a_{1}$ is the genus amplitude, $a_{0,2}$ are the first order corrections of order $a_{0,2} \sim {\cal O}(\sigma_{0})$ and we can expect $a_{3}$ will be induced at second order $a_{3} \sim {\cal O}(\sigma_{0}^{2})$. We therefore use this term as a proxy to estimate the magnitude of higher order corrections to the genus amplitude. 

\begin{figure}
  \includegraphics[width=0.45\textwidth]{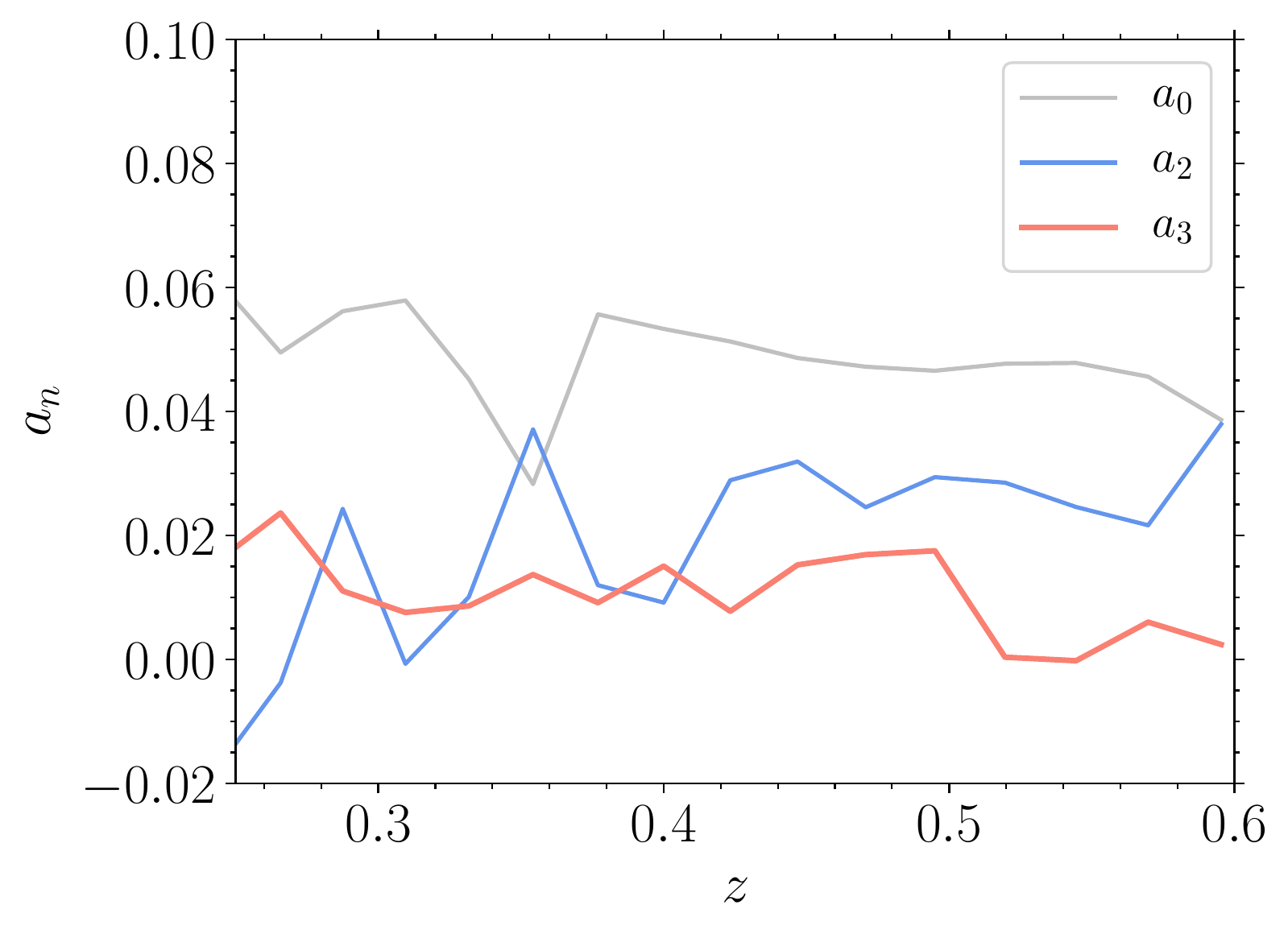}
  \caption{The Hermite polynomial coefficients $a_{0}, a_{2}, a_{3}$ (grey, blue, red) obtained from twenty shells of all-sky mock galaxy data. There is no strong evidence of evolution of $a_{3}$.   }
  \label{fig:app4}
\end{figure}

We extract $a_{0}, a_{2}, a_{3}$ from the twenty all-sky lightcone shells of Horizon Run 4, in redshift space, by integrating the genus curve using

\begin{equation} a_{n} ={1 \over n!} { \int_{-4}^{4} d\nu_{\rm A} g_{\rm 2D}(\nu_{\rm A}) H_{n}(\nu_{\rm A}) \over \int_{-4}^{4} d\nu_{\rm A} g_{\rm 2D}(\nu_{\rm A}) H_{1}(\nu_{\rm A})}  , \end{equation} 

\noindent taking $\nu_{0} =4$. In Figure \ref{fig:app4} we present $a_{0}, a_{2}, a_{3}$ (grey, blue, red). The red curve is the next to leading order correction term $a_{3}$. There is some suggestion that $a_{3}$ is increasing with decreasing redshift, from $\sim 0.02$ at $z=0.25$ to $\sim 0.01$ at $z=0.6$. Although the effect is small and the statistical uncertainty large, the higher order, non-linear corrections require further study. The $a_{3}$ term is present at the $1\%$ level at the scales probed. 

The $a_{0},a_{2}$ coefficients are  perturbatively small at the scales studied in this work. These terms can be interpreted as integrals over the Bispectrum, and contain complementary information to the amplitude studied in this work.  $a_{2}$ exhibits some evidence of evolution over the redshift range under consideration.

\subsection*{4 - L\lowercase{ack of high threshold critical points}}

The first three issues described above are physical effects. The fourth is a purely spurious systematic that can be introduced into the analysis if we improperly select the $\nu_{\rm A}$ threshold range. Specifically, one can observe evolution of the genus amplitude with redshift if we measure the genus over threshold values that are too high. The reason for this lies in the relation between $\nu_{\rm A}$ and $\nu$. The $\nu_{\rm A}$ parameterisation of the genus curve selects thresholds that have the same area fraction as a Gaussian random field. However, since the galaxy catalogs occupy a finite area, high threshold peaks will not be represented within the observed domain, and the area fraction will be systematically under-represented compared to a hypothetical Gaussian random field of arbitrarily large extent. This leads to an increase in the genus curve in the high $\nu_{\rm A}$ tails, which increases the genus amplitude. This can introduce spurious redshift evolution because the area of the data at low redshift is smaller than at high redshift, and so the low-$z$ regime will contain fewer high threshold peaks.

\begin{figure}[b!]
  \includegraphics[width=0.45\textwidth]{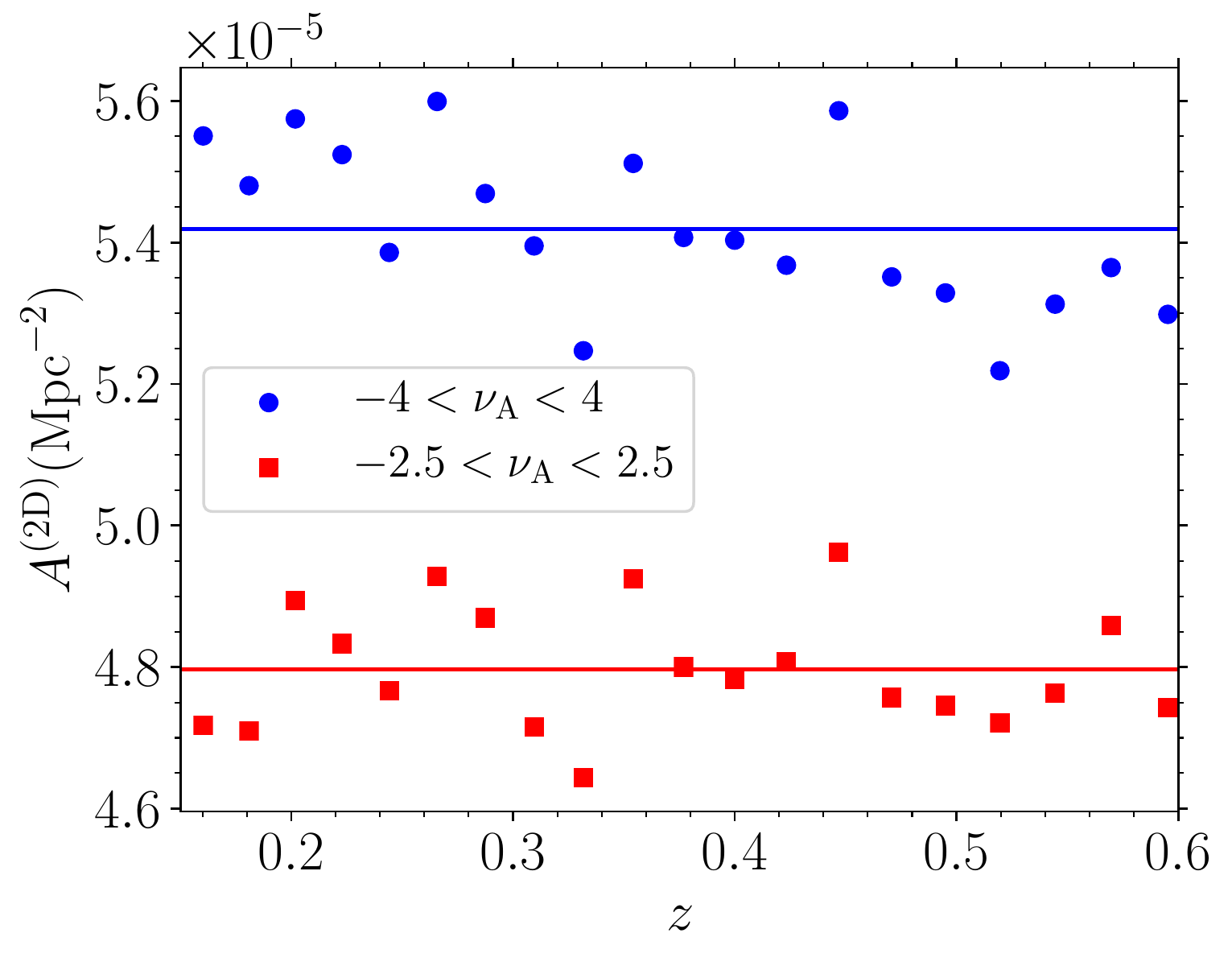}
  \caption{The amplitude proxy $A^{(\rm 2D)}$ defined in equation ($\ref{eq:inte}$), measured from the twenty shells of lightcone data. The red/blue points correspond to $\nu_{0} = 2.5$, $\nu_{0} = 4$ respectively.  The solid horizontal lines are the mean values of the respective points. One can clearly observe a systematic evolution in the blue points, due to the lack of high threshold maxima/minima at low-$z$.  }
  \label{fig:app3}
\end{figure}

We can eliminate this effect by restricting our analysis to $\nu_{\rm A}$ threshold values that are well sampled at each redshift. To present the effect, we take the twenty all-sky lightcone mock galaxy shells from the Horizon Run 4 simulation in real space, smooth them and then apply a mask, only keeping a spherical cap of data of radius $\theta_{\rm cap} = \pi/(2\sqrt{2}) \, {\rm rad}$. We select this value as the area fraction of such a cap roughly matches the area of the BOSS mask. We then measure the genus of this subset of data over the threshold ranges $-4 < \nu_{\rm A} < 4$ and $-2.5 < \nu_{\rm A} < 2.5$. As a proxy for the genus amplitude, we use the following integral

\begin{equation}\label{eq:inte} A^{(\rm 2D)} \simeq {1 \over \sqrt{2\pi}}  \int_{-\nu_{0}}^{\nu_{0}} g_{\rm 2D} \nu_{\rm A} d\nu_{\rm A} , \end{equation} 

\noindent with $\nu_{0} = 2.5, 4$. As $\nu_{0} \to \infty$, the integral ($\ref{eq:inte}$) approaches the exact genus curve amplitude. In Figure \ref{fig:app3} we present $A^{(\rm 2D)}$ for $\nu_{0} = 4$ (blue points) and $\nu_{0} = 2.5$ (red squares) from the twenty slices. We also show the mean value of the points as similarly coloured horizontal lines. The exact value of $A^{(\rm 2D)}$ is not relevant to our discussion, the important point is the clear redshift evolution in the blue points, which is due to selecting a large value $\nu_{0} = 4$. For the more conservative choice $\nu_{0} = 2.5$, no redshift evolution is detected relative to the mean value (red points). This indicates that peaks in the range $-2.5 < \nu_{\rm A} < 2.5$ are suitably well represented over the range $0.2 < z < 0.6$ and motivates our choice $-2.5 < \nu_{\rm A} < 2.5$ in the main body of the text.

\section*{Appendix B -- variation of $\Delta$}

Finally, we check that the genus amplitudes extracted from the data are insensitive to the small variations in shell thickness $\Delta$ induced by selecting different cosmological models to infer the distance-redshift relation. Although the genus is a function of the thickness $\Delta$, we will argue that for large $\Delta$ this sensitivity is low and can be neglected. 

\begin{figure}
  \includegraphics[width=0.45\textwidth]{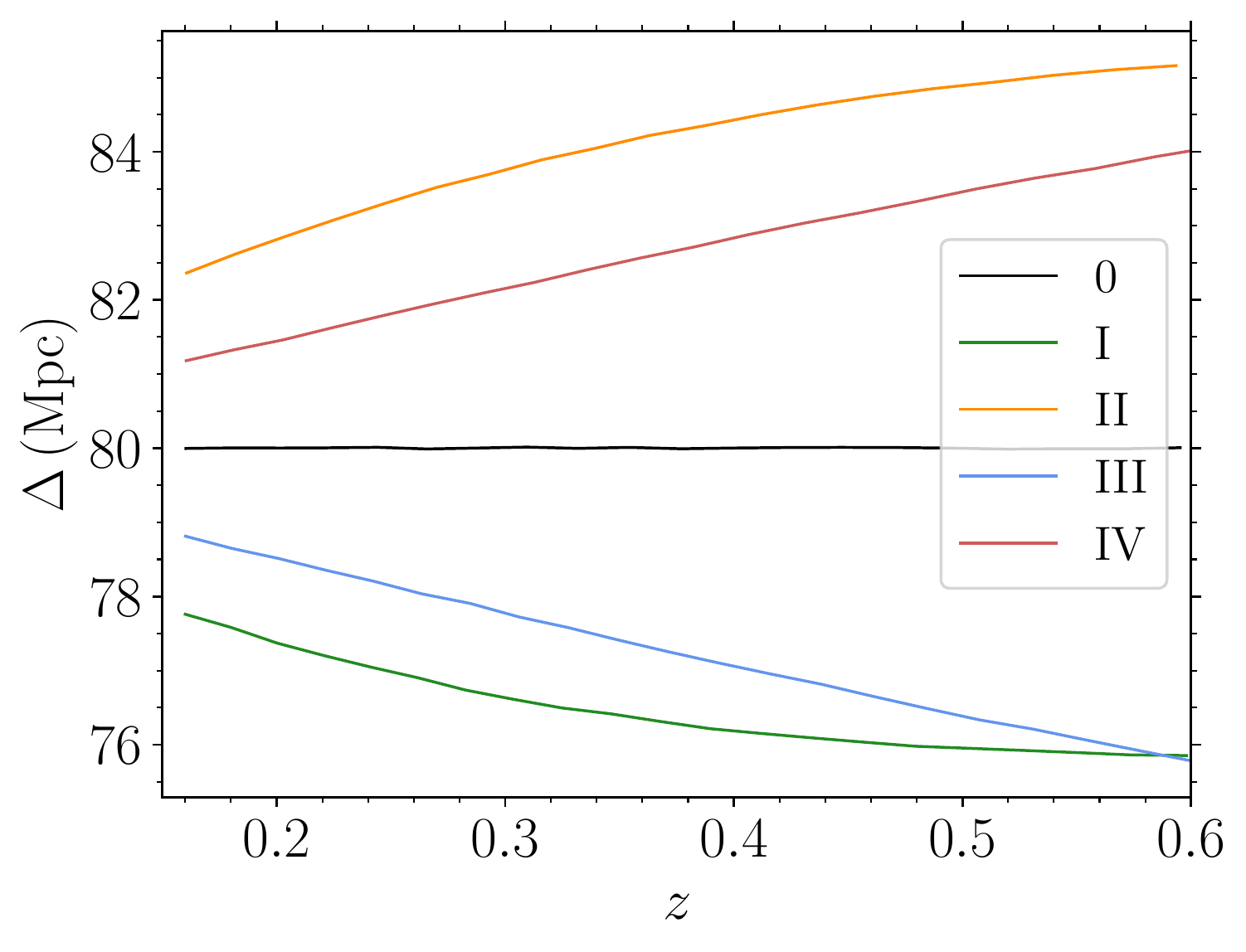}
    \includegraphics[width=0.45\textwidth]{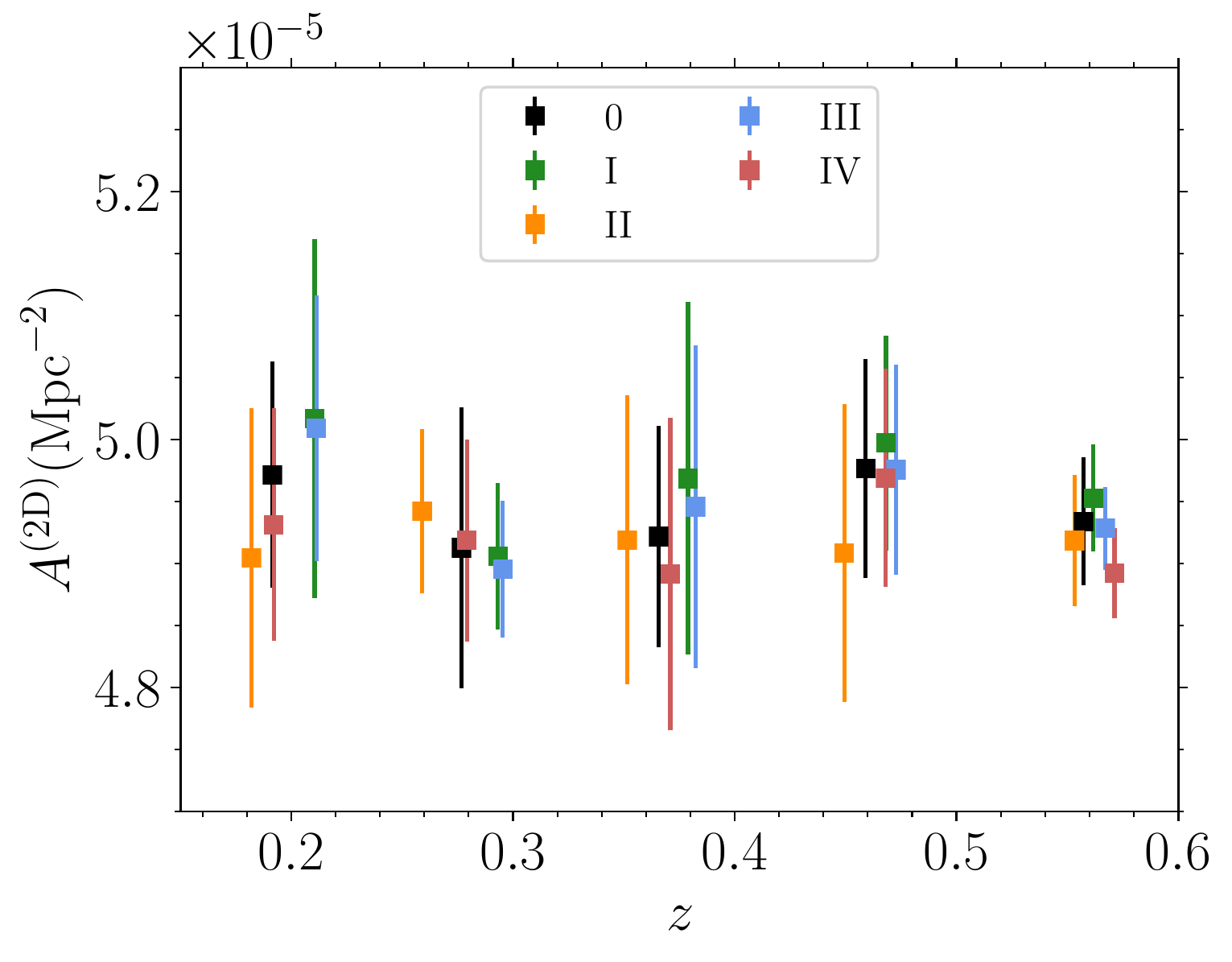}
  \caption{[Top panel] The redshift evolution of the shell thickness $\Delta$, if we use an incorrect cosmological model to infer the redshift limits of the shells. The $0$ model is the fiducial (correct) model of the simulation, and I-IV are incorrect models with parameters given in Table \ref{tab:appb}. [Bottom panel] The measured genus amplitudes of all-sky lightcone shells of the mock galaxy data, using the shell widths presented in the top panel. No systematic evolution of the genus amplitude is found as a result of selecting an incorrect slice thickness.  }
  \label{fig:appb}
\end{figure}

To show this, we take the Horizon Run 4 all-sky mock galaxy lightcone, and use four different cosmological models to fix the redshift boundaries of the shells. For each cosmology we select redshift limits of the shells $z_{\rm min}$, $z_{\rm max}$ such that the comoving distance $d_{\rm cm}(z_{\rm max},\tilde{\Omega}_{\rm m}, \tilde{w}_{\rm de}) - d_{\rm cm}(z_{\rm min},\tilde{\Omega}_{\rm m}, \tilde{w}_{\rm de})  = \tilde{\Delta} = 80 {\rm Mpc}$, where tildes indicate incorrect cosmological parameters that are presented in Table \ref{tab:appb}, with model $0$ being the correct, fiducial cosmology of the simulation. The true values of the slice thicknesses are given by $\Delta = d_{\rm cm}(z_{\rm max},\Omega_{\rm m}, w_{\rm de}) - d_{\rm cm}(z_{\rm min},\Omega_{\rm m}, w_{\rm de})$ with $\Omega_{\rm m} = 0.26$, $w_{\rm de} = -1$.  In Figure \ref{fig:appb} (top panel) we present $\Delta(z)$ as a function of $z$ for each of the cosmological models used to infer the distance redshift relations of the shells. For each cosmological model we have selected redshift limits such that $\tilde{\Delta} = 80 {\rm Mpc}$, independent of redshift, but the true value of $\Delta$ (obtained by using the true cosmology) is evolving.

After fixing the redshift shell limits using the incorrect cosmological models, we proceed to calculate the genus in the twenty data shells using the correct cosmological model. We do this as we wish to isolate the effect of a systematically evolving $\Delta$ thickness. We measure the genus curves and extract the amplitudes. In Figure \ref{fig:appb} (bottom panel) we present the genus amplitude $A^{(\rm 2D)}(\tilde{\Delta})$. For clarity we plot the average and standard deviation of every four shells. One can observe no systematic evolution with redshift for any of the cosmological models selected, and the statistical uncertainty is dominant. This insensitivity is because we are using relatively thick slices $\Delta \sim 80 {\rm Mpc}$; thinner slices will exhibit stronger cosmological parameter sensitivity.

\begin{table}
\begin{center}
 \begin{tabular}{|| c  c  c  ||}
 \hline
 Model \, & \, $\tilde{\Omega}_{\rm m}$ \, & \, $\tilde{w}_{\rm de}$  \\ [0.5ex] 
 \hline\hline
 0 \, & \, $0.26$ \, & \, $-1$   \\ 
 I \, & \, $0.26$ \, & \, $-1.2$ \\ 
 II \, & \, $0.26$ \, & \, $-0.8$ \\ 
 III \, & \, $0.2$ \, & \, $-1$ \\ 
 IV \,  & \, $0.32$ \, & \, $-1$ \\
 \hline
\end{tabular}
\caption{\label{tab:appb}The four models used in Appendix B to test the effect of variable $\Delta$ slice thickness on the genus amplitude. The $0$ model is the fiducial model of the simulation, and yields a constant $\Delta = 80 {\rm Mpc}$ slice thickness.}
\end{center} 
\end{table}

\newpage 

\bibliographystyle{ApJ}
\bibliography{biblio}{}

\begin{thebibliography}{}
\expandafter\ifx\csname natexlab\endcsname\relax\def\natexlab#1{#1}\fi

\bibitem[{{Abazajian} {et~al.}(2009){Abazajian}, {Adelman-McCarthy},
  {Ag{\"u}eros}, {Allam}, {Allende Prieto}, {An}, {Anderson}, {Anderson},
  {Annis}, {Bahcall}, \& et~al.}]{2009ApJS..182..543A}
{Abazajian}, K.~N., {Adelman-McCarthy}, J.~K., {Ag{\"u}eros}, M.~A., {et~al.}
  2009, ApJS., 182, 543

\bibitem[{Adler(1981)}]{Adler}
Adler, R. 1981, The Geometry of Random Fields (Wiley)

\bibitem[{Aghanim {et~al.}(2020)}]{Aghanim:2018eyx}
Aghanim, N., {et~al.} 2020, Astron. Astrophys., 641, A6

\bibitem[{{Alam} {et~al.}(2015){Alam}, {Albareti}, {Allende Prieto}, {Anders},
  {Anderson}, {Anderton}, {Andrews}, {Armengaud}, {Aubourg}, {Bailey}, \&
  et~al.}]{2015ApJS..219...12A}
{Alam}, S., {Albareti}, F.~D., {Allende Prieto}, C., {et~al.} 2015, ApJS., 219,
  12

\bibitem[{Appleby {et~al.}(2018{\natexlab{a}})Appleby, Chingangbam, Park, Hong,
  Kim, \& Ganesan}]{Appleby:2017uvb}
Appleby, S., Chingangbam, P., Park, C., {et~al.} 2018{\natexlab{a}}, ApJ., 858,
  87

\bibitem[{Appleby {et~al.}(2018{\natexlab{b}})Appleby, Chingangbam, Park,
  Yogendran, \& Joby}]{Appleby:2018tzk}
Appleby, S., Chingangbam, P., Park, C., Yogendran, K.~P., \& Joby, P.~K.
  2018{\natexlab{b}}, ApJ., 863, 200

\bibitem[{Appleby {et~al.}(2017)Appleby, Park, Hong, \& Kim}]{Appleby:2017ahh}
Appleby, S., Park, C., Hong, S.~E., \& Kim, J. 2017, ApJ., 836, 45

\bibitem[{Appleby {et~al.}(2018{\natexlab{c}})Appleby, Park, Hong, \&
  Kim}]{Appleby:2018jew}
Appleby, S., Park, C., Hong, S.~E., \& Kim, J. 2018{\natexlab{c}}, ApJ., 853,
  17

\bibitem[{Appleby {et~al.}(2020)Appleby, Park, Hong, Hwang, \&
  Kim}]{Appleby:2020pem}
Appleby, S.~A., Park, C., Hong, S.~E., Hwang, H.~S., \& Kim, J. 2020, ApJ.,
  896, 145

\bibitem[{Beisbart {et~al.}(2001{\natexlab{a}})Beisbart, Buchert, \&
  Wagner}]{Beisbart:2001gk}
Beisbart, C., Buchert, T., \& Wagner, H. 2001{\natexlab{a}}, Physica, A293, 592

\bibitem[{Beisbart {et~al.}(2001{\natexlab{b}})Beisbart, Valdarnini, \&
  Buchert}]{Beisbart:2001vb}
Beisbart, C., Valdarnini, R., \& Buchert, T. 2001{\natexlab{b}}, Astron.
  Astrophys., 379, 412

\bibitem[{Blake {et~al.}(2014)Blake, James, \& Poole}]{Blake:2013noa}
Blake, C., James, J.~B., \& Poole, G.~B. 2014, MNRAS, 437, 2488

\bibitem[{Blake {et~al.}(2011)}]{Blake:2011wn}
Blake, C., {et~al.} 2011, MNRAS, 415, 2892

\bibitem[{{Blanton} {et~al.}(2005){Blanton}, {Schlegel}, {Strauss},
  {Brinkmann}, {Finkbeiner}, {Fukugita}, {Gunn}, {Hogg}, {Ivezi{\'c}}, {Knapp},
  {Lupton}, {Munn}, {Schneider}, {Tegmark}, \& {Zehavi}}]{2005AJ....129.2562B}
{Blanton}, M.~R., {Schlegel}, D.~J., {Strauss}, M.~A., {et~al.} 2005, AJ, 129,
  2562

\bibitem[{Buchert {et~al.}(2017)Buchert, France, \& Steiner}]{Buchert:2017uup}
Buchert, T., France, M.~J., \& Steiner, F. 2017, Class. Quant. Grav., 34,
  094002

\bibitem[{Chingangbam {et~al.}(2017)Chingangbam, Ganesan, Yogendran, \&
  Park}]{Chingangbam:2017sap}
Chingangbam, P., Ganesan, V., Yogendran, K.~P., \& Park, C. 2017, Phys. Lett.,
  B771, 67

\bibitem[{Choi {et~al.}(2010{\natexlab{a}})Choi, Han, \& Kim}]{articleyyc}
Choi, Y.-Y., Han, D.-H., \& Kim, S. 2010{\natexlab{a}}, JKAS, 43, 191

\bibitem[{Choi {et~al.}(2013)Choi, Kim, Rossi, Kim, \& Lee}]{Choi:2013eej}
Choi, Y.-Y., Kim, J., Rossi, G., Kim, S.~S., \& Lee, J.-E. 2013, ApJS., 209, 19

\bibitem[{Choi {et~al.}(2010{\natexlab{b}})Choi, Park, Kim, Gott, Weinberg,
  Vogeley, \& Kim}]{Choi:2010sx}
Choi, Y.-Y., Park, C., Kim, J., {et~al.} 2010{\natexlab{b}}, ApJS., 190, 181

\bibitem[{Codis {et~al.}(2013)Codis, Pichon, Pogosyan, Bernardeau, \&
  Matsubara}]{Codis:2013exa}
Codis, S., Pichon, C., Pogosyan, D., Bernardeau, F., \& Matsubara, T. 2013,
  MNRAS, 435, 531

\bibitem[{Codis {et~al.}(2018)Codis, Pogosyan, \& Pichon}]{Codis:2018niz}
Codis, S., Pogosyan, D., \& Pichon, C. 2018, MNRAS, 479, 973

\bibitem[{{Colless} {et~al.}(2001){Colless}, {Dalton}, {Maddox}, {Sutherland},
  {Norberg}, {Cole}, {Bland-Hawthorn}, {Bridges}, {Cannon}, {Collins}, {Couch},
  {Cross}, {Deeley}, {De Propris}, {Driver}, {Efstathiou}, {Ellis}, {Frenk},
  {Glazebrook}, {Jackson}, {Lahav}, {Lewis}, {Lumsden}, {Madgwick}, {Peacock},
  {Peterson}, {Price}, {Seaborne}, \& {Taylor}}]{2001MNRAS.328.1039C}
{Colless}, M., {Dalton}, G., {Maddox}, S., {et~al.} 2001, MNRAS, 328, 1039

\bibitem[{{de Vaucouleurs} {et~al.}(1991){de Vaucouleurs}, {de Vaucouleurs},
  {Corwin}, {Buta}, {Paturel}, \& {Fouqu{\'e}}}]{1991rc3..book.....D}
{de Vaucouleurs}, G., {de Vaucouleurs}, A., {Corwin}, Jr., H.~G., {et~al.}
  1991, {Third Reference Catalogue of Bright Galaxies, Springer, New York, USA}

\bibitem[{{Doroshkevich}(1970)}]{1970Ap......6..320D}
{Doroshkevich}, A.~G. 1970, Astrophysics, 6, 320

\bibitem[{Dubinski {et~al.}(2004)Dubinski, Huhan, Park, \&
  Humble}]{Dubinski:2003fq}
Dubinski, J., Huhan, K., Park, C., \& Humble, R. 2004, New Astron., 9, 111

\bibitem[{Ducout {et~al.}(2013)Ducout, Bouchet, Colombi, Pogosyan, \&
  Prunet}]{Ducout:2012it}
Ducout, A., Bouchet, F., Colombi, S., Pogosyan, D., \& Prunet, S. 2013, MNRAS,
  429, 2104

\bibitem[{{Falco} {et~al.}(1999){Falco}, {Kurtz}, {Geller}, {Huchra}, {Peters},
  {Berlind}, {Mink}, {Tokarz}, \& {Elwell}}]{1999PASP..111..438F}
{Falco}, E.~E., {Kurtz}, M.~J., {Geller}, M.~J., {et~al.} 1999, PASP, 111, 438

\bibitem[{Feldbrugge {et~al.}(2019)Feldbrugge, van Engelen, van~de Weygaert,
  Pranav, \& Vegter}]{Feldbrugge:2019tal}
Feldbrugge, J., van Engelen, M., van~de Weygaert, R., Pranav, P., \& Vegter, G.
  2019, JCAP, 1909, 052

\bibitem[{Ganesan \& Chingangbam(2017)}]{Ganesan:2016jdk}
Ganesan, V., \& Chingangbam, P. 2017, JCAP, 1706, 023

\bibitem[{Gay {et~al.}(2012)Gay, Pichon, \& Pogosyan}]{Gay:2011wz}
Gay, C., Pichon, C., \& Pogosyan, D. 2012, Phys. Rev., D85, 023011

\bibitem[{Gorski {et~al.}(2005)Gorski, Hivon, Banday, Wandelt, Hansen,
  Reinecke, \& Bartelman}]{Gorski:2004by}
Gorski, K.~M., Hivon, E., Banday, A.~J., {et~al.} 2005, ApJ., 622, 759

\bibitem[{{Gott} {et~al.}(1990){Gott}, {Park}, {Juszkiewicz}, {Bies},
  {Bennett}, {Bouchet}, \& {Stebbins}}]{1990ApJ...352....1G}
{Gott}, J.~Richard, I., {Park}, C., {Juszkiewicz}, R., {et~al.} 1990, ApJ.,
  352, 1

\bibitem[{Gott {et~al.}(2009)Gott, Choi, Park, \& Kim}]{Gott:2008kk}
Gott, J.~R., Choi, Y.-Y., Park, C., \& Kim, J. 2009, ApJ., 695, L45

\bibitem[{Gott {et~al.}(1986)Gott, Dickinson, \& Melott}]{Gott:1986uz}
Gott, J.~R., Dickinson, M., \& Melott, A.~L. 1986, ApJ., 306, 341

\bibitem[{{Gott} {et~al.}(1987){Gott}, {Weinberg}, \&
  {Melott}}]{1987ApJ...319....1G}
{Gott}, J.~R., {Weinberg}, D.~H., \& {Melott}, A.~L. 1987, ApJ., 319, 1

\bibitem[{{Gott} {et~al.}(1992){Gott}, {Mao}, {Park}, \&
  {Lahav}}]{1992ApJ...385...26G}
{Gott}, III, J.~R., {Mao}, S., {Park}, C., \& {Lahav}, O. 1992, ApJ., 385, 26

\bibitem[{Gott {et~al.}(2008)Gott, Hambrick, Vogeley, Kim, Park, Choi, Cen, \&
  Ostriker}]{Gott:2006yy}
Gott, J. R.~I., Hambrick, D.~C., Vogeley, M.~S., {et~al.} 2008, ApJ., 675, 16

\bibitem[{Hamilton {et~al.}(1986)Hamilton, Gott, \& Weinberg}]{Hamilton:1986}
Hamilton, J. S.~A., Gott, J.~R., \& Weinberg, D. 1986, {\apj}, 309, 1

\bibitem[{Hikage {et~al.}(2008)Hikage, Coles, Grossi, Moscardini, Dolag,
  Branchini, \& Matarrese}]{10.1111/j.1365-2966.2008.12944.x}
Hikage, C., Coles, P., Grossi, M., {et~al.} 2008, MNRAS, 385, 1613

\bibitem[{Hikage {et~al.}(2006)Hikage, Komatsu, \& Matsubara}]{Hikage:2006fe}
Hikage, C., Komatsu, E., \& Matsubara, T. 2006, ApJ., 653, 11

\bibitem[{Hikage {et~al.}(2001)Hikage, Taruya, \& Suto}]{Hikage_2001}
Hikage, C., Taruya, A., \& Suto, Y. 2001, ApJ., 556, 641

\bibitem[{Hikage {et~al.}(2002)Hikage, Suto, Kayo, Taruya, Matsubara, Vogeley,
  Hoyle, Gott, \& Brinkmann}]{Hikage:2002ki}
Hikage, C., Suto, Y., Kayo, I., {et~al.} 2002, Publ. Astron. Soc. Jap., 54, 707

\bibitem[{Hikage {et~al.}(2003)Hikage, Schmalzing, Buchert, Suto, Kayo, Taruya,
  Vogeley, Hoyle, Gott, \& Brinkmann}]{Hikage:2003fc}
Hikage, C., Schmalzing, J., Buchert, T., {et~al.} 2003, Publ. Astron. Soc.
  Jap., 55, 911

\bibitem[{Hong {et~al.}(2020)Hong, Jeong, Hwang, Kim, Hong, Park, Dey,
  Milosavljevic, Gebhardt, \& Lee}]{10.1093/mnras/staa566}
Hong, S., Jeong, D., Hwang, H.~S., {et~al.} 2020, MNRAS, 493, 5972

\bibitem[{Hong {et~al.}(2016)Hong, Park, \& Kim}]{Hong:2016hsd}
Hong, S.~E., Park, C., \& Kim, J. 2016, ApJ., 823, 103

\bibitem[{Howlett {et~al.}(2015)Howlett, Ross, Samushia, Percival, \&
  Manera}]{Howlett:2014opa}
Howlett, C., Ross, A., Samushia, L., Percival, W., \& Manera, M. 2015, MNRAS,
  449, 848

\bibitem[{James {et~al.}(2009)James, Colless, Lewis, \&
  Peacock}]{10.1111/j.1365-2966.2008.14358.x}
James, J.~B., Colless, M., Lewis, G.~F., \& Peacock, J.~A. 2009, MNRAS, 394,
  454

\bibitem[{Jiang {et~al.}(2008)Jiang, Jing, Faltenbacher, Lin, \&
  Li}]{Jiang:2007xd}
Jiang, C.~Y., Jing, Y.~P., Faltenbacher, A., Lin, W.~P., \& Li, C. 2008, ApJ.,
  675, 1095

\bibitem[{Joby {et~al.}(2019)Joby, Chingangbam, Ghosh, Ganesan, \&
  Ravikumar}]{K.:2018wpn}
Joby, P.~K., Chingangbam, P., Ghosh, T., Ganesan, V., \& Ravikumar, C.~D. 2019,
  JCAP, 1901, 009

\bibitem[{Kapahtia {et~al.}(2019)Kapahtia, Chingangbam, \&
  Appleby}]{Kapahtia:2019ksk}
Kapahtia, A., Chingangbam, P., \& Appleby, S. 2019, JCAP, 09, 053

\bibitem[{Kapahtia {et~al.}(2018)Kapahtia, Chingangbam, Appleby, \&
  Park}]{Kapahtia:2017qrg}
Kapahtia, A., Chingangbam, P., Appleby, S., \& Park, C. 2018, JCAP, 1810, 011

\bibitem[{Kim {et~al.}(2015)Kim, Park, L'Huillier, \& Hong}]{Kim:2015yma}
Kim, J., Park, C., L'Huillier, B., \& Hong, S.~E. 2015, JKAS, 48, 213

\bibitem[{Kim {et~al.}(2014)Kim, Choi, Kim, Kim, Lee, Shin, \&
  Kim}]{Kim:2014axe}
Kim, Y.-R., Choi, Y.-Y., Kim, S.~S., {et~al.} 2014, ApJS., 212, 22

\bibitem[{Kitaura {et~al.}(2015)Kitaura, Gil-Marín, Scóccola, Chuang,
  Müller, Yepes, \& Prada}]{10.1093/mnras/stv645}
Kitaura, F.-S., Gil-Marín, H., Scóccola, C.~G., {et~al.} 2015, MNRAS, 450,
  1836

\bibitem[{{Kitaura} {et~al.}(2014){Kitaura}, {Yepes}, \&
  {Prada}}]{2014MNRAS.439L..21K}
{Kitaura}, F.-S., {Yepes}, G., \& {Prada}, F. 2014, MNRAS, 439, L21

\bibitem[{{Kitaura} {et~al.}(2016){Kitaura}, {Rodr{\'{\i}}guez-Torres},
  {Chuang}, {Zhao}, {Prada}, {Gil-Mar{\'{\i}}n}, {Guo}, {Yepes}, {Klypin},
  {Sc{\'o}ccola}, {Tinker}, {McBride}, {Reid}, {S{\'a}nchez},
  {Salazar-Albornoz}, {Grieb}, {Vargas-Magana}, {Cuesta}, {Neyrinck},
  {Beutler}, {Comparat}, {Percival}, \& {Ross}}]{2016MNRAS.456.4156K}
{Kitaura}, F.-S., {Rodr{\'{\i}}guez-Torres}, S., {Chuang}, C.-H., {et~al.}
  2016, MNRAS, 456, 4156

\bibitem[{Kraljic {et~al.}(2020)}]{Kraljic:2019acs}
Kraljic, K., {et~al.} 2020, MNRAS, 491, 4294

\bibitem[{L'Huillier {et~al.}(2014)L'Huillier, Park, \&
  Kim}]{L'Huillier:2014dpa}
L'Huillier, B., Park, C., \& Kim, J. 2014, New Astron., 30, 79

\bibitem[{Li {et~al.}(2017)Li, Park, Sabiu, Park, Cheng, Kim, \&
  Hong}]{Li:2017nzs}
Li, X.-D., Park, C., Sabiu, C.~G., {et~al.} 2017, ApJ, 844, 91

\bibitem[{Li {et~al.}(2016)Li, Park, Sabiu, Park, Weinberg, Schneider, Kim, \&
  Hong}]{Li:2016wbl}
Li, X.-D., Park, C., Sabiu, C.~G., {et~al.} 2016, ApJ., 832, 103

\bibitem[{Li {et~al.}(2018)Li, Sabiu, Park, Wang, Zhao, Park, Shafieloo, Kim,
  \& Hong}]{Li:2018nlh}
Li, X.-D., Sabiu, C.~G., Park, C., {et~al.} 2018, ApJ, 856, 88

\bibitem[{Matsubara(1994{\natexlab{a}})}]{Matsubara:1994wn}
Matsubara, T. 1994{\natexlab{a}}, ApJ., 434, L43

\bibitem[{Matsubara(1994{\natexlab{b}})}]{Matsubara:1994we}
Matsubara, T. 1994{\natexlab{b}}, arXiv:astro-ph/9501076

\bibitem[{{Matsubara}(1996)}]{1996ApJ...457...13M}
{Matsubara}, T. 1996, ApJ., 457, 13

\bibitem[{{Matsubara}(2000)}]{2000astro.ph..6269M}
{Matsubara}, T. 2000, astro-ph/0006269

\bibitem[{Matsubara \& Suto(1996)}]{Matsubara:1995dv}
Matsubara, T., \& Suto, Y. 1996, ApJ., 460, 51

\bibitem[{Matsubara \& Yokoyama(1996)}]{Matsubara:1995ns}
Matsubara, T., \& Yokoyama, J. 1996, ApJ., 463, 409

\bibitem[{Mecke {et~al.}(1994)Mecke, Buchert, \& Wagner}]{Mecke:1994ax}
Mecke, K.~R., Buchert, T., \& Wagner, H. 1994, Astron. Astrophys., 288, 697

\bibitem[{{Melott} {et~al.}(1989){Melott}, {Cohen}, {Hamilton}, {Gott}, \&
  {Weinberg}}]{1989ApJ...345..618M}
{Melott}, A.~L., {Cohen}, A.~P., {Hamilton}, A.~J.~S., {Gott}, J.~R., \&
  {Weinberg}, D.~H. 1989, ApJ., 345, 618

\bibitem[{{Melott} {et~al.}(1988){Melott}, {Weinberg}, \&
  {Gott}}]{1988ApJ...328...50M}
{Melott}, A.~L., {Weinberg}, D.~H., \& {Gott}, J.~R. 1988, ApJ., 328, 50

\bibitem[{{Padmanabhan} {et~al.}(2008){Padmanabhan}, {Schlegel}, {Finkbeiner},
  {Barentine}, {Blanton}, {Brewington}, {Gunn}, {Harvanek}, {Hogg},
  {Ivezi{\'c}}, {Johnston}, {Kent}, {Kleinman}, {Knapp}, {Krzesinski}, {Long},
  {Neilsen}, {Nitta}, {Loomis}, {Lupton}, {Roweis}, {Snedden}, {Strauss}, \&
  {Tucker}}]{2008ApJ...674.1217P}
{Padmanabhan}, N., {Schlegel}, D.~J., {Finkbeiner}, D.~P., {et~al.} 2008, ApJ.,
  674, 1217

\bibitem[{{Parihar} {et~al.}(2014){Parihar}, {Vogeley}, {Gott}, {Choi}, {Kim},
  {Kim}, {Speare}, {Brownstein}, \& {Brinkmann}}]{2014ApJ...796...86P}
{Parihar}, P., {Vogeley}, M.~S., {Gott}, III, J.~R., {et~al.} 2014, ApJ., 796,
  86

\bibitem[{{Park} \& {Gott}(1991)}]{1991ApJ...378..457P}
{Park}, C., \& {Gott}, J.~R. 1991, ApJ., 378, 457

\bibitem[{{Park} {et~al.}(2001){Park}, {Gott}, \& {Choi}}]{2001ApJ...553...33P}
{Park}, C., {Gott}, J.~R., \& {Choi}, Y.~J. 2001, ApJ., 553, 33

\bibitem[{{Park} {et~al.}(1992){Park}, {Gott}, {Melott}, \&
  {Karachentsev}}]{1992ApJ...387....1P}
{Park}, C., {Gott}, J.~R., {Melott}, A.~L., \& {Karachentsev}, I.~D. 1992,
  ApJ., 387, 1

\bibitem[{{Park} {et~al.}(2005){Park}, {Kim}, \& {Gott}}]{2005ApJ...633....1P}
{Park}, C., {Kim}, J., \& {Gott}, J.~R. 2005, ApJ., 633, 1

\bibitem[{Park \& Kim(2010)}]{Park:2009ja}
Park, C., \& Kim, Y.-R. 2010, ApJ., 715, L185

\bibitem[{Park {et~al.}(2005)Park, Choi, Vogeley, Gott, Kim, Hikage, Mastubara,
  Park, Suto, \& Weinberg}]{Park:2005fk}
Park, C., Choi, Y.-Y., Vogeley, M., {et~al.} 2005, ApJ., 633, 11

\bibitem[{Park {et~al.}(2013)Park, Pranav, Chingangbam, van~de Weygaert, Jones,
  Vegter, Kim, Hidding, \& Hellwing}]{Park:2013dga}
Park, C., Pranav, P., Chingangbam, P., {et~al.} 2013, JKAS, 46, 125

\bibitem[{Park {et~al.}(2019)Park, Park, Sabiu, Li, Hong, Kim, Tonegawa, \&
  Zheng}]{Park:2019mvn}
Park, H., Park, C., Sabiu, C.~G., {et~al.} 2019, ApJ, 881, 146

\bibitem[{Petri {et~al.}(2013)Petri, Haiman, Hui, May, \&
  Kratochvil}]{Petri:2013ffb}
Petri, A., Haiman, Z., Hui, L., May, M., \& Kratochvil, J.~M. 2013, Phys. Rev.,
  D88, 123002

\bibitem[{Pogosyan {et~al.}(2009)Pogosyan, Gay, \& Pichon}]{Pogosyan:2009rg}
Pogosyan, D., Gay, C., \& Pichon, C. 2009, Phys. Rev., D80, 081301

\bibitem[{Pranav {et~al.}(2019{\natexlab{a}})Pranav, Adler, Buchert,
  Edelsbrunner, Jones, Schwartzman, Wagner, \& van~de
  Weygaert}]{Pranav:2018lox}
Pranav, P., Adler, R.~J., Buchert, T., {et~al.} 2019{\natexlab{a}}, Astron.
  Astrophys., 627, A163

\bibitem[{Pranav {et~al.}(2017)Pranav, Edelsbrunner, van~de Weygaert, Vegter,
  Kerber, Jones, \& Wintraecken}]{Pranav:2016gwr}
Pranav, P., Edelsbrunner, H., van~de Weygaert, R., {et~al.} 2017, MNRAS, 465,
  4281

\bibitem[{Pranav {et~al.}(2019{\natexlab{b}})Pranav, van~de Weygaert, Vegter,
  Jones, Adler, Feldbrugge, Park, Buchert, \& Kerber}]{Pranav:2018pnu}
Pranav, P., van~de Weygaert, R., Vegter, G., {et~al.} 2019{\natexlab{b}},
  MNRAS, 485, 4167

\bibitem[{Reid {et~al.}(2016)}]{Reid:2015gra}
Reid, B., {et~al.} 2016, MNRAS, 455, 1553

\bibitem[{{Rodr{\'{\i}}guez-Torres} {et~al.}(2016){Rodr{\'{\i}}guez-Torres},
  {Chuang}, {Prada}, {Guo}, {Klypin}, {Behroozi}, {Hahn}, {Comparat}, {Yepes},
  {Montero-Dorta}, {Brownstein}, {Maraston}, {McBride}, {Tinker},
  {Gottl{\"o}ber}, {Favole}, {Shu}, {Kitaura}, {Bolton}, {Scoccimarro},
  {Samushia}, {Schlegel}, {Schneider}, \& {Thomas}}]{2016MNRAS.460.1173R}
{Rodr{\'{\i}}guez-Torres}, S.~A., {Chuang}, C.-H., {Prada}, F., {et~al.} 2016,
  MNRAS, 460, 1173

\bibitem[{Ross {et~al.}(2015)Ross, Samushia, Howlett, Percival, Burden, \&
  Manera}]{Ross:2014qpa}
Ross, A.~J., Samushia, L., Howlett, C., {et~al.} 2015, MNRAS, 449, 835

\bibitem[{Ryden {et~al.}(1989)Ryden, Melott, Craig, Gott, Weinberg, Scherrer,
  Bhavsar, \& Miller}]{Ryden:1988rk}
Ryden, B.~S., Melott, A.~L., Craig, D.~A., {et~al.} 1989, ApJ., 340, 647

\bibitem[{Saunders {et~al.}(2000)}]{Saunders:2000af}
Saunders, W., {et~al.} 2000, MNRAS, 317, 55

\bibitem[{Schmalzing \& Buchert(1997)}]{Schmalzing:1997aj}
Schmalzing, J., \& Buchert, T. 1997, Astrophys. J., 482, L1

\bibitem[{Schmalzing \& Gorski(1998)}]{Schmalzing:1997uc}
Schmalzing, J., \& Gorski, K.~M. 1998, MNRAS, 297, 355

\bibitem[{{Shin} {et~al.}(2017){Shin}, {Kim}, {Pichon}, {Jeong}, \&
  {Park}}]{2017ApJ...843...73S}
{Shin}, J., {Kim}, J., {Pichon}, C., {Jeong}, D., \& {Park}, C. 2017, ApJ, 843,
  73

\bibitem[{Shivshankar {et~al.}(2015)Shivshankar, Pranav, Natarajan, van~de
  Weygaert, Bos, \& Rieder}]{Shivshankar:2015aza}
Shivshankar, N., Pranav, P., Natarajan, V., {et~al.} 2015, Comput. Graphics, 1,
  1

\bibitem[{Sousbie {et~al.}(2011)Sousbie, Pichon, \&
  Kawahara}]{10.1111/j.1365-2966.2011.18395.x}
Sousbie, T., Pichon, C., \& Kawahara, H. 2011, MNRAS, 414, 384

\bibitem[{Sullivan {et~al.}(2019)Sullivan, Wiegand, \&
  Eisenstein}]{Sullivan:2017mhr}
Sullivan, J.~M., Wiegand, A., \& Eisenstein, D.~J. 2019, MNRAS, 485, 1708

\bibitem[{Tegmark {et~al.}(2004)}]{Tegmark:2003uf}
Tegmark, M., {et~al.} 2004, ApJ., 606, 702

\bibitem[{Tomita(1986)}]{10.1143/PTP.76.952}
Tomita, H. 1986, Progress of Theoretical Physics, 76, 952

\bibitem[{Tonegawa {et~al.}(2020)Tonegawa, Park, Zheng, Park, Hong, Hwang, \&
  Kim}]{article_moto}
Tonegawa, M., Park, C., Zheng, Y., {et~al.} 2020, ApJ., 897, 17

\bibitem[{van~de Weygaert {et~al.}(2011)}]{vandeWeygaert:2011ip}
van~de Weygaert, R., {et~al.} 2011, arXiv:1110.5528

\bibitem[{Wang {et~al.}(2015)Wang, Xu, Wu, Chen, Wang, Kim, Park, Lee, \&
  Cen}]{Wang:2015eua}
Wang, Y., Xu, Y., Wu, F., {et~al.} 2015, PoS, AASKA14, 033

\bibitem[{{Weinberg} {et~al.}(1987){Weinberg}, {Gott}, \&
  {Melott}}]{1987ApJ...321....2W}
{Weinberg}, D.~H., {Gott}, J.~R., \& {Melott}, A.~L. 1987, ApJ., 321, 2

\bibitem[{Wiegand {et~al.}(2014)Wiegand, Buchert, \&
  Ostermann}]{Wiegand:2013xfa}
Wiegand, A., Buchert, T., \& Ostermann, M. 2014, MNRAS, 443, 241

\bibitem[{Wiegand \& Eisenstein(2017)}]{Wiegand:2016ezl}
Wiegand, A., \& Eisenstein, D.~J. 2017, MNRAS, 467, 3361

\bibitem[{{York} {et~al.}(2000){York}, {Adelman}, {Anderson}, {Anderson},
  {Annis}, {Bahcall}, {Bakken}, {Barkhouser}, {Bastian}, {Berman}, {Boroski},
  {Bracker}, {Briegel}, {Briggs}, {Brinkmann}, {Brunner}, {Burles}, {Carey},
  {Carr}, {Castander}, {Chen}, {Colestock}, {Connolly}, {Crocker}, {Csabai},
  {Czarapata}, {Davis}, {Doi}, {Dombeck}, {Eisenstein}, {Ellman}, {Elms},
  {Evans}, {Fan}, {Federwitz}, {Fiscelli}, {Friedman}, {Frieman}, {Fukugita},
  {Gillespie}, {Gunn}, {Gurbani}, {de Haas}, {Haldeman}, {Harris}, {Hayes},
  {Heckman}, {Hennessy}, {Hindsley}, {Holm}, {Holmgren}, {Huang}, {Hull},
  {Husby}, {Ichikawa}, {Ichikawa}, {Ivezi{\'c}}, {Kent}, {Kim}, {Kinney},
  {Klaene}, {Kleinman}, {Kleinman}, {Knapp}, {Korienek}, {Kron}, {Kunszt},
  {Lamb}, {Lee}, {Leger}, {Limmongkol}, {Lindenmeyer}, {Long}, {Loomis},
  {Loveday}, {Lucinio}, {Lupton}, {MacKinnon}, {Mannery}, {Mantsch}, {Margon},
  {McGehee}, {McKay}, {Meiksin}, {Merelli}, {Monet}, {Munn}, {Narayanan},
  {Nash}, {Neilsen}, {Neswold}, {Newberg}, {Nichol}, {Nicinski}, {Nonino},
  {Okada}, {Okamura}, {Ostriker}, {Owen}, {Pauls}, {Peoples}, {Peterson},
  {Petravick}, {Pier}, {Pope}, {Pordes}, {Prosapio}, {Rechenmacher}, {Quinn},
  {Richards}, {Richmond}, {Rivetta}, {Rockosi}, {Ruthmansdorfer}, {Sandford},
  {Schlegel}, {Schneider}, {Sekiguchi}, {Sergey}, {Shimasaku}, {Siegmund},
  {Smee}, {Smith}, {Snedden}, {Stone}, {Stoughton}, {Strauss}, {Stubbs},
  {SubbaRao}, {Szalay}, {Szapudi}, {Szokoly}, {Thakar}, {Tremonti}, {Tucker},
  {Uomoto}, {Vanden Berk}, {Vogeley}, {Waddell}, {Wang}, {Watanabe},
  {Weinberg}, {Yanny}, {Yasuda}, \& {SDSS Collaboration}}]{2000AJ....120.1579Y}
{York}, D.~G., {Adelman}, J., {Anderson}, Jr., J.~E., {et~al.} 2000, AJ, 120,
  1579

\bibitem[{Zhang {et~al.}(2019{\natexlab{a}})Zhang, Huang, \&
  Li}]{Zhang:2018jfu}
Zhang, X., Huang, Q.-G., \& Li, X.-D. 2019{\natexlab{a}}, MNRAS, 483, 1655

\bibitem[{Zhang {et~al.}(2010)Zhang, Springel, \& Yang}]{Zhang:2010tha}
Zhang, Y., Springel, V., \& Yang, X. 2010, AJ, 722, 812

\bibitem[{Zhang {et~al.}(2019{\natexlab{b}})Zhang, Gu, Wang, Li, Sabiu, Park,
  Miao, Luo, Fang, \& Li}]{Zhang:2019jsu}
Zhang, Z., Gu, G., Wang, X., {et~al.} 2019{\natexlab{b}}, ApJ, 878, 137

\bibitem[{Zunckel {et~al.}(2011)Zunckel, Gott, \&
  Lunnan}]{doi:10.1111/j.1365-2966.2010.18015.x}
Zunckel, C., Gott, III, J.~R., \& Lunnan, R. 2011, MNRAS, 412, 1401

\end{thebibliography}

\end{document}